\documentclass[reprint,amsmath,amssymb,aps]{revtex4-2}
\usepackage{graphicx}
\usepackage{bm}
\usepackage{tabstackengine}
\usepackage{eucal}
\usepackage{bm}
\usepackage{booktabs}
\usepackage{xcolor}
\usepackage{amsmath, graphics, epsfig, color, verbatim, amsfonts, tensor}
\usepackage{hhline}
\usepackage{multirow}
\usepackage{siunitx}
\usepackage{tabularx}
\makeatletter
\renewcommand*\env@matrix[1][c]{\hskip -\arraycolsep
  \let\@ifnextchar\new@ifnextchar
  \array{*\c@MaxMatrixCols #1}}
\makeatother

\newcommand{\lntxn}{Ln$_3$XN}

\begin{document}
\title{
The stability and topological behaviors in lanthanide antiperovskite nitrides: a high-throughput study}
\author{Shuxiang Zhou$^1$}
\email{shuxiang.zhou@inl.gov}
\author{Kevin Vallejo$^1$} 
\author{Krzysztof Gofryk$^1$}
\affiliation{$^1$Center for Quantum Actinide Science and Technology, Idaho National Laboratory, Idaho Falls, ID 83415, USA}

\begin{abstract}
Antiperovskite (APV) nitrides exhibit a diverse range of electronic properties, including superconductivity, magnetic effects, and nontrivial topological behaviors. In this study, we propose a new family of APV nitrides by incorporating 4$f$-electron metals, known for strong electron correlations, localized magnetic moments, and spin-orbit coupling, to further explore the unique properties of APVs. A high-throughput density functional theory (DFT) calculation was utilized to identify stable lanthanide APV nitride compounds. To address the challenge of strong electron correlation, we developed a double-screening framework that assumes either a fully itinerant or localized nature of the $f$-electrons during calculations. Using this approach, we systematically identified 37 stable lanthanide APV nitride compounds from both thermodynamic and dynamical perspectives. Furthermore, we report nontrivial topological behaviors observed among these stable lanthanide APV nitride compounds, as computed by DFT. Notably, Dirac and semi-Dirac cones are observed near the Fermi level for Er$_3$TlN. This study opens a pathway to investigate lanthanide APVs, revealing potential novel physical properties by leveraging the rich physics of both APVs and $f$-electrons.

\end{abstract}

\maketitle

\section{\label{sec:intro}Introduction}

\begin{figure*}[t]
    \centering
    \includegraphics[width=0.85\linewidth]{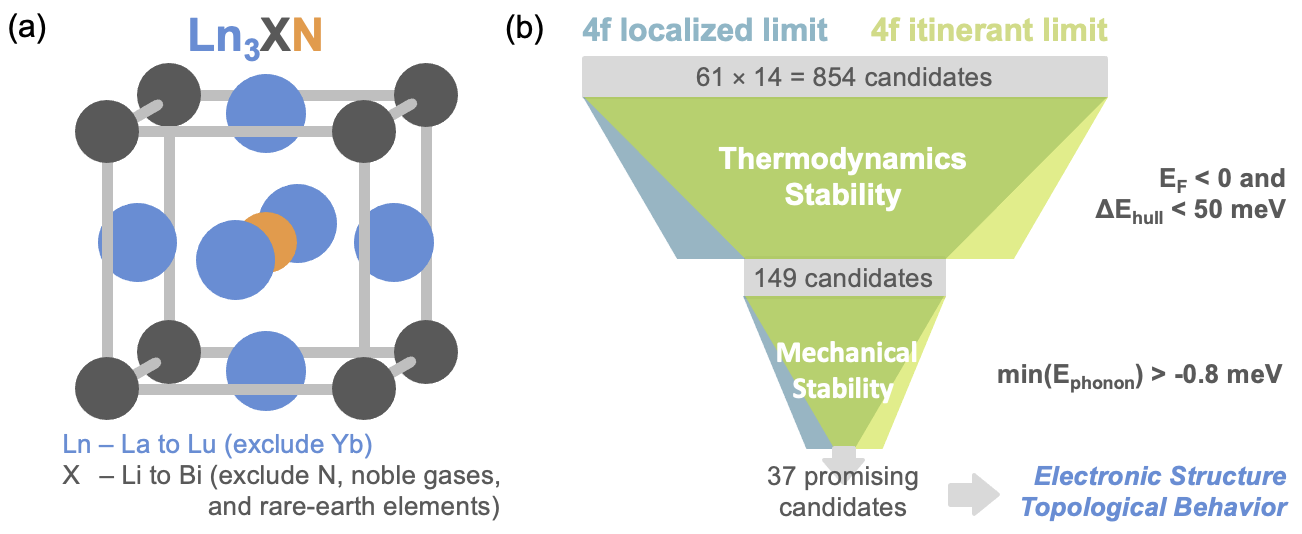}
    \caption{(a) Crystal structure of lanthanide antiperovskite nitrides \lntxn . (b) Schematic representation of the screening funnel used to identify promising \lntxn{} candidates. The pseudo-potentials of \textit{Ln} and \textit{Ln\_3}, representing the itinerant and localized limits of 4\textit{f} electrons respectively, were used independently. A candidate passes the screening only if both itinerant and localized conditions are met.}
    \label{fig:framework}
\end{figure*}

Antiperovskite (APV) nitrides, represented by the formula M$_3$XN, where M typically includes alkaline metals, alkaline-earth metals, and transition metals, and X is a $p$-group element, have attracted substantial research interests due to their rich physical properties, including superconductivity (Ni$_3$CuN \cite{he_cunni3_2013} and Ni$_3$ZnN \cite{masatomo_new_2009}),  magnetostriction (Mn$_3$CuN \cite{asano_magnetostriction_2008}), complex magnetism (Mn$_3$ZnN \cite{masrour_magnetic_2021}), anamolous Hall effect \cite{boldrin_anomalous_2019, gurung_anomalous_2019}, magnetocaloric effect (Fe$_3$AlN \cite{amraoui_magnetic_2021}), negative thermal expansion (Mn$_3$Cu$_{1-x}$Ge$_x$N \cite{iikubo_local_2008} and Mn$_3$Zn$_{1-x}$Sn$_x$N \cite{sun_negative_2010, takenaka_tailoring_2012}), optoelectronic and thermoelectric properties \cite{tahir_farid_optoelectronic_2021}, high ionic conductivity \cite{xia_antiperovskite_2022, yu_optimized_2021, kim_multivalent_2021}, Dirac semimetallic behavior (Cu$_3$PdN  \cite{yu_topological_2015}), and topological insulating properties (strained Ca$_3$BiN \cite{sun_new_2010,goh_coemergence_2018} and Sr$_3$BiN \cite{sun_new_2010}). 
The remarkable properties of APV nitrides can be attributed to their unique perovskite crystal structure and the flexibility of their chemical composition, particularly the variety of cations that can occupy the M site.

To further expand the materials within the APV nitrides family, it is natural to explore cations beyond the $d$-block transition metals, venturing into the realm of 4$f$-electron rare-earth metals, or lanthanides. The incorporation of lanthanides introduces the intriguing characteristics of $f$-orbitals, known for strong electron correlation, localized magnetic moments, and strong spin-orbit coupling, which hosts fascinating quantum phenomena, including heavy fermion behavior, quantum criticality, and unconventional superconductivity \cite{coleman_heavy_2007, si_heavy_2010}. Combining the rich physics inherent in both APVs and $f$-electrons, lanthanide APV nitrides also hold immense potential for uncovering novel physical properties. Recent studies have identified several lanthanide-based APVs with notable properties, such as noncollinear magnetic structures in Ce$_3$InN \cite{gabler_magnetic_2008}, Dirac nodes near the Fermi level in Yb$_3$SnO and Yb$_3$PbO \cite{pertsova_computational_2019}, magnetocaloric effects in Eu$_3$SnO \cite{guillou_metamagnetic_2020}, and the coexistence of ferromagnetism and heavy fermion behaviors in Gd$_3$SnC \cite{bang_antiperovskite_2021}. Despite these advancements, some fundamental questions for lanthanide APVs, such as the identification of stable compounds, are yet to be fully answered due to the lack of systematic studies.

High-throughput computational techniques based on density functional theory (DFT) have emerged as an effective and systematic approach for screening materials with targeted properties \cite{jain_high-throughput_2011}. Owing to the wide variety of cations that can occupy the M site, high-throughput computation has been widely utilized to screen APVs for diverse properties, including magnetic behavior in 3$d$ transition metal APVs \cite{singh_high-throughput_2018}, structure distortion in alkaline-earth metal APVs \cite{mochizuki_theoretical_2020}, superconductivity \cite{hoffmann_superconductivity_2022}, spin Hall effect in 4$d$ and 5$d$ transition metal APVs \cite{xu_high-throughput_2025}, and solid-state electrolytes in alkaline metal APVs \cite{lin_compositional_2025}. Despite the growing success of high-throughput computational screening, its application to lanthanide-based materials remains relatively limited due to the inherent challenges of modeling strongly correlated $f$-electrons within conventional DFT. Standard approximations to the exchange-correlation functional, such as the local density approximation (LDA) and generalized gradient approximation (GGA), are well known to inadequately capture the strong electron correlations characteristic of $f$-electron systems. To address these limitations, several advanced methods have been developed, including hybrid functionals (HF) \cite{becke_new_1993}, self-interaction correction (SIC) \cite{perdew_self-interaction_1981}, DFT plus Hubbard $U$ (DFT+$U$) \cite{anisimov_band_1991, dudarev_effect_1997}, and DFT combined with dynamical mean-field theory (DFT+DMFT) \cite{georges_dynamical_1996, kotliar_electronic_2006}. While these approaches offer improved accuracy in describing localized electrons, their direct application in high-throughput frameworks remains challenging. This is primarily due to their computational cost (as in HF, SIC, and DMFT) or the need for material-specific parameters (e.g., the $U$ value in DFT+$U$), which limits their scalability to the large number of calculations required for high-throughput screening.

In this work, we propose a high-throughput double-screening framework designed specifically for f-electron systems, in the complexity level of conventional DFT. The core concept of our framework is straightforward: the screening process is performed twice, separately assuming either fully itinerant or localized nature of the $f$-electrons, hence this framework is called double-screening. We demonstrate this double-screening framework to filter stable lanthanide APV nitride compounds, from both thermodynamical and dynamical perspectives. Furthermore, for these screened stable lanthanide APV nitride compounds, we report the observation for nontrivial topological behaviors from DFT.   

\section{\label{sec:method}Methods}

\textbf{High-throughput calculation.} In this work, we employed a high-throughput computational framework to screen thermodynamically and dynamically stable lanthanide APV nitrides, denoted as \lntxn. Here, Ln represents a lanthanide element from La to Lu, excluding Yb (14 elements in total), and X represents any element from Li to Bi, excluding N, noble gases, and lanthanide elements (61 elements in total). The thermodynamical stability of each \lntxn{} compound was evaluated using the formation energy $E_F$ and the energy above the convex hull $\Delta E_\mathrm{hull}$.
The formation energy describes the energy absorbed when a compound is synthesized from its constituent elements and is defined as
$E_F ($\lntxn$)=E($\lntxn$) - 3E(\mathrm{Ln}) - E(\mathrm{X}) - E(\mathrm{N})$,
where $E$ denotes the total energy, and Ln, X, and N represent their most stable elemental phases.
$\Delta E_\mathrm{hull}$ is the energy difference between the compound \lntxn{} and the convex hull at the same position of the corresponding ternary phase diagram Ln–X–N. 
To construct the convex hull for each ternary phase diagram Ln–X–N, all relevant phases with $\Delta E_\mathrm{hull} < 50$ meV/atom based on data from the Materials Project database \cite{horton_accelerated_2025} were computed, involving approximately 3500 relevant phases to establish the convex hulls for 854 ternary phase diagrams. 
Additionally, to address the total energy errors in compounds containing N anions, a correction of -0.316 eV/atom was utilized for each N anions, following Ref.\cite{wang_framework_2021} and the Materials Project database \cite{horton_accelerated_2025}.
Thermodynamically stable compounds were identified using the criteria $E_F < 0$ and $\Delta E_\mathrm{hull} < 50$ meV/atom, taking the finite temperature effect as well as DFT accuracy into account  following Ref.\cite{singh_high-throughput_2018}. For these thermodynamically stable compounds, dynamical stability was assessed using phonon calculations, where imaginary phonon energies indicate dynamic instability. Following the convention used in the Phonopy package \cite{togo_implementation_2023}, imaginary phonon energies were treated as negative real values. Accordingly, we applied a dynamical stability criterion requiring the minimum phonon energy to exceed $-0.8$ meV. Please note that, Yb-containing APVs were excluded due to the temporary deprecation of all Yb compounds in the Materials Project database at the time this work was conducted. Nevertheless, our findings for other lanthanide elements are consistent and can be extended to include Yb. Finally, we accounted for the strong correlation of 4\textit{f} electrons for all lanthanide elements except La, which lacks 4\textit{f} electrons. For compounds containing lanthanide elements from Ce to Lu, high-throughput screening was performed independently with two pseudo-potentials, \textit{Ln} and \textit{Ln\_3}, representing the itinerant and localized limits of the 4\textit{f} electrons, respectively. In the \textit{Ln} potentials, 4\textit{f} electrons are treated as valence states without localization, whereas in the \textit{Ln\_3} potentials, 4\textit{f} electrons are frozen in the core, simulating a fully localized configuration.
Effectively, one could also apply DFT+$U$ with $U=0$ and a reasonably large $U$ to simulate the itinerant and localized limits, respectively, when a fully localized pseudo-potential is not available (e.g., actinides).
In our high-throughput screening, a compound was considered stable only if it remained stable under both itinerant and localized limits, ensuring robustness regardless of the degree of 4\textit{f} localization (see a schematic representation in Fig.~\ref{fig:framework}(b)). This conservative approach may exclude some truly stable compounds, e.g., a strongly localized compound which appears unstable under itinerant assumptions, but it guarantees that all selected candidates are reliable. The actual family of stable lanthanide nitride–based APVs is likely even larger than reported here.

\begin{figure*}[ht]
    \centering
    \includegraphics[width=0.85\linewidth]{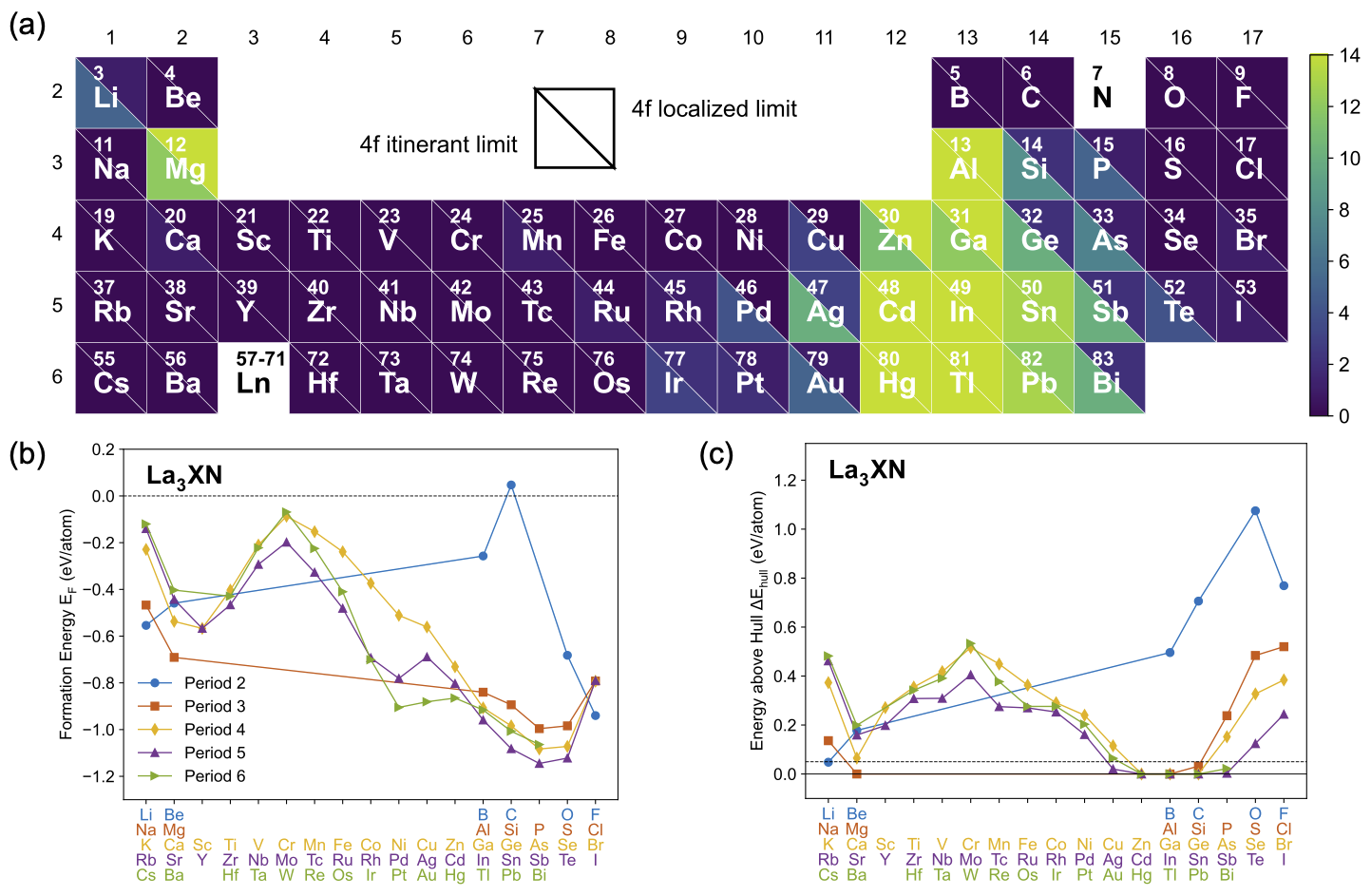}
    \caption{(a) Colormap showing the number of thermodynamically stable \lntxn{} compounds as a function of the X element. The lower-left and upper-right triangles correspond to results obtained using the itinerant and localized limits of the 4\textit{f} electrons, respectively. (b) Calculated formation energies and (c) energies above the convex hull for La$_3$XN compounds. The dashed lines represent the criteria for formation energy and energy above the convex hull, respectively.}
    \label{fig:thermo}
\end{figure*}

\begin{figure*}[ht]
    \centering
    \includegraphics[width=0.85\linewidth]{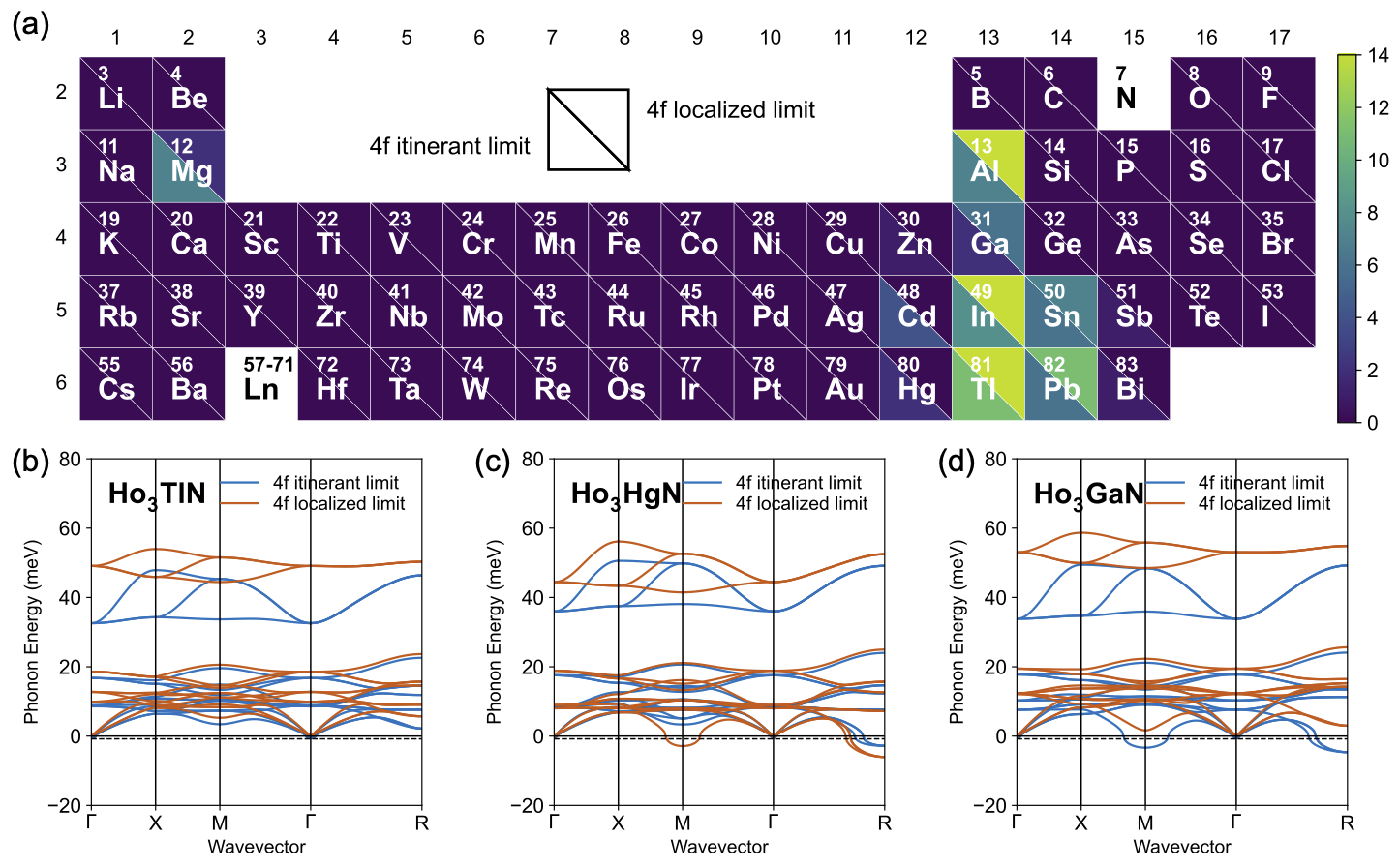}
    \caption{(a) Colormap showing the number of dynamically stable \lntxn{} compounds as a function of the X element. The lower-left and upper-right triangles correspond to results obtained using the itinerant and localized limits of the 4\textit{f} electrons, respectively. Calculated phonon dispersions of (b) Ho$_3$TlN, (c) Ho$_3$HgN, and (d) Ho$_3$GaN, in both itinerant and localized limits. The dashed lines represent the criterion for phonon energy.}
    \label{fig:phonon}
\end{figure*}

\textbf{Computational details.} Lanthanide APV nitrides \lntxn{} crystallize in space group 221, $Pm\bar{3}m$, and the crystal structure is illustrated in Fig.~\ref{fig:framework}. All DFT calculations were performed using the projector augmented-wave (PAW) method \cite{blochl_projector_1994, kresse_ultrasoft_1999} as implemented in the Vienna Ab initio Simulation Package (VASP) \cite{kresse_ab_1993, kresse_efficient_1996}. The generalized gradient approximation (GGA) formulated by Perdew, Burke, and Ernzerhof (PBE) \cite{perdew_generalized_1996} was employed as the exchange–correlation functional. A plane-wave cutoff energy of 650 eV was used, and the energy convergence criterion was $10^{-6}$ eV/atom. Spin polarization is always utilized, and the magnetic ordering is initialized using a ferromagnetic configuration.
$\Gamma$-centered \emph{k}-point meshes with densities of approximately $2000/N_\mathrm{atom}$ \AA$^{-3}$ and $10000/N_\mathrm{atom}$ \AA$^{-3}$ were employed for structure optimization and total-energy calculations, respectively, where $N_\mathrm{atom}$ is the number of atoms in the compound. The generation of VASP input files and postprocessing of thermodynamical data utilized the Python Materials Genomics (pymatgen) library \cite{jain_high-throughput_2011}. Phonon calculations were performed using density functional perturbation theory (DFPT) as implemented in VASP with a $2\times2\times2$ supercell (40 atoms) and a $7\times7\times7$ $\Gamma$-centered \emph{k}-point mesh. Postprocessing of phonon data utilized the Phonopy package \cite{togo_implementation_2023}. Electronic calculations  were performed using  $15\times15\times15$ $\Gamma$-centered \emph{k}-point mesh with spin-orbit coupling (SOC). 
The analyses of irreducible representations of electronic states and topological classification were conducted using the IRVSP package \cite{gao_irvsp_2021}. For the mathematical definitions of topological classifications and irreducible representations of electronic states, please refer to Refs.~\cite{vergniory_complete_2019, gao_irvsp_2021}. Finally, the electronic band structures and density of states (DOS) calculations were carried out using DFT+$U$ in the simplified
rotationally invariant approach \cite{dudarev_electron-energy-loss_1998}. $U$ = 10 eV was applied following Ref.~\cite{pertsova_computational_2019}, to ensure that the 4$f$ bands are positioned within the valence band, and varying $U$ by 2 eV showed negligible effects near the Fermi level.

\section{\label{sec:result}Results and discussions}

\textbf{Thermodynamical stability.} We begin with the high-throughput screening results for thermodynamical stability. The number of thermodynamically stable \lntxn{} compounds as a function of the X element is shown in Fig.~\ref{fig:thermo}(a). Compounds containing La were computed only in the itinerant limit using the \textit{La} pseudo-potential, because the 4\textit{f} orbital of La is approximately empty \cite{kadri_investigation_2023}, where the itinerant and localized limits should yield identical results. Therefore, the itinerant-limit results for La are also counted under the localized limit. Examining the formation energies and energies above the convex hull for \lntxn{} compounds shows that the results are broadly consistent across different Ln elements and between the itinerant and localized limits of the 4\textit{f} electrons (see the results of all \lntxn{} compounds in Supplemental Materials [SM], Sections I and II). As an example, the formation energies and energies above the convex hull for La$_3$XN compounds are presented in Fig.~\ref{fig:thermo}(b) and (c), respectively. Most \lntxn{} compounds exhibit negative formation energies, indicating that their formation from the elemental constituents is exothermic. Generally, a similar trend is followed for all Ln: the formation energy decreases with increasing atomic number up to the group 3 elements, then increases until reaching the group 6 elements, followed by a decline to a minimum near the group 15 elements and finally an increase towards the end of each period. A similar trend of formation energy has also been reported in transition metal APV nitrides \cite{singh_high-throughput_2018}, explained by the similar trend of ionic radii change in each period. 
However, in transition metal APV nitrides, the formation energies have similar values when X is in groups 4 and 13, whereas for lanthanide APV nitrides, X in group 13 has a much lower formation energy compared to X in group 4.
Since lanthanide elements have smaller electronegativity (1.1--1.3) compared to transition metals (1.6--1.8), they have a greater tendency to form compounds with strongly electronegative ions, such as those in groups 13--17, which contributes to the lower formation energy.

Consistently, a comparable trend is observed for the energies above the convex hull, where X elements near group 13 show the lowest $\Delta E_\mathrm{hull}$ values, with the exception of period 2 elements. These deviations for period 2 elements is likely due to the large atomic-size mismatch with the lanthanide elements. 
Consequently, the major candidates of thermodynamically stable \lntxn{}  are formed with X from group 13 elements: Al, Ga, In, and Tl, and their neighboring elements. Accounting for 4\textit{f} localization yields a stricter distribution; the localized limit confines the stable X elements to group 13 and their immediate neighbors,  predicting 157 stable compounds, whereas the itinerant limit produces a broader distribution extending to some second-nearest neighbors of group 13, such as Ag and Sb, predicting 223 stable compounds. In total, 149 \lntxn{} candidates satisfy the thermodynamical stability criteria simultaneously in both the itinerant and localized limits; these candidates move forward for dynamical stability screening.

\textbf{Dynamical stability.} The number of dynamically stable \lntxn{} compounds as a function of the X element is shown in Fig.~\ref{fig:phonon}(a), with a total of 37 compounds. A complete list of these 37 \lntxn{} compounds is presented in SM Section IV \cite{Supplemental_Material}. The dynamically stable \lntxn{} compounds can only be formed by seven X elements: Mg, Al, Ga, In, Sn, Tl, and Pb, consisting of four group 13 elements and three of their immediate neighbors. Among these seven X elements, Tl has the largest number of dynamically stable \lntxn{} compounds, with a total of 11.
Comparing the distribution of thermodynamically stable \lntxn{} compounds with its dynamically stable counterpart (Fig.~\ref{fig:thermo}(a) with Fig.~\ref{fig:thermo}(b)), a few X elements exhibit a significant reduction in stable compounds, such as group 12 elements Cd and Hg, both decreasing from being stable in all 14 compounds to none, and group 13 element Ga, decreasing from 12 compounds to only 1.

To explain this reduction, three examples are provided in Fig.~\ref{fig:phonon}(b--d), for Ho$_3$TlN, Ho$_3$HgN, and Ho$_3$GaN, respectively, while all phonon dispersions for thermodynamically stable compounds are presented in SM Section III \cite{Supplemental_Material}. The blue and red curves represent the phonon dispersions computed under the itinerant and localized limits of 4\textit{f} electrons, respectively. The dashed line represents the $-0.8$ meV criterion for dynamical stability. As \lntxn{} compounds contain five atoms in a primitive cell, the phonon dispersion contains 15 phonon modes. The three optical phonon modes with the highest energies, primarily contributed to by the vibrations of N atoms, exhibit a significant energy gap from all other phonon modes.
Except for these three optical phonon modes with the highest energies, usually for a dynamically stable \lntxn{} compound, all other phonons have consistent behaviors under the itinerant and localized limits of 4\textit{f} electrons, with only a small energy difference (see Fig.~\ref{fig:phonon}(b) for Ho$_3$TlN as an example). Generally, the localization of \textit{f} electrons increases the lattice parameters, therefore producing larger phonon energies; this behavior has also been demonstrated in 5\textit{f} electron systems \cite{zhou_capturing_2022}.
For Ln$_3$GaN, while six compounds are predicted as dynamically stable in the localized limit, only two are stable in the itinerant limit. As an example, Ho$_3$GaN is stable in the localized limit  but unstable in the itinerant limit (see Fig.~\ref{fig:phonon}(c)). While generally lanthanide element compounds are considered to host strongly localized 4\textit{f} electrons, the localized limit may provide a better reference point for \lntxn{} compounds. Consequently, the decrease in the number of stable Ln$_3$GaN compounds may be due to our conservative screening strategy, suggesting that additional Ln$_3$GaN compound could, in fact, be dynamically stable. On the other hand, the number of dynamically stable Ln$_3$HgN compounds is limited to two under the itinerant limit and zero under the localized limit. For instance, Ho$_3$HgN is unstable at the $R$ point regardless of the degree of 4\textit{f} localization (see Fig.~\ref{fig:phonon}(d)). This indicates that the scarcity of stable Ln$_3$HgN compounds is not an artifact of our screening approach.
Since imaginary phonon energies indicate dynamical instability, the presence of imaginary phonons at $\Gamma$ points or other $Q$-points suggests that a competing phase with lower energy exists in a primitive cell or a supercell, respectively.
Although Ho$_3$HgN is unstable in the perfect cubic perovskite structure (e.g., the space group of $Pm\bar{3}m$), given the small imaginary components in its phonon dispersion, it may stabilize in a distorted perovskite phase, such as the orthorhombic $Pbnm$ structure \cite{mochizuki_theoretical_2020}. Expanding the structural space of lanthanide APV nitrides to include such distortions could reveal additional stable compounds, particularly those incorporating groups 13 and 14 elements like Ga, Cd, and Hg, thereby broadening the known lanthanide APV nitride family.

\begin{figure}
    \centering
    \includegraphics[width=0.95\linewidth]{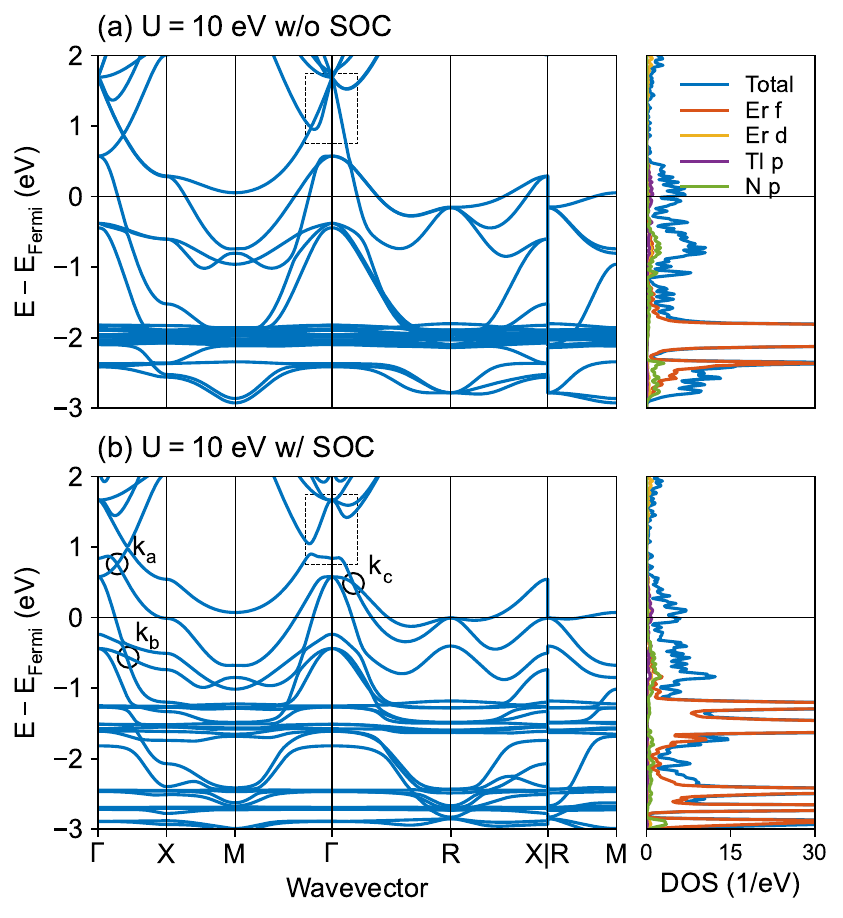}
    \caption{ Computed electronic structures and density of states for Er$_3$TlN using DFT+$U$ ($U = 10$ eV) (a) without spin-orbit coupling and (b) with spin-orbit coupling. Three symmetry enforced band crossings are marked by circles in the panel (b).}
    \label{fig:er3tln}
\end{figure}

\begin{figure*}[ht]
    \centering
    \includegraphics[width=0.68\linewidth]{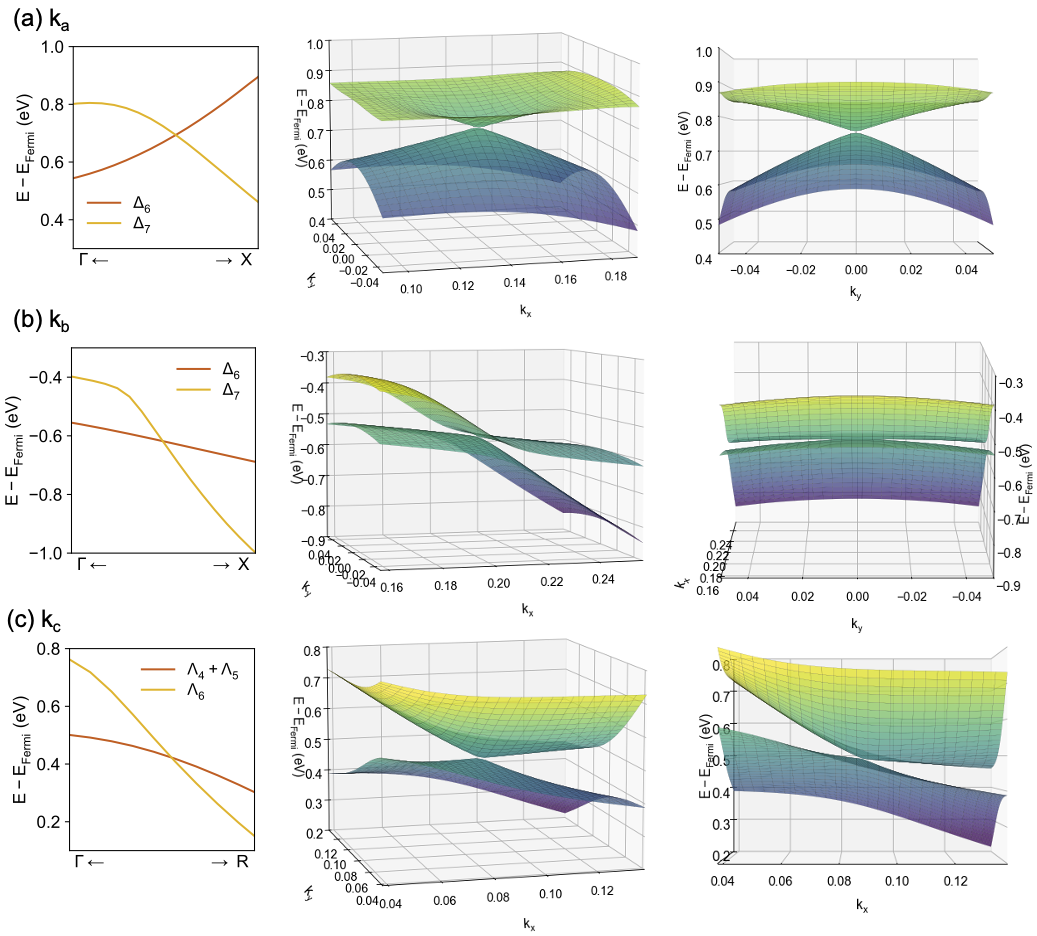}
    \caption{ Computed symmetry enforced band crossing for Er$_3$TlN at three different $k$-points, (a) $\mathbf{k}_a=(0.141, 0, 0)$, (b) $\mathbf{k}_b=(0.207, 0, 0)$, and (c) $\mathbf{k}_c=(0.088, 0.088, 0.088)$. For each $k$-point, one 2D plot along the high-symmetry path and two 3D plots from different viewing angles are provided. Along the $\Gamma$--$X$ path ($\zeta00$), $k_y$ and $k_z$ are equivalent, while $k_x$, $k_y$, and $k_z$ are all equivalent along the $\Gamma$--$R$ path ($\zeta\zeta\zeta$).}
    \label{fig:topo}
\end{figure*}

\textbf{Topological behavior.} Spin–orbit coupling plays a crucial role in topological phenomena, and the inherently strong SOC of 4$f$ electrons in lanthanide elements makes them promising candidates for hosting diverse topological behaviors. Here, we applied the irreducible representations of electronic states \cite{gao_irvsp_2021} on SOC calculations to classify the 37 stable lanthanide APV nitrides \lntxn{} (see SM Section IV for all results). Remarkably, among these 37 compounds, only two, Ho$_3$PbN and Ho$_3$TlN, were found to be trivial. The remaining 35 compounds exhibited nontrivial topological features in their electronic bands and were categorized into three topological subclasses: enforced semimetals (3 compounds), enforced semimetals with Fermi degeneracy (30 compounds), and split elementary band representation (SEBR, 2 compounds) (see Ref.~\cite{vergniory_complete_2019} for definitions). This distribution highlights the exceptional potential of lanthanide APV nitrides as a rich platform for exploring topological quantum materials.

Among the three topological subclasses, SEBR is notable because insulating (gapped) phases of SEBR compounds must be topological, indicating a topological insulator. Therefore, further DFT+$U$ calculations are carried out to examine the electronic structure of one compound in the SEBR subclass, Er$_3$TlN (see Fig.~\ref{fig:er3tln}). However, no gapped phase is observed, and Er$_3$TlN is nearly semimetallic in nature. Comparing the computed electronic band structures with and without SOC, the 4$f$ orbitals split from a single strong peak into two spreading peaks. Despite this splitting, all 4$f$ contributions remain well within the valence bands, thanks to the large $U$. The conduction bands show a small density of states similar to semimetals; in this region, a small gap is opened at the $\Gamma$ point by the SOC (see the dashed boxes in Fig.~\ref{fig:er3tln}). 

Although Er$_3$TlN is not a topological insulator, the strong SOC and the crystal symmetry of APV still lead to several symmetry-enforced band crossings, three of which are within a 1 eV range of the Fermi level (see the circles marked as $\mathbf{k}_a$, $\mathbf{k}_b$, and $\mathbf{k}_c$ in Fig.~\ref{fig:er3tln}(b) and the irreducible representations of the crossed bands in Fig.~\ref{fig:topo}). The electronic bands from different irreducible representations cannot mix, thus the band crossing is protected by symmetry \cite{gao_irvsp_2021}. In Fig.~\ref{fig:topo}, 3D plots are also provided, computed using a denser $k$-point mesh near the crossing. Along the $\Gamma$--$X$ path ($\zeta00$), $k_y$ and $k_z$ are equivalent, while $k_x$, $k_y$, and $k_z$ are all equivalent along the $\Gamma$--$R$ path ($\zeta\zeta\zeta$). The electronic bands cross linearly in all directions $k_x$, $k_y$, and $k_z$ at $\mathbf{k}_a$ and $\mathbf{k}_c$, indicating a Type-I Dirac cone at $\mathbf{k}_a$ and a Type-II Dirac cone at $\mathbf{k}_c$. At $\mathbf{k}_b$, the bands cross linearly along $k_x$ but exhibit parabolic dispersion along $k_y$ and $k_z$, forming a semi-Dirac cone. These two Dirac cones and one semi-Dirac cone, all located within 1 eV of the Fermi level, underscore the rich topological landscape of Er$_3$TlN and highlight the broader potential of lanthanide APV nitrides as a platform for exploring diverse topological phases.

\section{\label{sec:conclusion}Conclusion}

In this work, we present a systematic study of a newly identified sub-family of antiperovskite nitrides: lanthanide antiperovskite nitrides. We developed a double-screening framework to identify stable lanthanide antiperovskite nitride compounds, assuming either fully itinerant or fully localized limits for 4$f$ electrons during the DFT calculations. By applying this framework to screen lanthanide antiperovskite nitrides \lntxn{} (where Ln represents a lanthanide element from La to Lu, excluding Yb, and X represents any element from Li to Bi, excluding N, noble gases, and lanthanide elements, a total of 854 compounds), we identified 149 thermodynamically stable compounds, and 37 were further confirmed to be dynamically stable, revealing the potential breadth of this material family. We then performed a topological classification of these 37 stable compounds. With only two exhibiting trivial behavior, the remaining 35 compounds show nontrivial topological behaviors. Using Er$_3$TlN as a representative example, we identified a Type-I Dirac cone, a Type-II Dirac cone, and a semi-Dirac cone within 1 eV of the Fermi level. These findings highlight the new family of lanthanide antiperovskite nitrides as a promising platform for exploring rich topological phenomena and other emergent quantum properties.

\section{Acknowledgments}
This work is supported through the INL Laboratory Directed Research and Development (LDRD) Program under DOE Idaho Operations Office Contract DE-AC07-05ID14517. KG acknowledges support from the US Department of Energy, Basic Energy Sciences, Materials Sciences, and Engineering Division. This research made use of Idaho National Laboratory’s High Performance Computing systems located at the Collaborative Computing Center and supported by the Office of Nuclear Energy of the U.S. Department of Energy and the Nuclear Science User Facilities under Contract No. DE-AC07-05ID14517.


\normalsize

\bibliographystyle{apsrev4-1} 

\end{document}


\renewcommand{\thefigure}{S\arabic{figure}}
\renewcommand{\thetable}{S\arabic{table}}
\renewcommand\cellalign{cl}

\raggedright{\textbf{\Large Supplemental materials to}}
\title{The stability and topological behaviors in lanthanide antiperovskite nitrides: a high-throughput study}
\author{Shuxiang Zhou$^1$}
\email{shuxiang.zhou@inl.gov}
\author{Kevin Vallejo$^1$} 
\author{Krzysztof Gofryk$^1$}
\affiliation{$^1$Center for Quantum Actinide Science and Technology, Idaho National Laboratory, Idaho Falls, ID 83415, USA}

\maketitle

\section{\label{sec:formation}Formation Energy}
\justifying

\begin{figure}[H]
    \centering
    \includegraphics[width=0.305\linewidth]{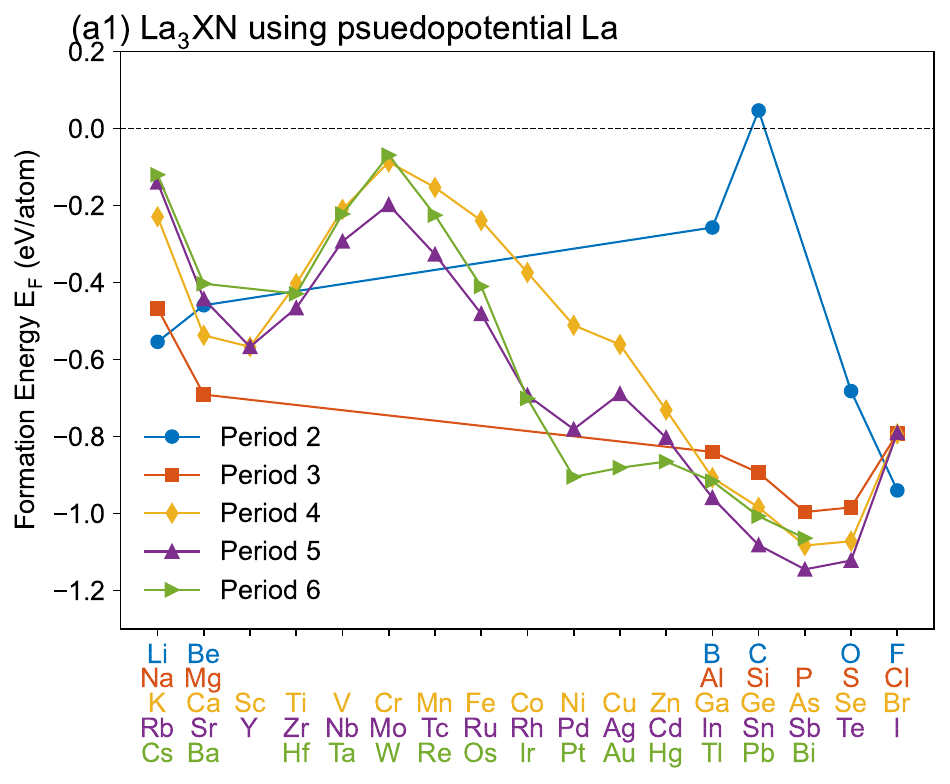}
    \includegraphics[width=0.305\linewidth]{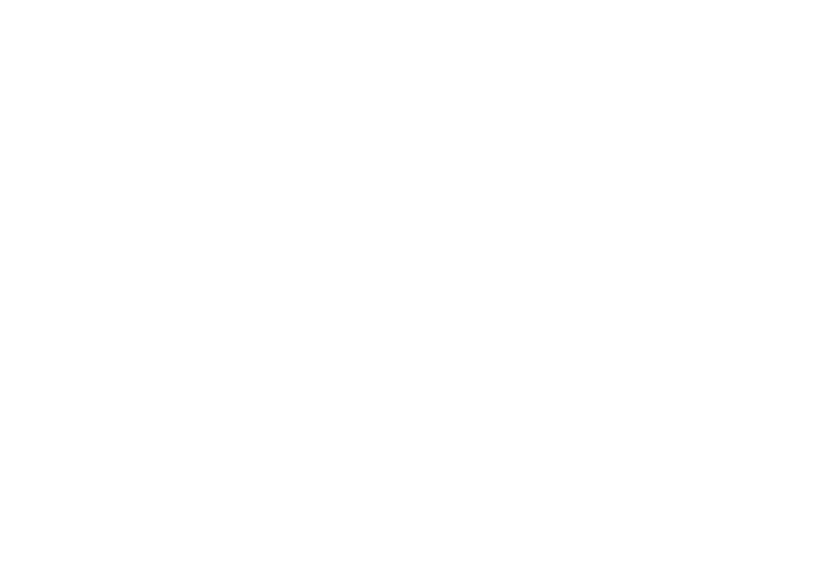}\\
    \includegraphics[width=0.305\linewidth]{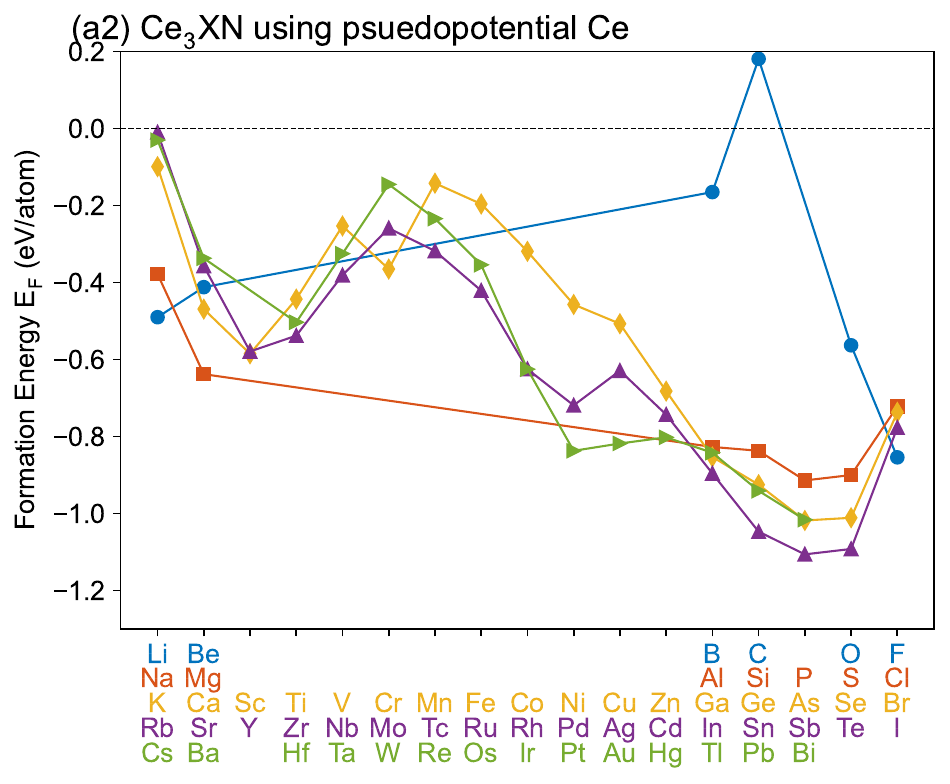}
    \includegraphics[width=0.305\linewidth]{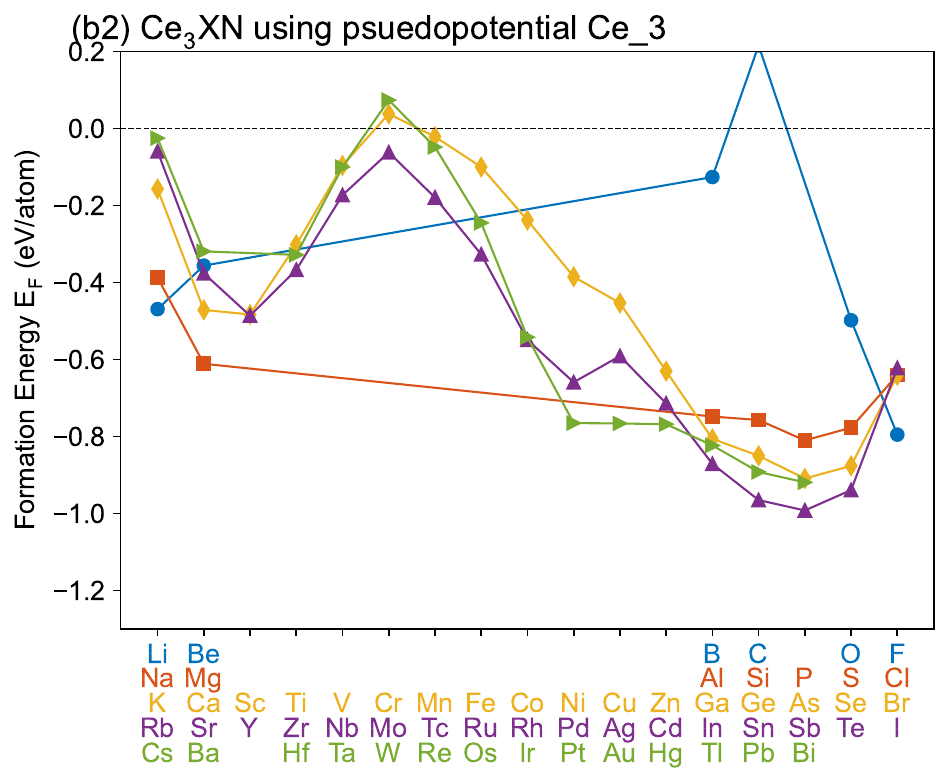}\\
    \includegraphics[width=0.305\linewidth]{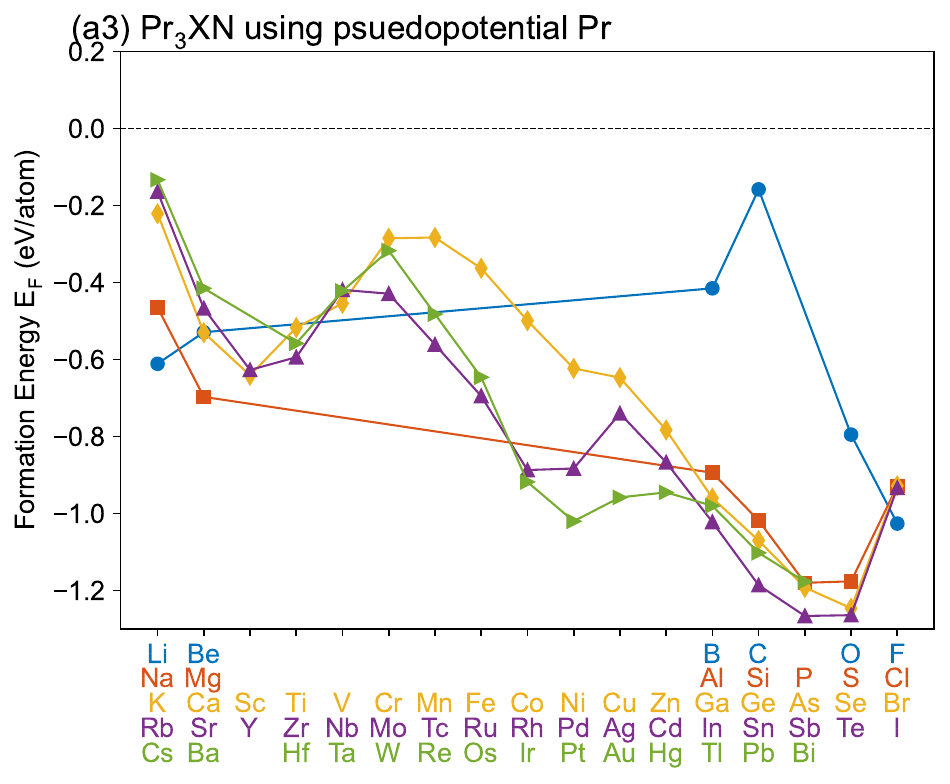}
    \includegraphics[width=0.305\linewidth]{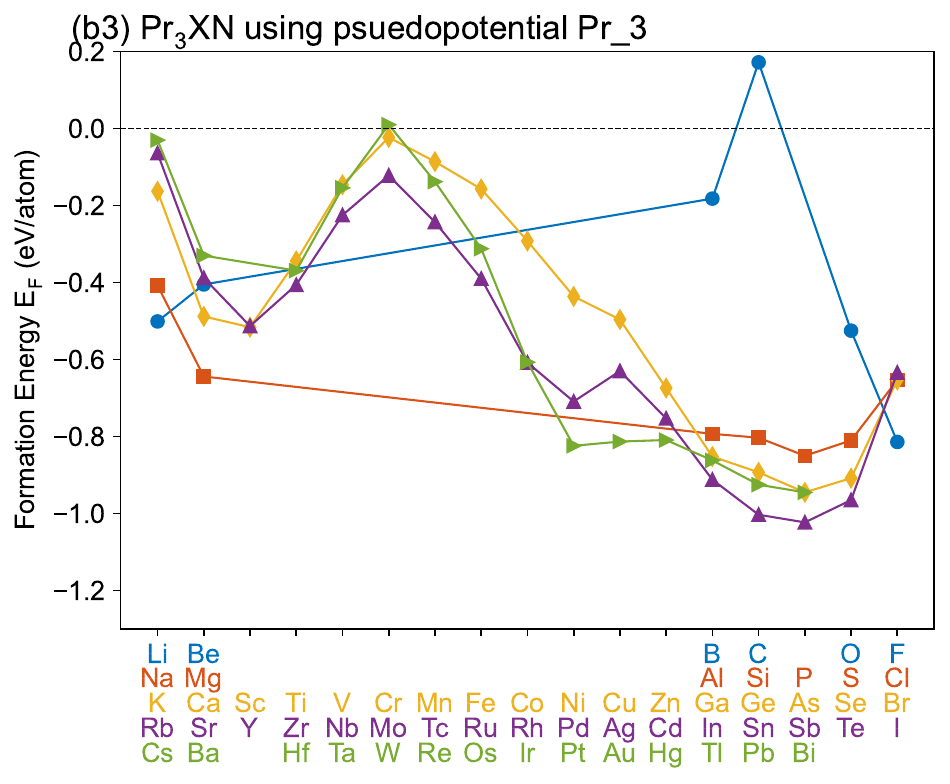}\\
\end{figure}
\clearpage
\begin{figure}[H]
    \centering
    \includegraphics[width=0.305\linewidth]{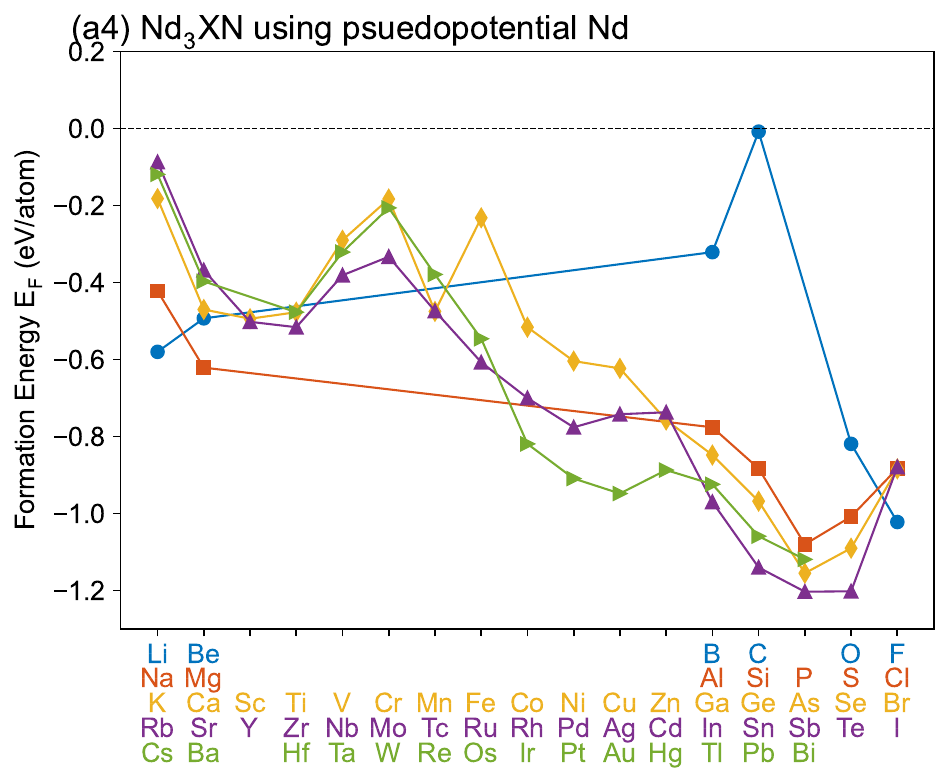}
    \includegraphics[width=0.305\linewidth]{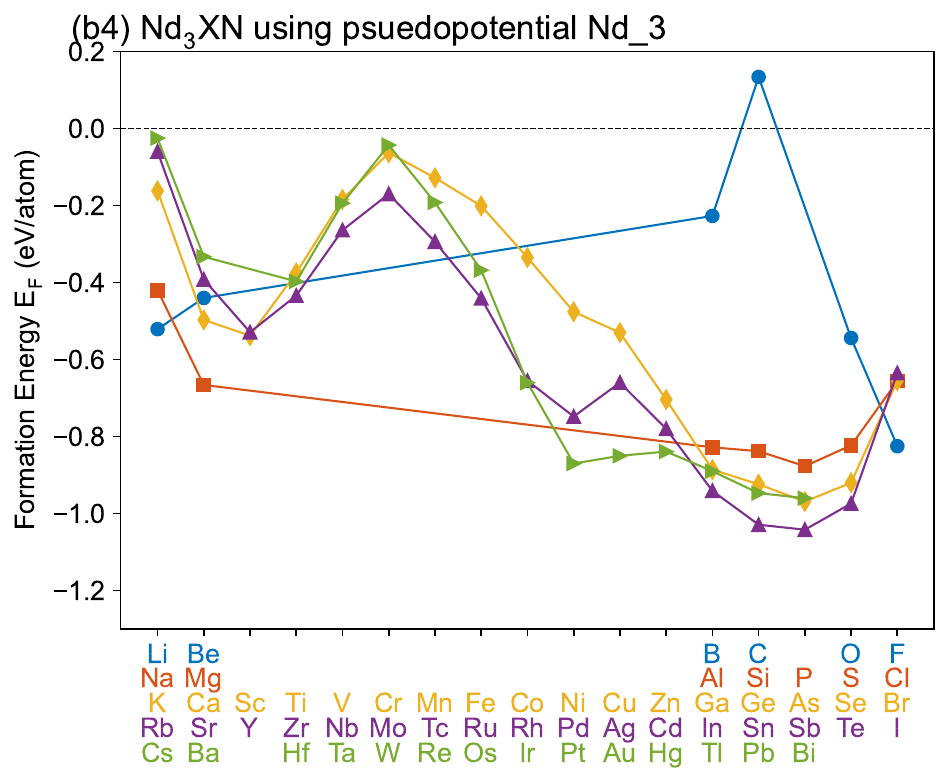}\\
    \includegraphics[width=0.305\linewidth]{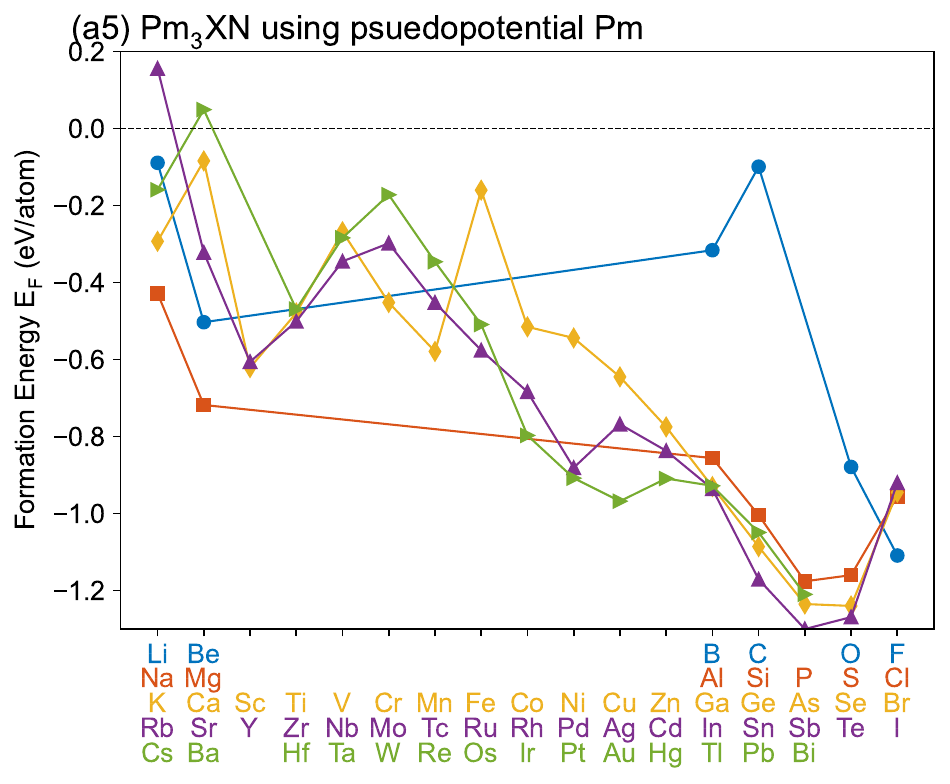}
    \includegraphics[width=0.305\linewidth]{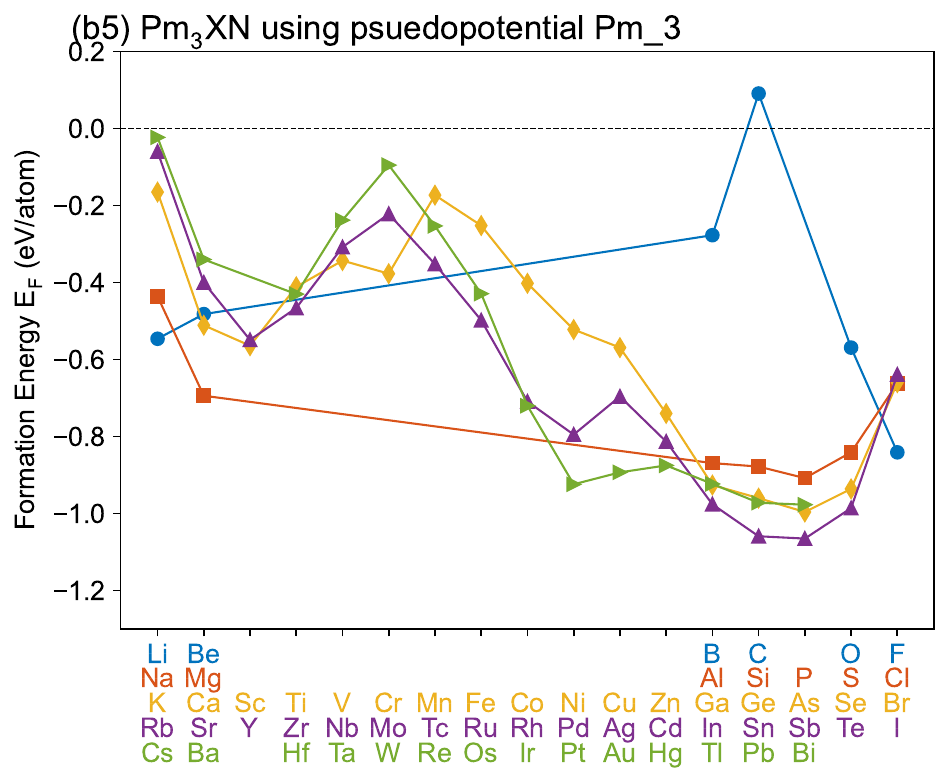}\\
    \includegraphics[width=0.305\linewidth]{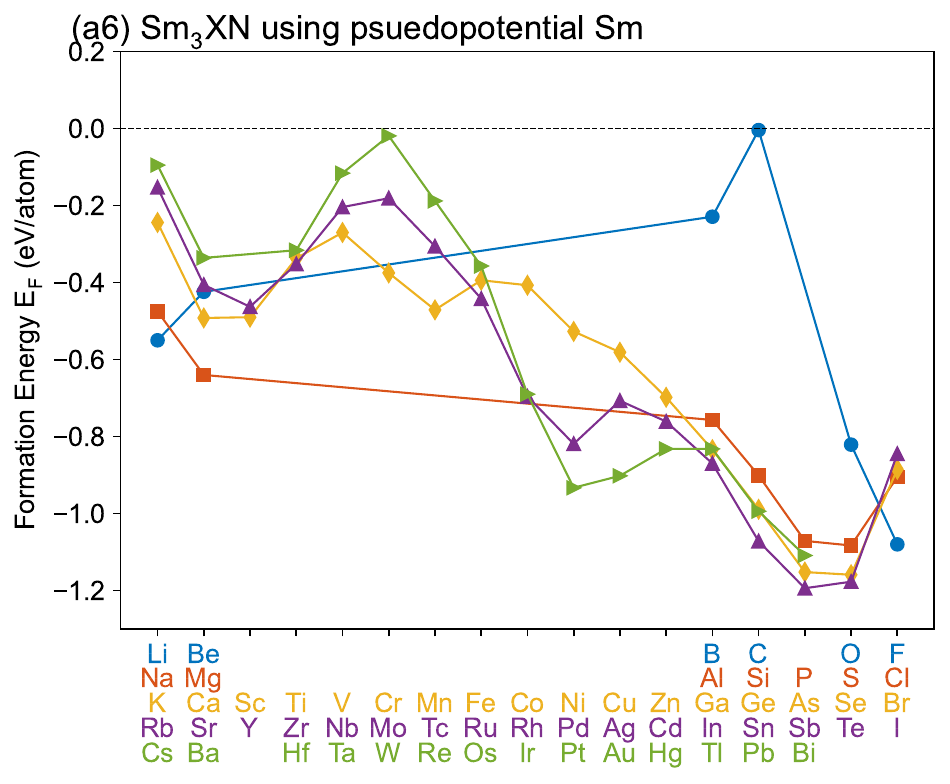}
    \includegraphics[width=0.305\linewidth]{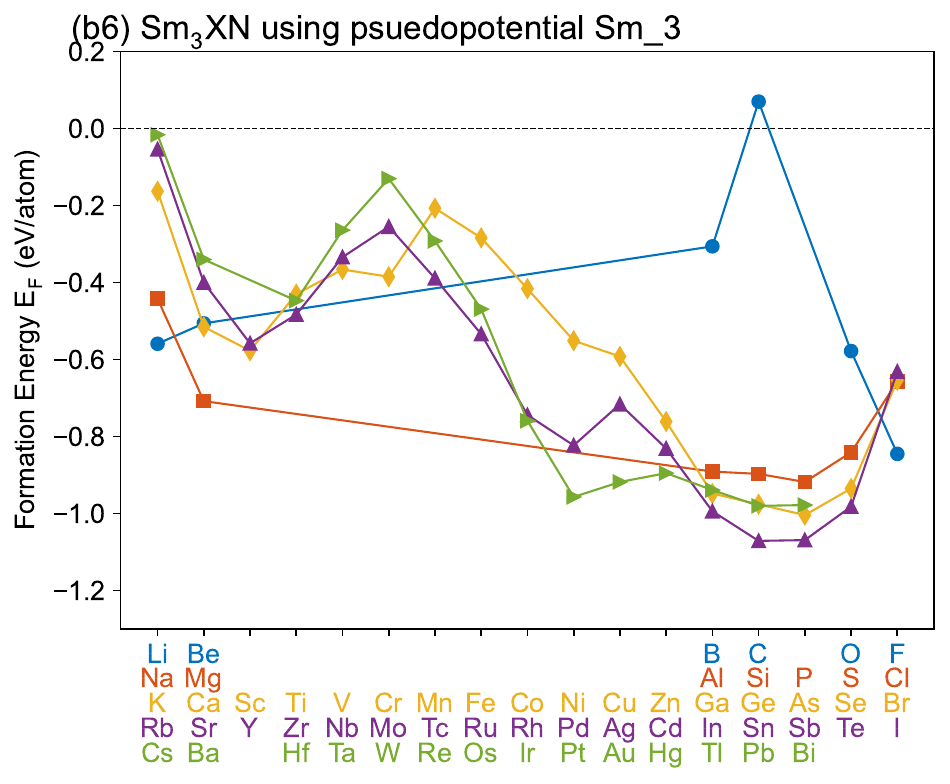}\\
    \includegraphics[width=0.305\linewidth]{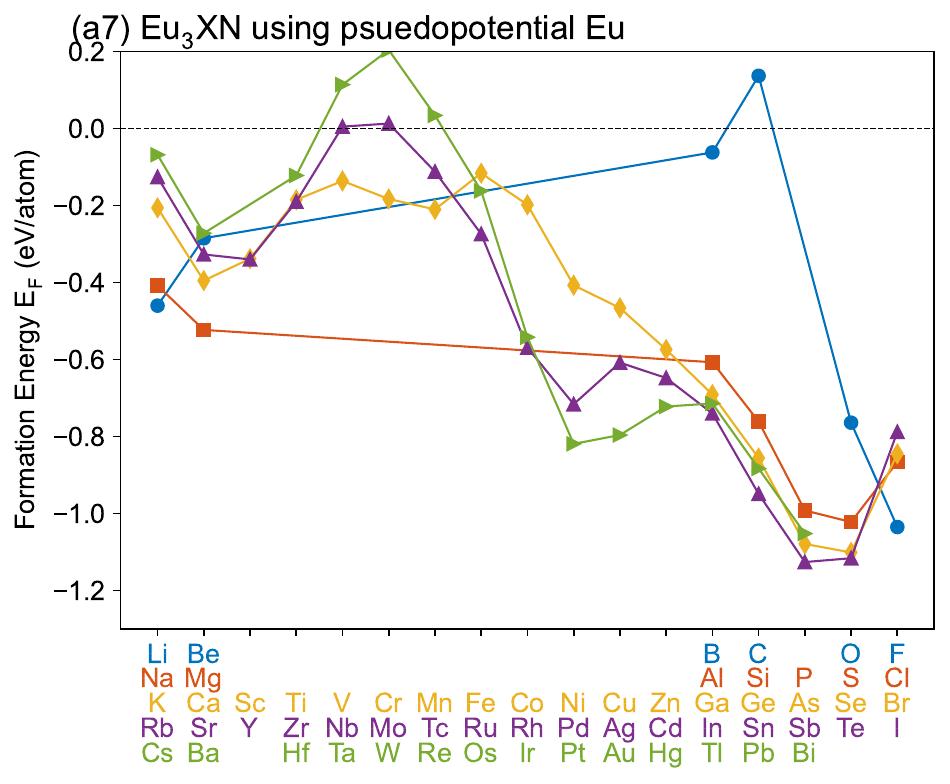}
    \includegraphics[width=0.305\linewidth]{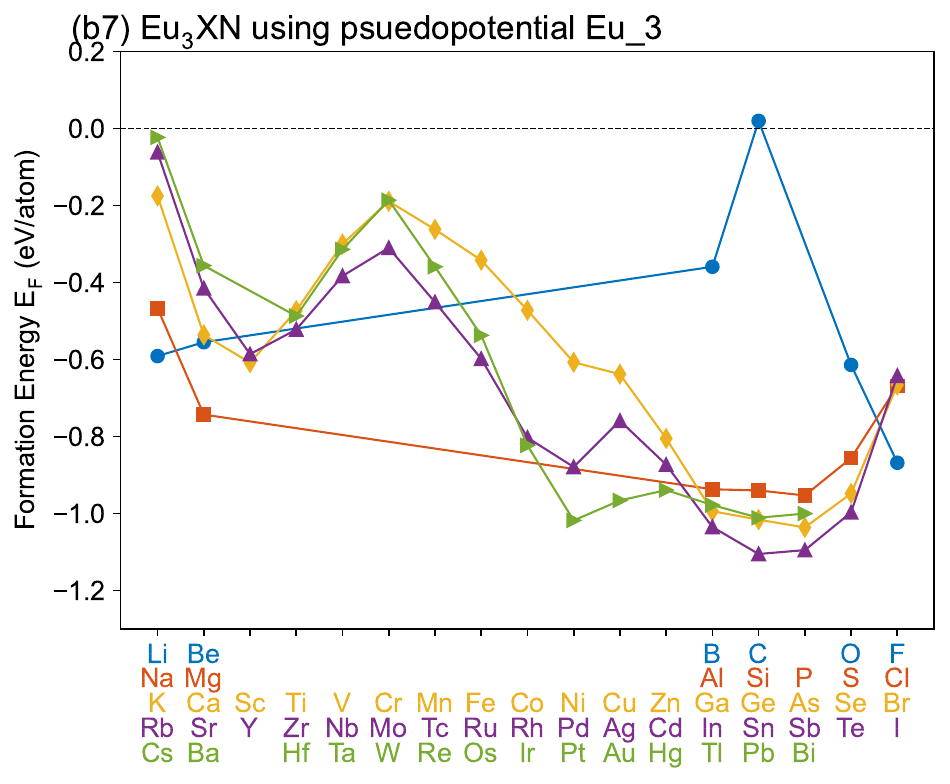}\\
    \includegraphics[width=0.305\linewidth]{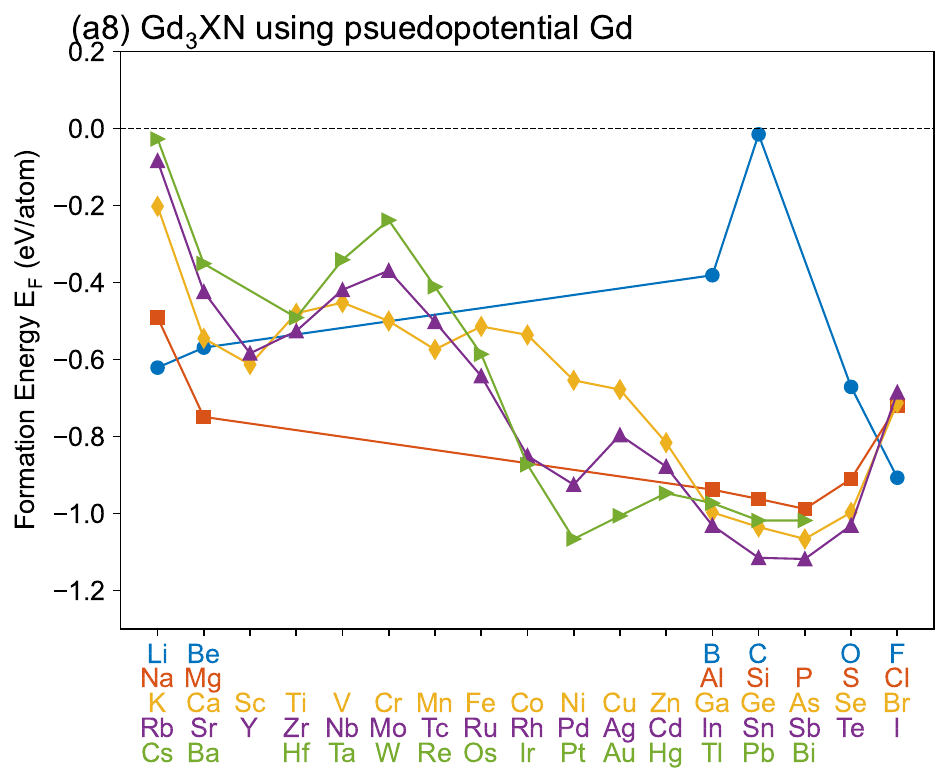}
    \includegraphics[width=0.305\linewidth]{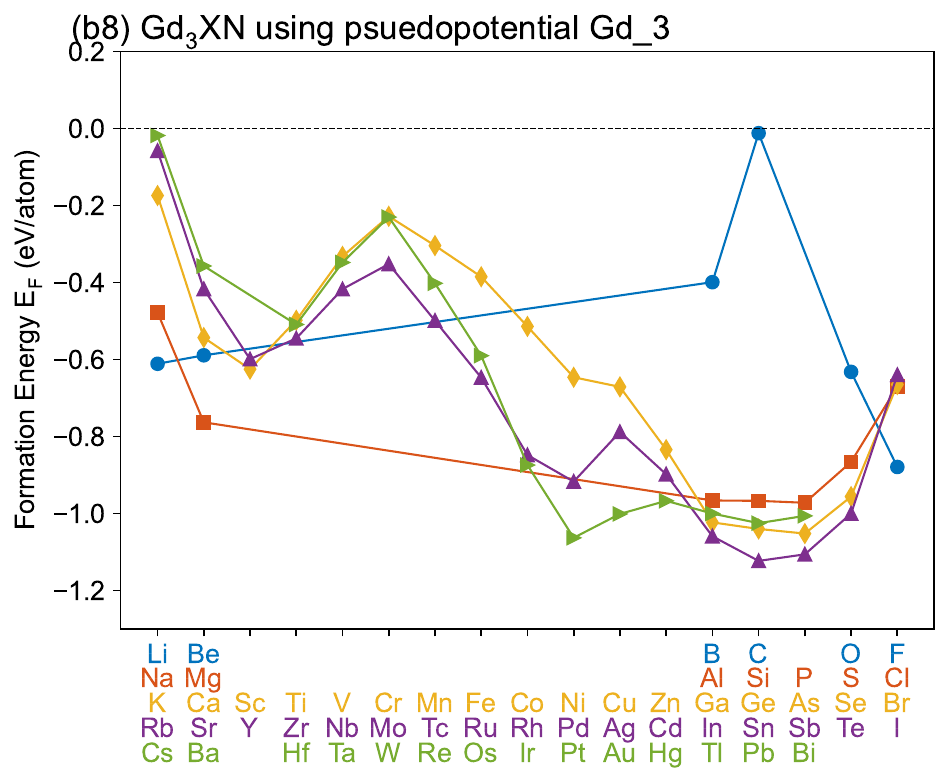}\\
\end{figure}
\clearpage
\begin{figure}[H]
\centering
    \includegraphics[width=0.305\linewidth]{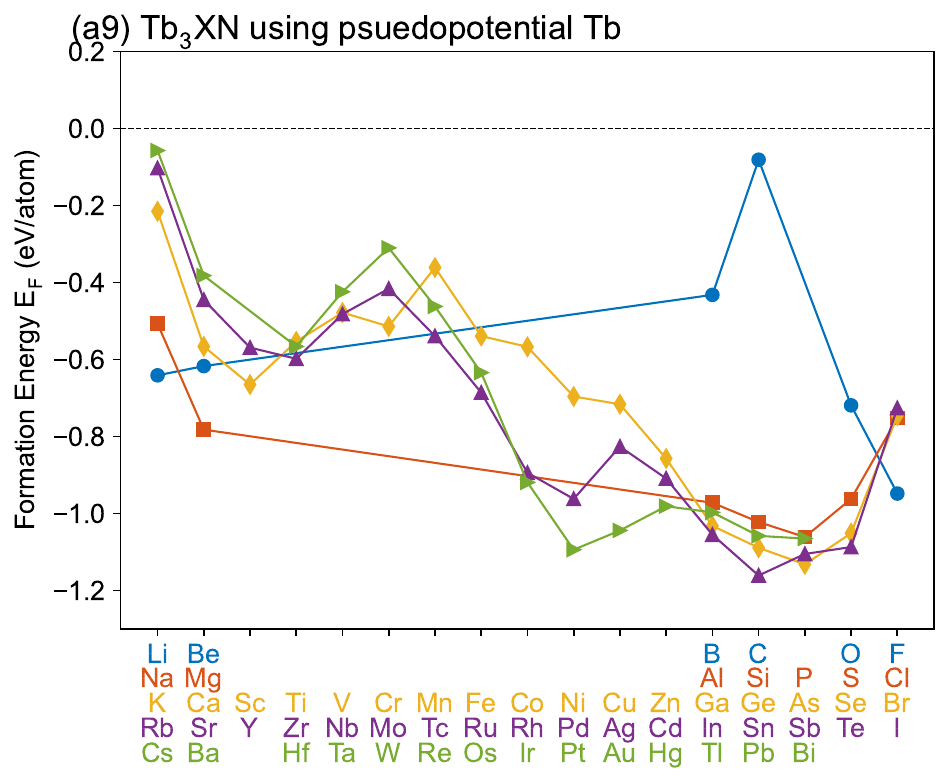}
    \includegraphics[width=0.305\linewidth]{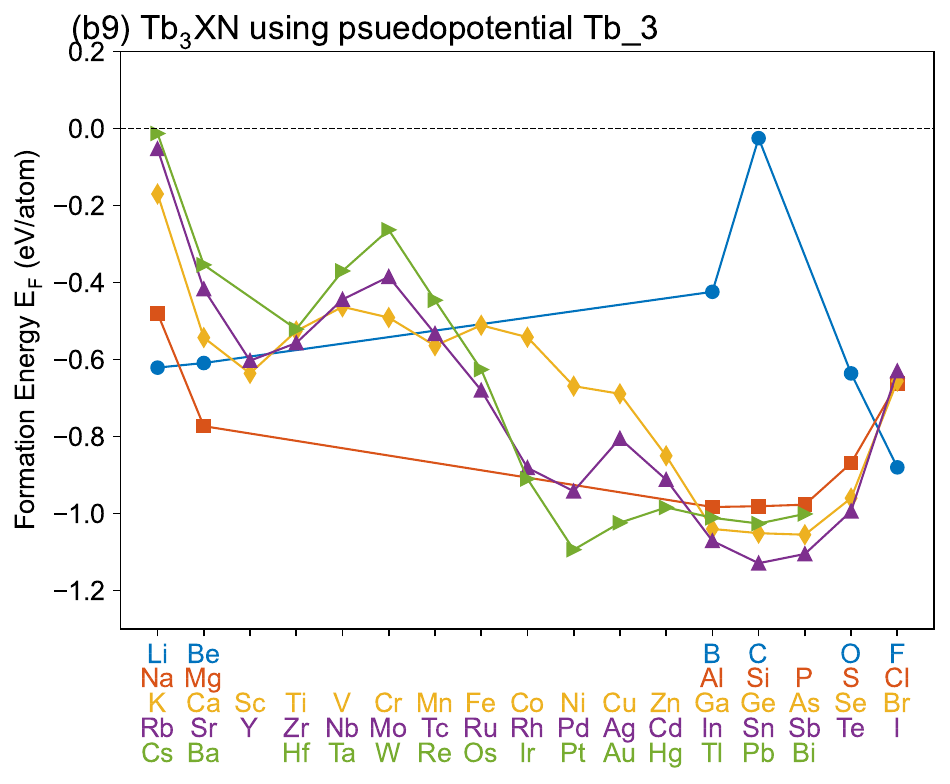}\\
    \includegraphics[width=0.305\linewidth]{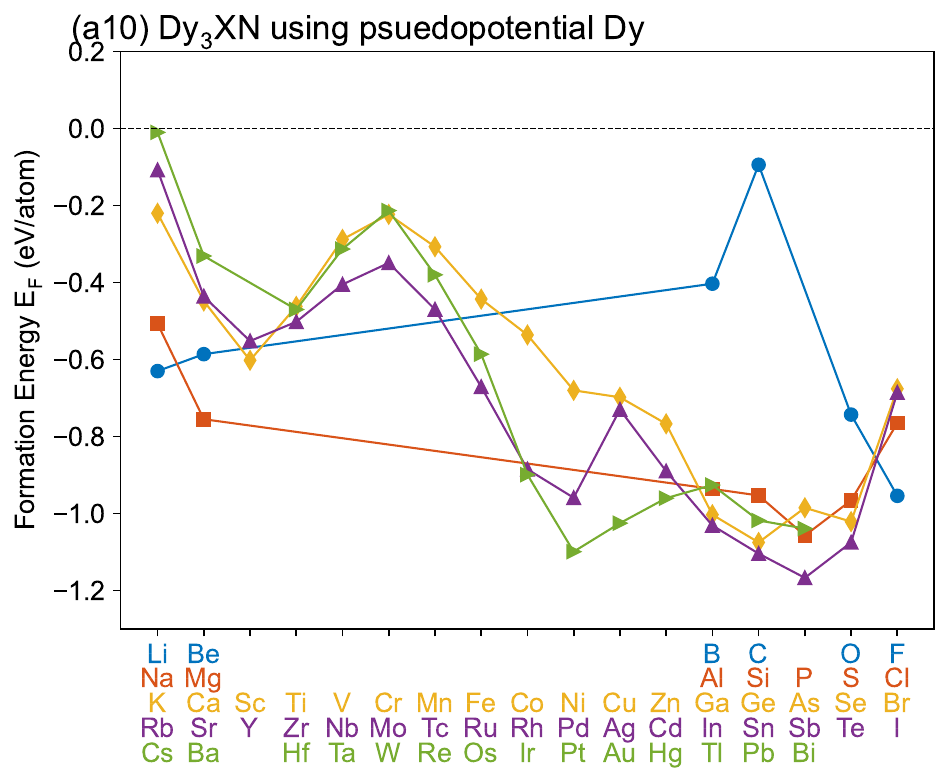}
    \includegraphics[width=0.305\linewidth]{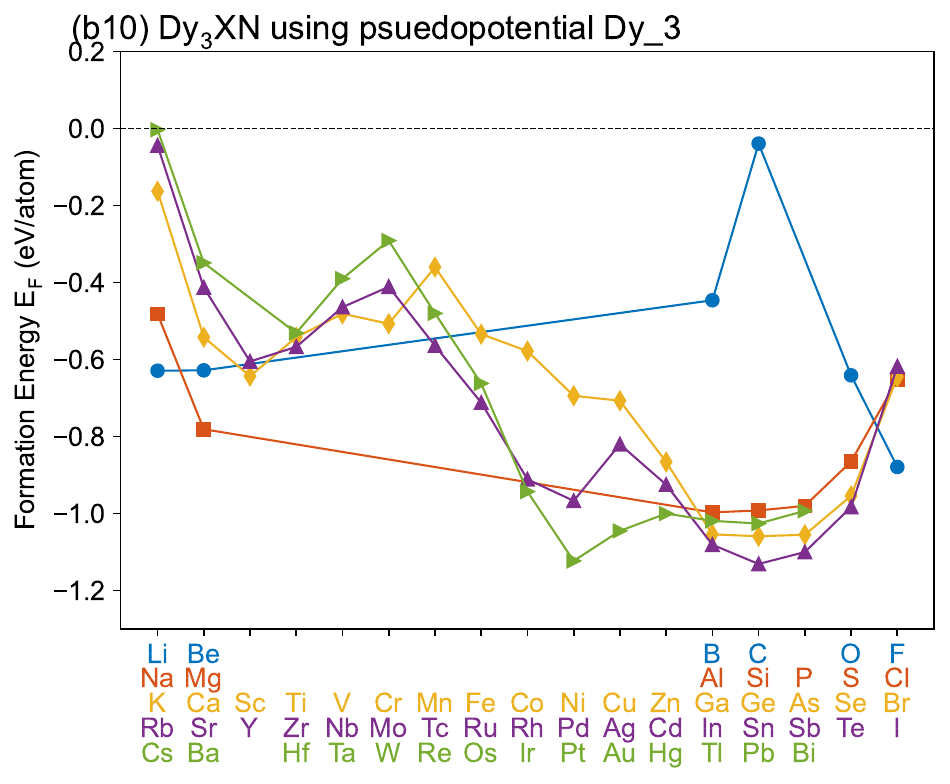}\\
    \includegraphics[width=0.305\linewidth]{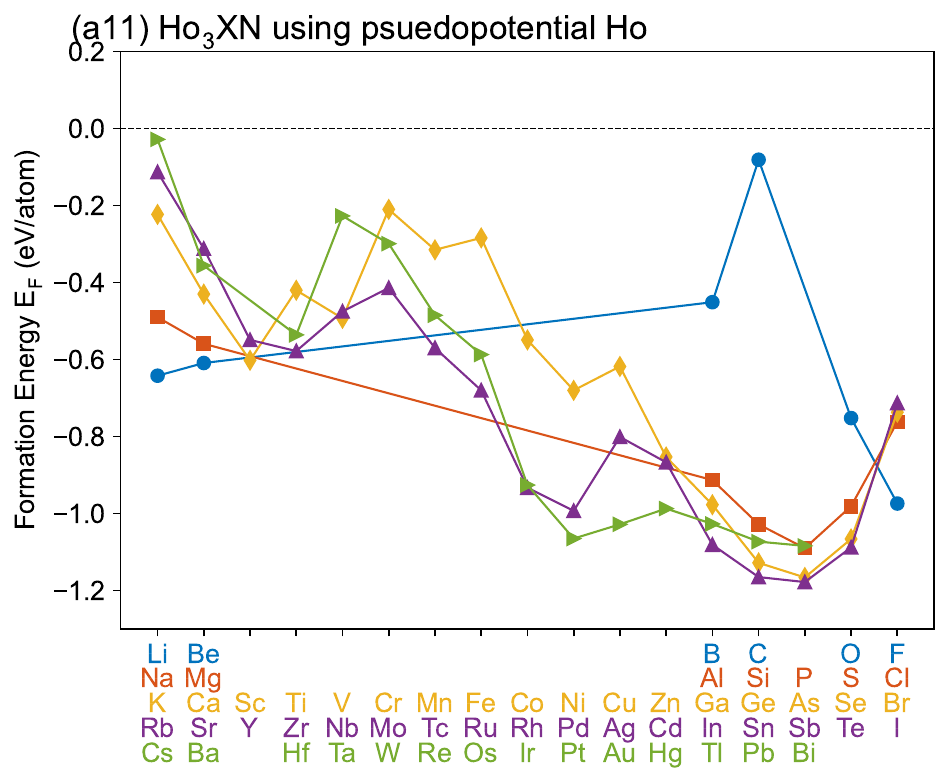}
    \includegraphics[width=0.305\linewidth]{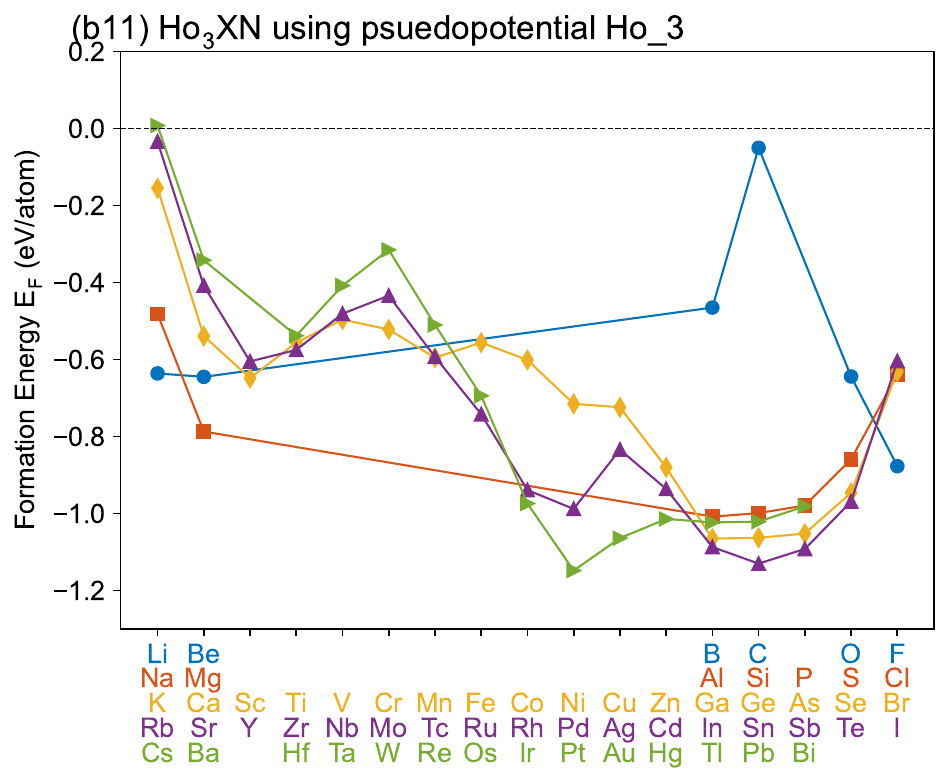}\\
    \includegraphics[width=0.305\linewidth]{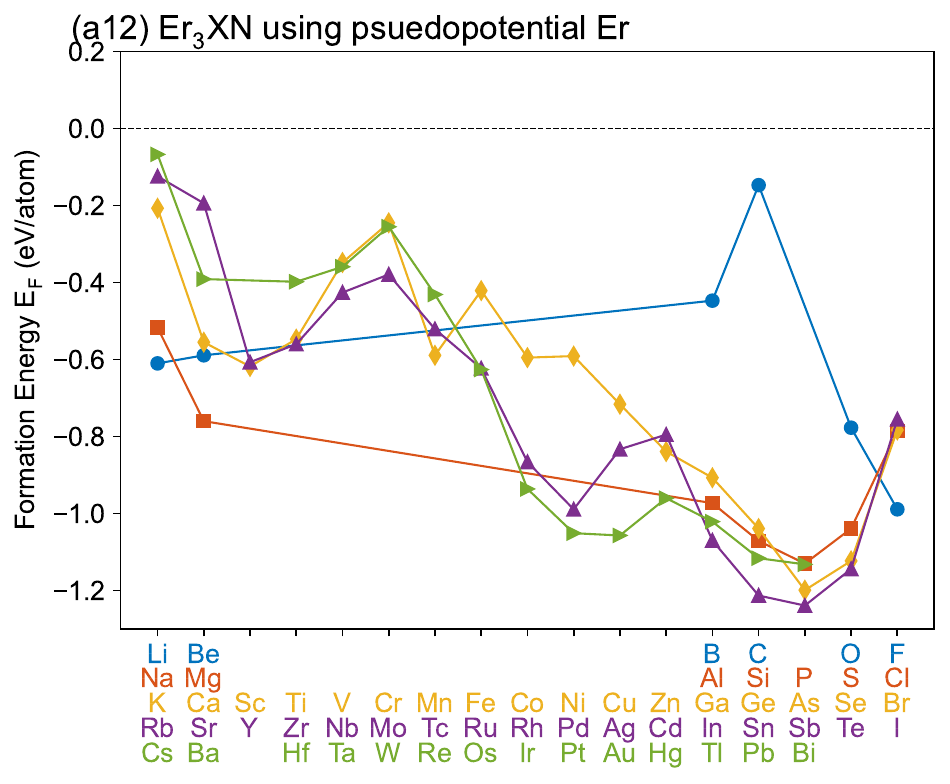}
    \includegraphics[width=0.305\linewidth]{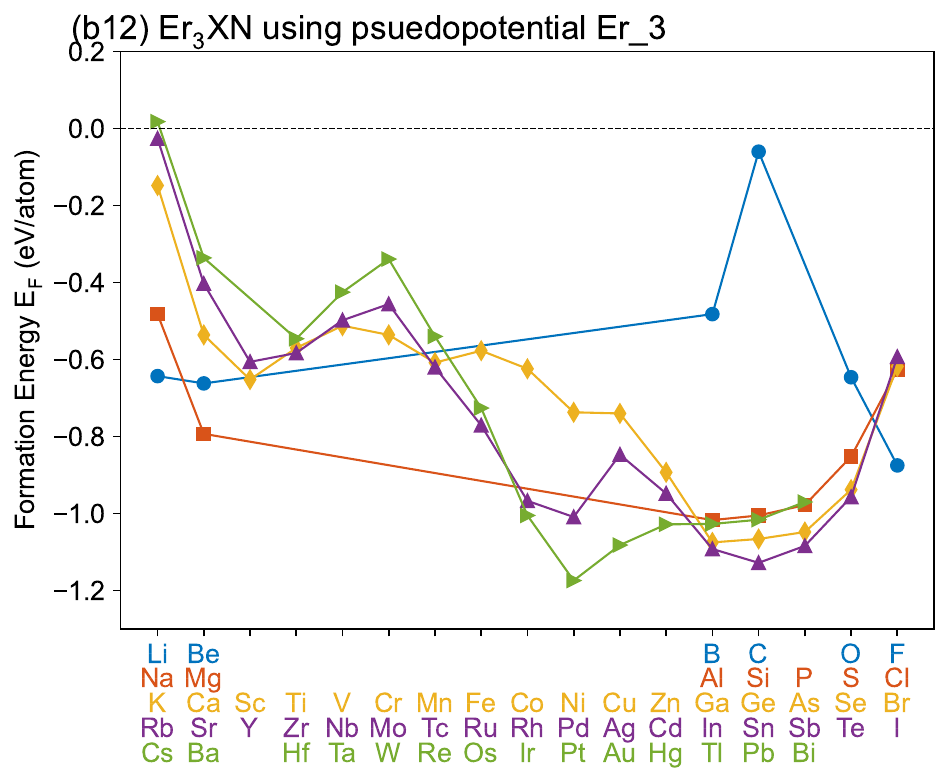}\\
    \includegraphics[width=0.305\linewidth]{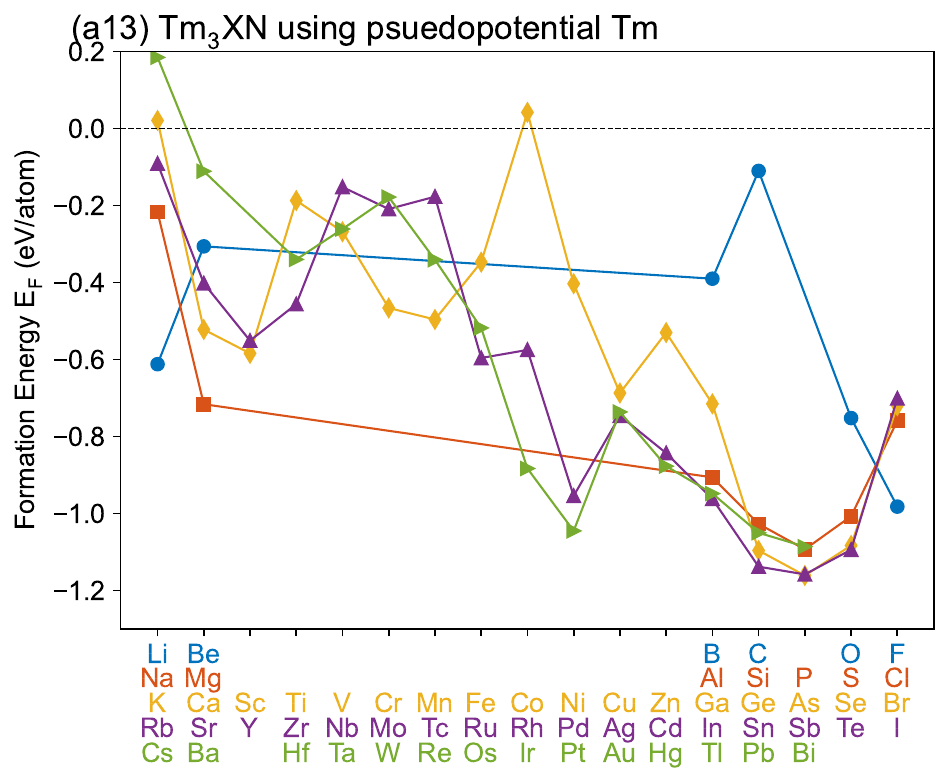}
    \includegraphics[width=0.305\linewidth]{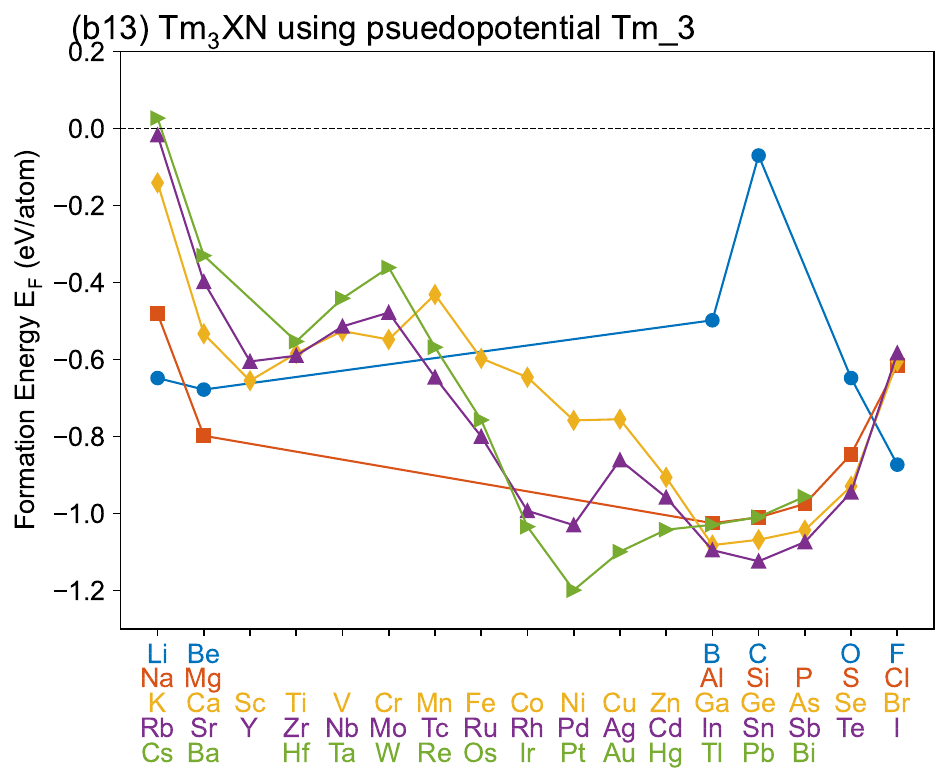}\\
\end{figure}
\clearpage
\begin{figure}[H]
\centering
    \includegraphics[width=0.305\linewidth]{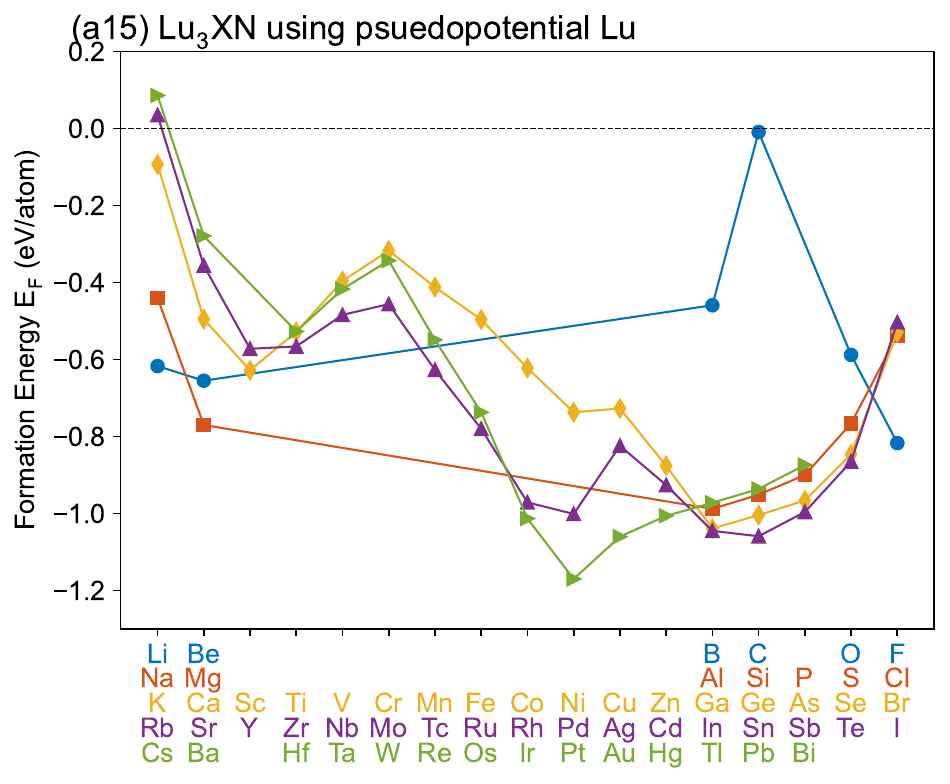}
    \includegraphics[width=0.305\linewidth]{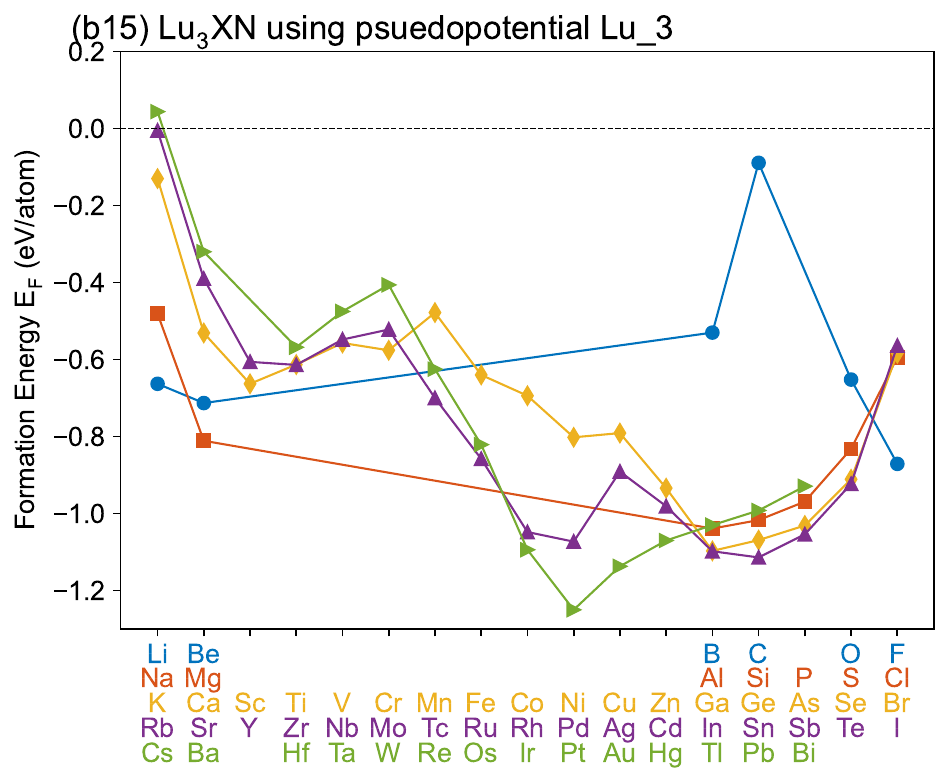}
    \caption{Calculated formation energies for lanthanide nitride antiperovskites Ln$_3$XN, using (a) itinerant psuedo-potentials and (b) localized psuedo-potentials. }
    \label{fig:SMFormation}
\end{figure}

\section{\label{sec:eabovehull}energy above the convex hull}
\justifying

\begin{figure}[H]
    \centering
    \includegraphics[width=0.305\linewidth]{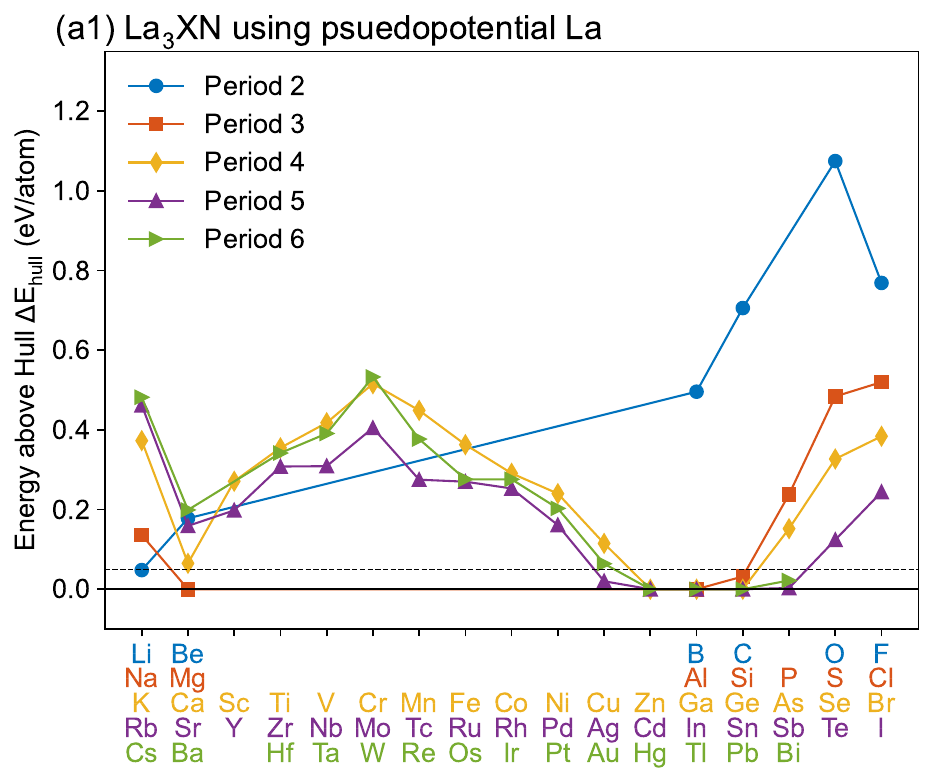}
    \includegraphics[width=0.305\linewidth]{figures/blank.pdf}\\
    \includegraphics[width=0.305\linewidth]{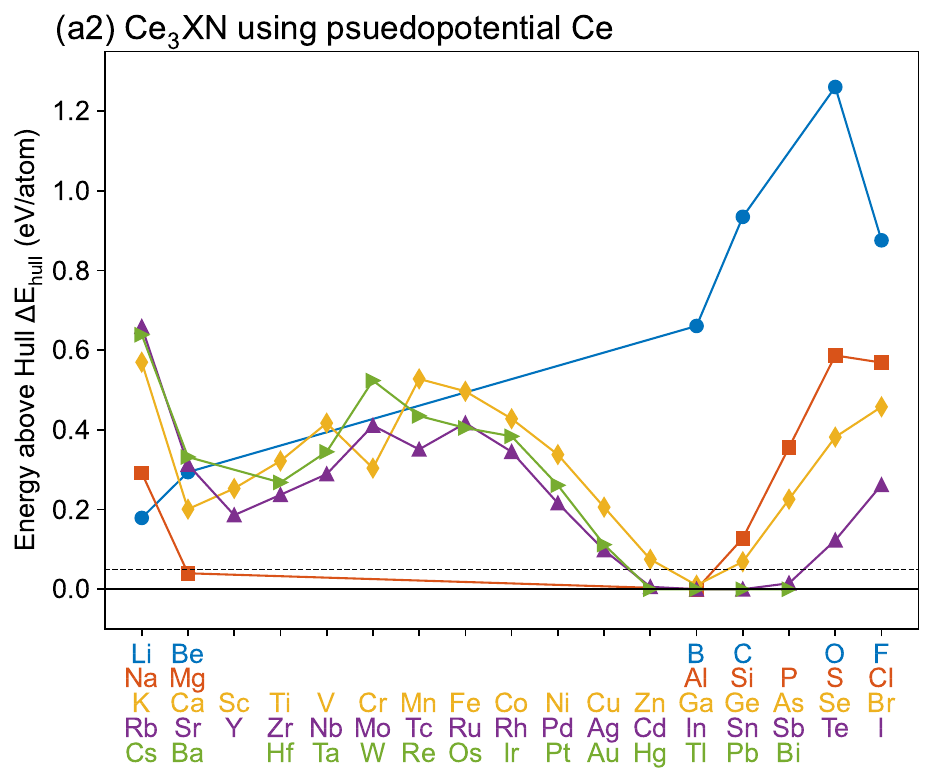}
    \includegraphics[width=0.305\linewidth]{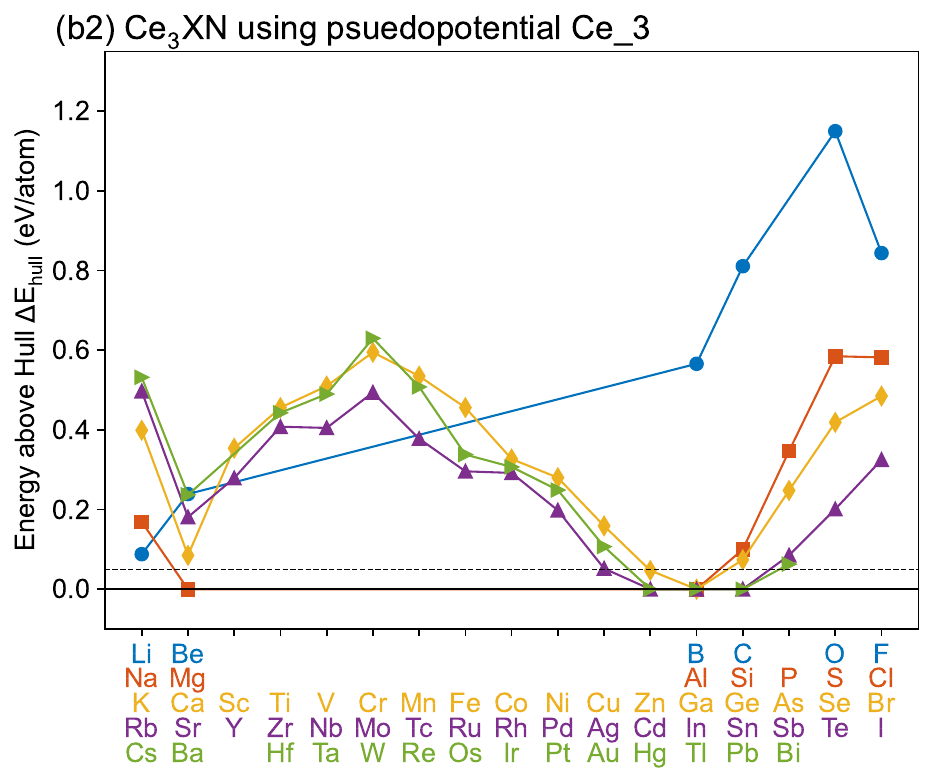}\\
    \includegraphics[width=0.305\linewidth]{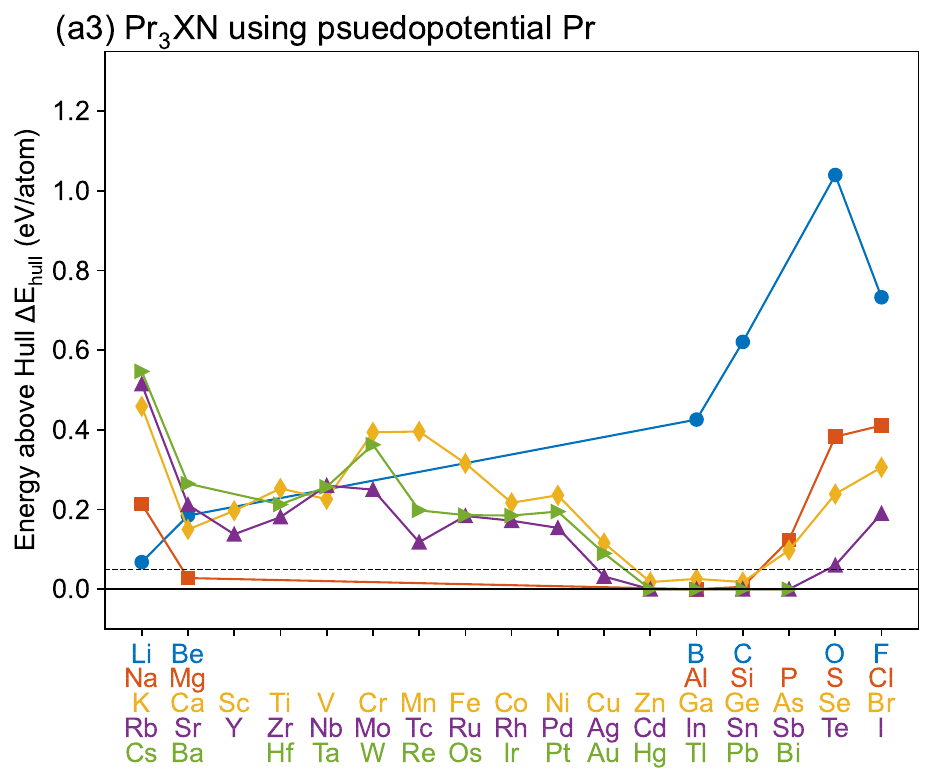}
    \includegraphics[width=0.305\linewidth]{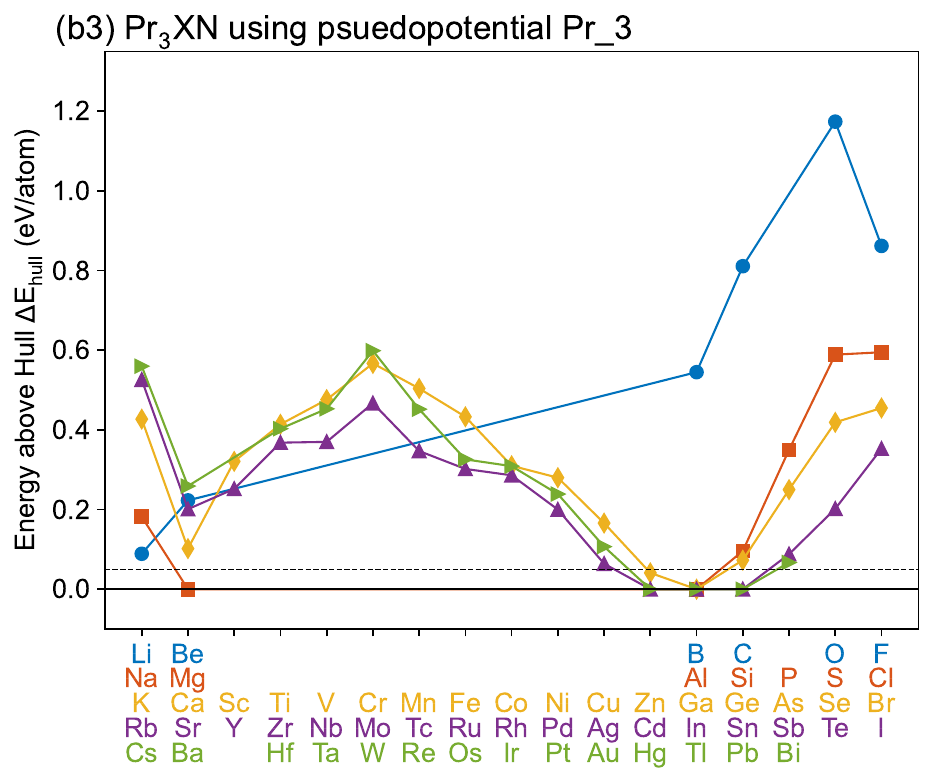}\\
\end{figure}
\clearpage
\begin{figure}[H]
\centering
    \includegraphics[width=0.305\linewidth]{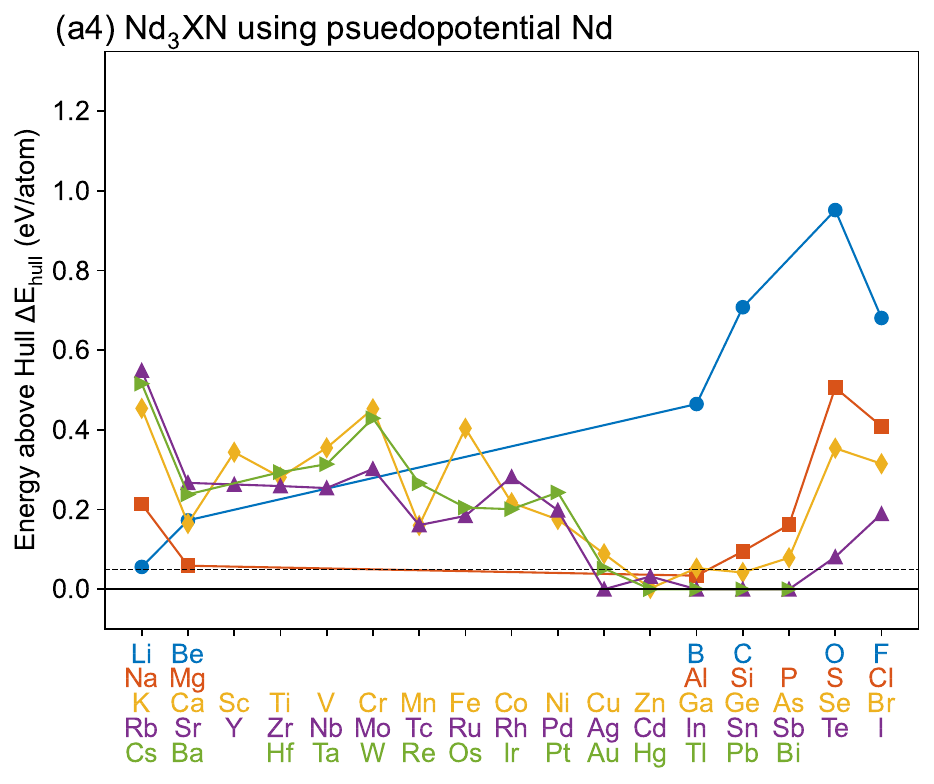}
    \includegraphics[width=0.305\linewidth]{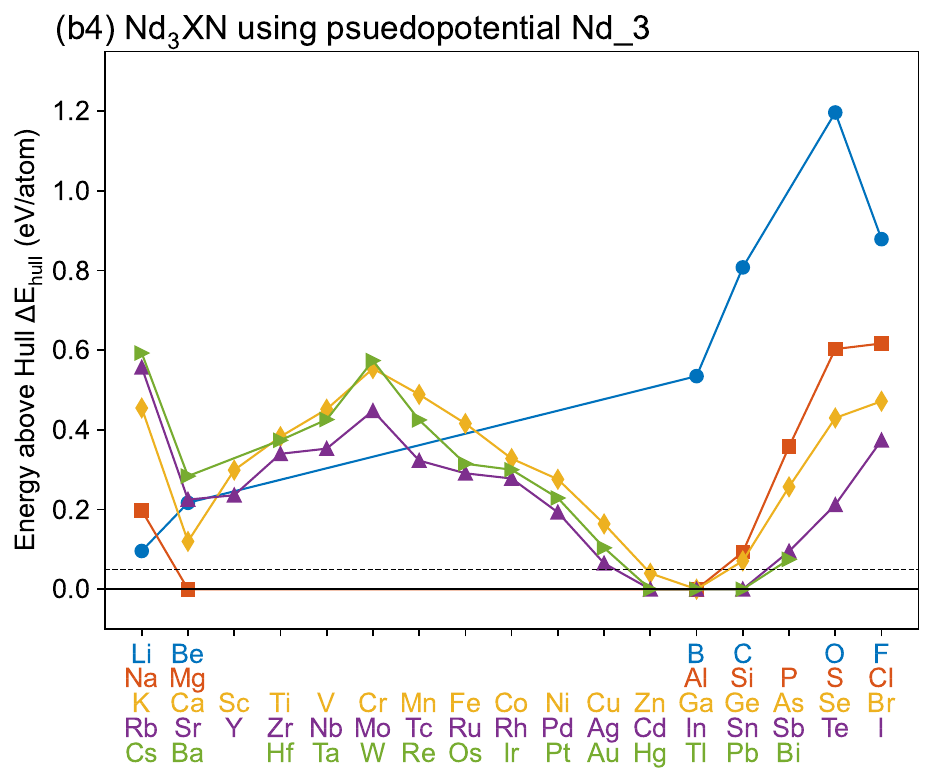}\\
    \includegraphics[width=0.305\linewidth]{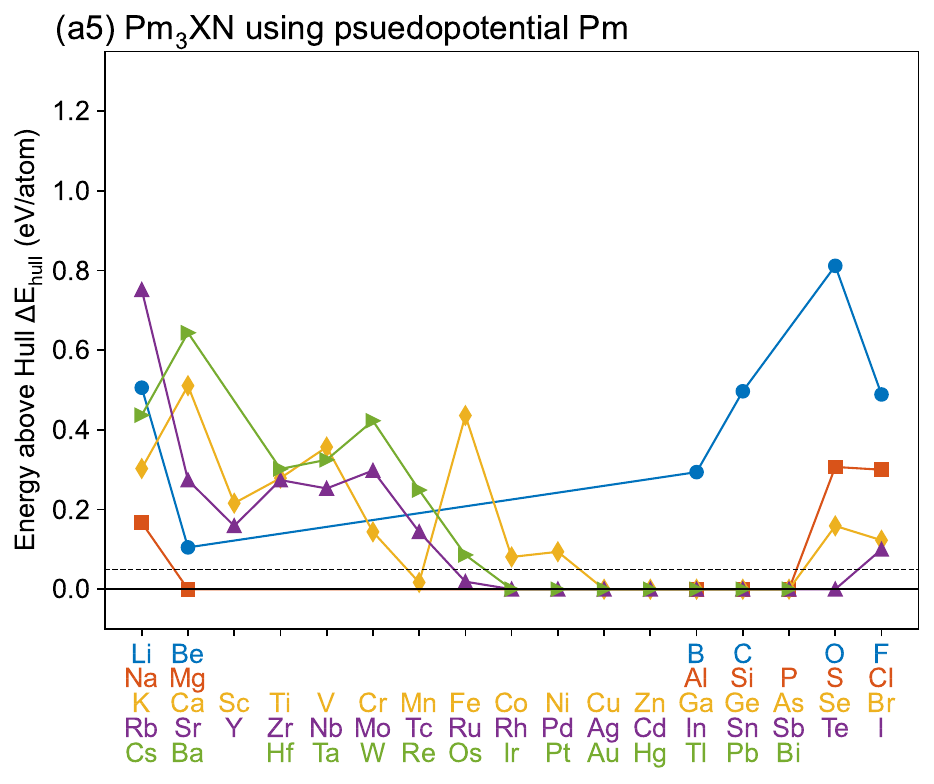}
    \includegraphics[width=0.305\linewidth]{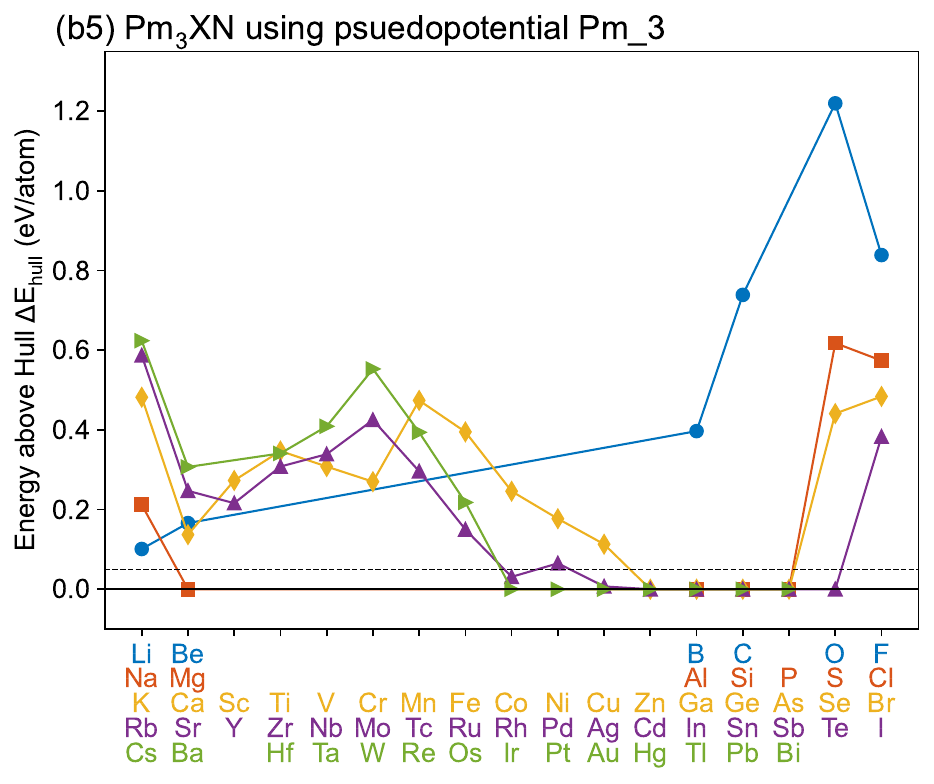}\\
    \includegraphics[width=0.305\linewidth]{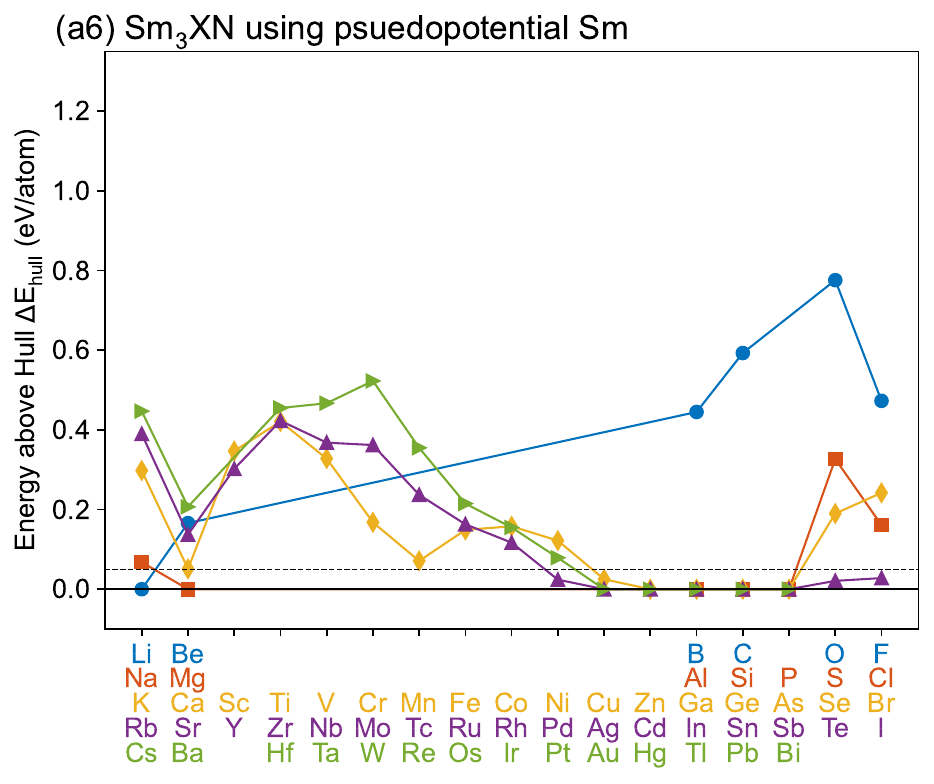}
    \includegraphics[width=0.305\linewidth]{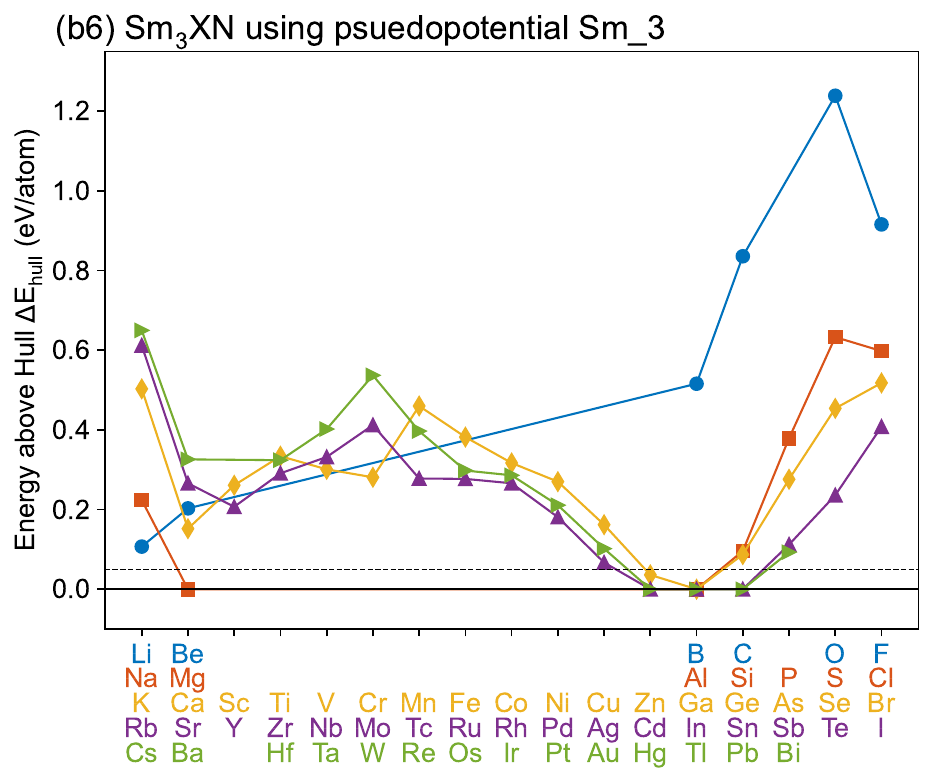}\\
    \includegraphics[width=0.305\linewidth]{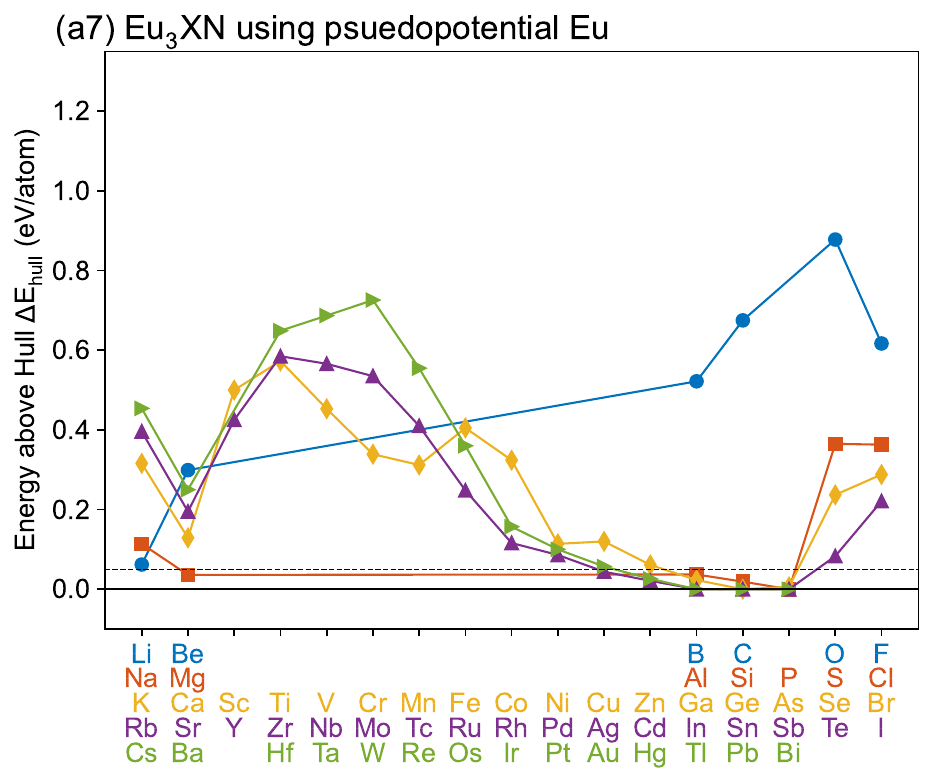}
    \includegraphics[width=0.305\linewidth]{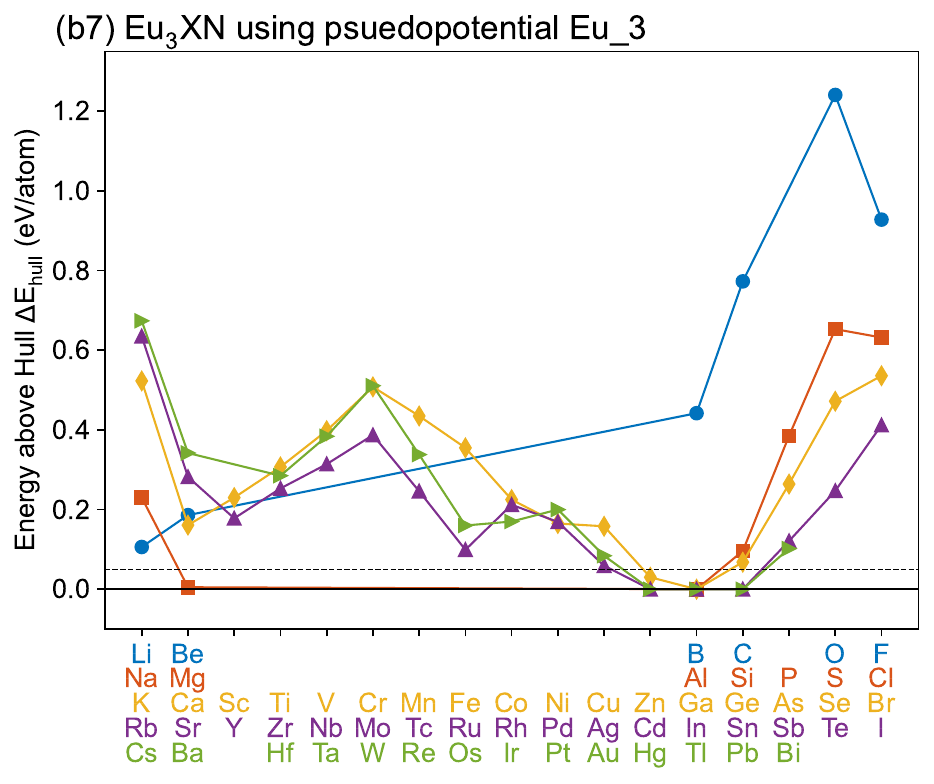}\\
    \includegraphics[width=0.305\linewidth]{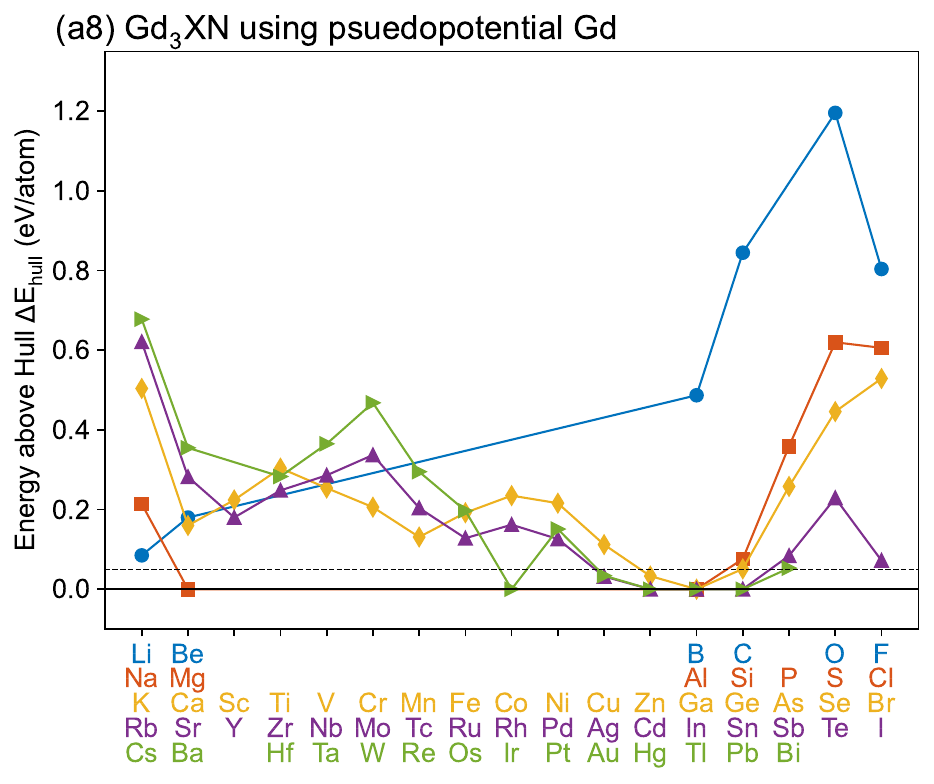}
    \includegraphics[width=0.305\linewidth]{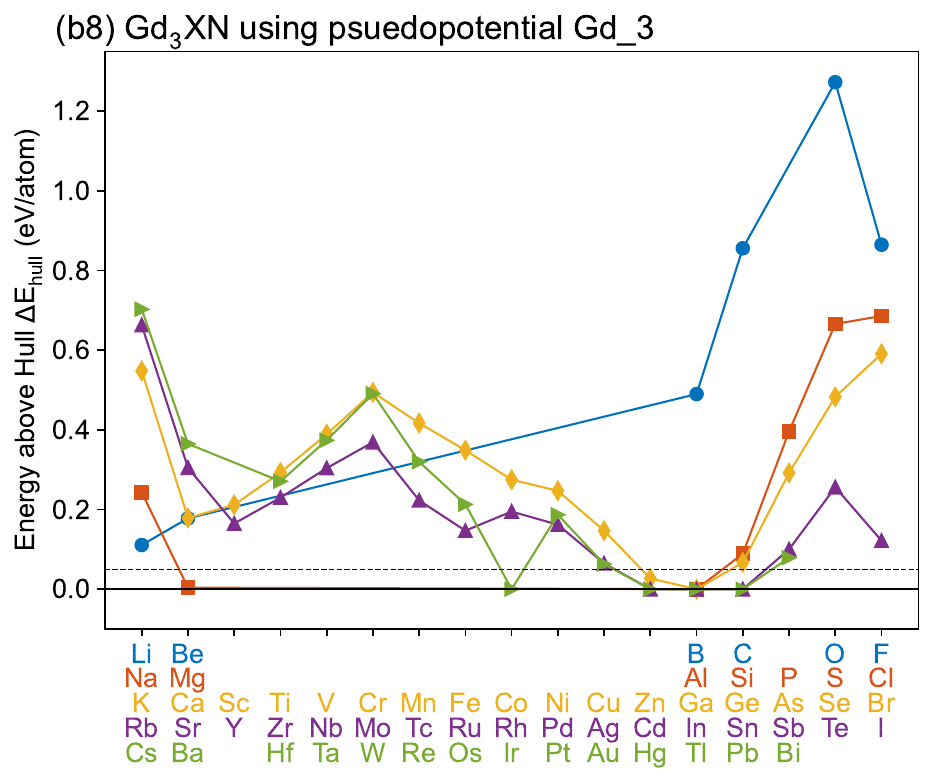}\\
\end{figure}
\clearpage
\begin{figure}[H]
\centering
    \includegraphics[width=0.305\linewidth]{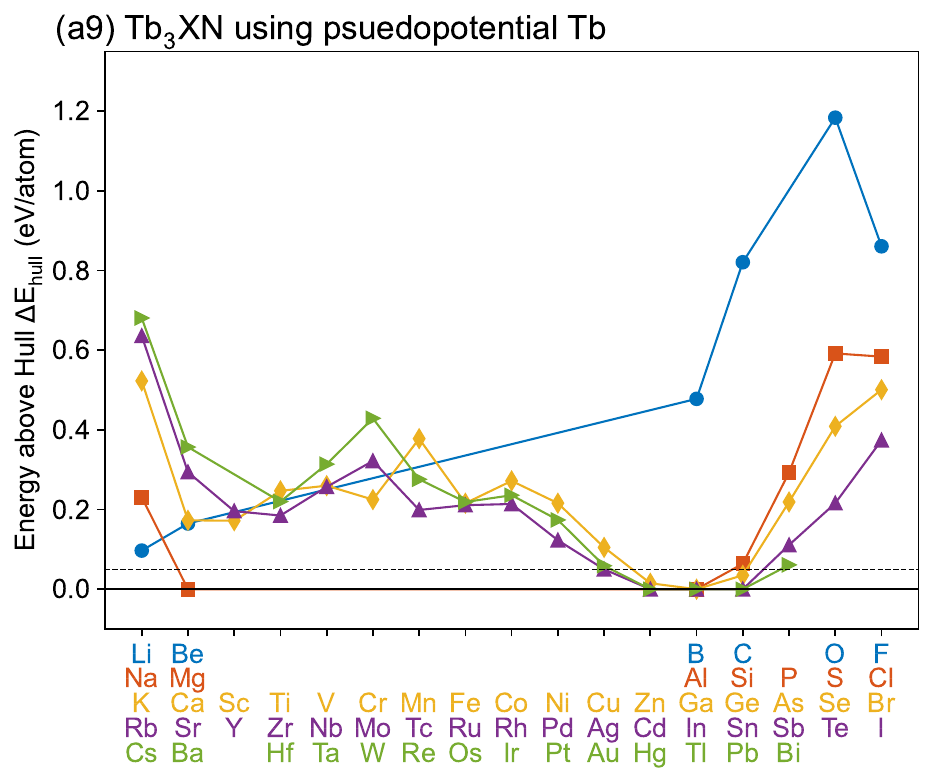}
    \includegraphics[width=0.305\linewidth]{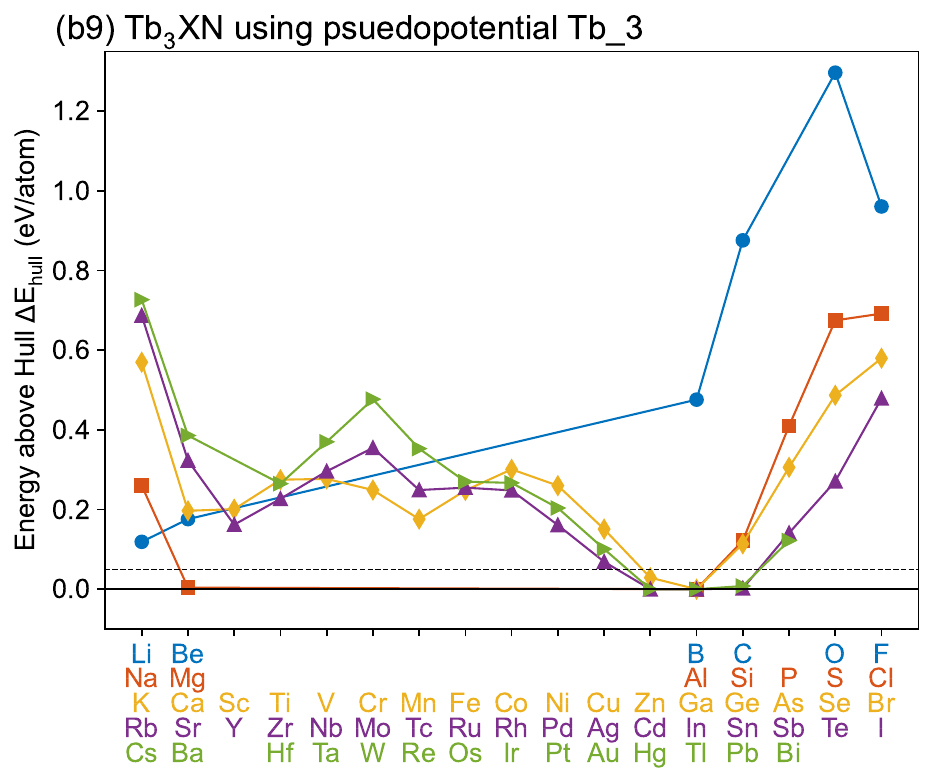}\\
    \includegraphics[width=0.305\linewidth]{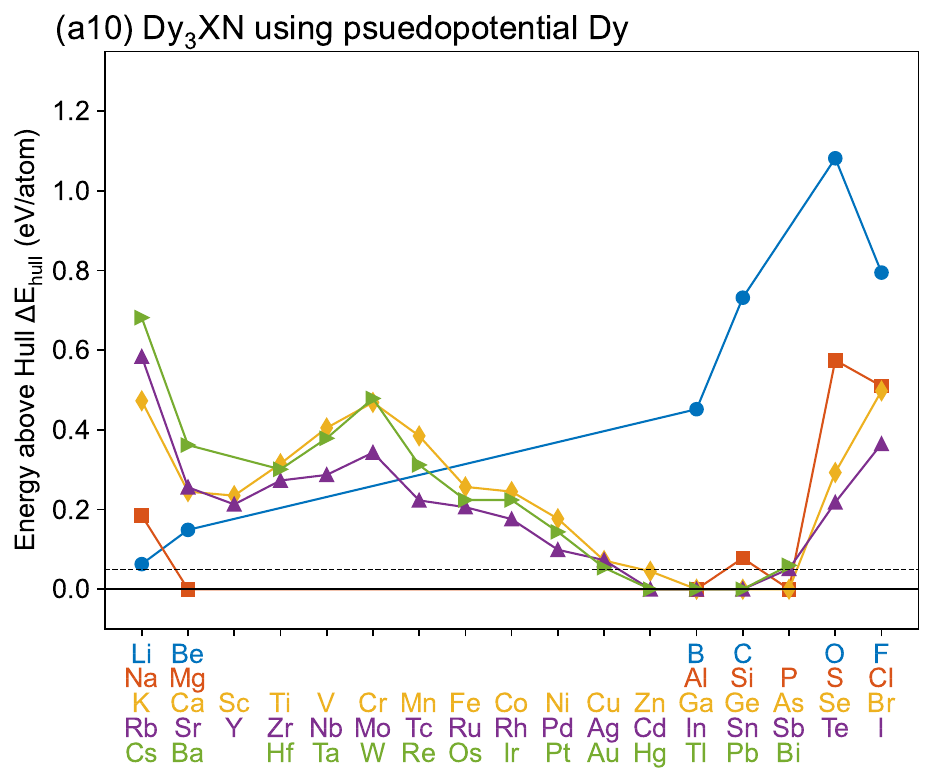}
    \includegraphics[width=0.305\linewidth]{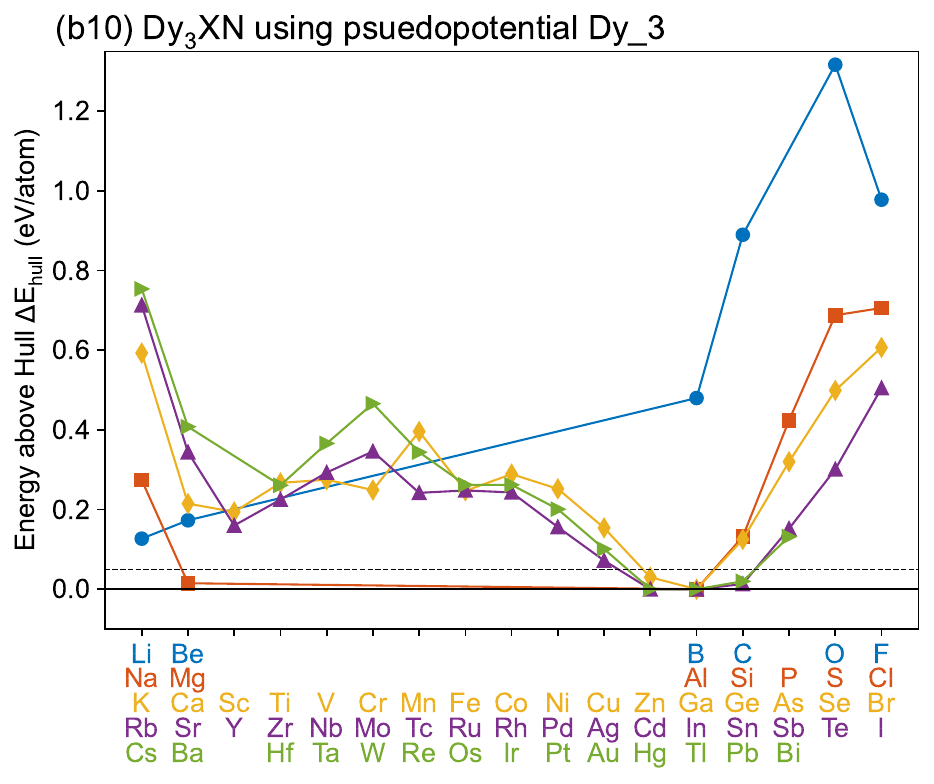}\\
    \includegraphics[width=0.305\linewidth]{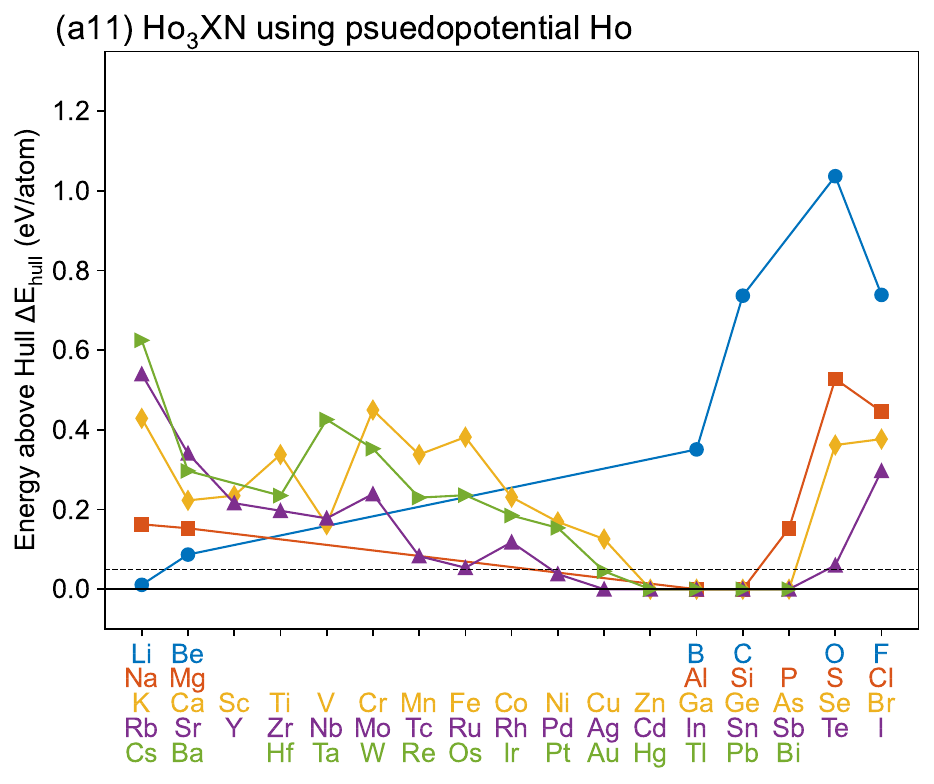}
    \includegraphics[width=0.305\linewidth]{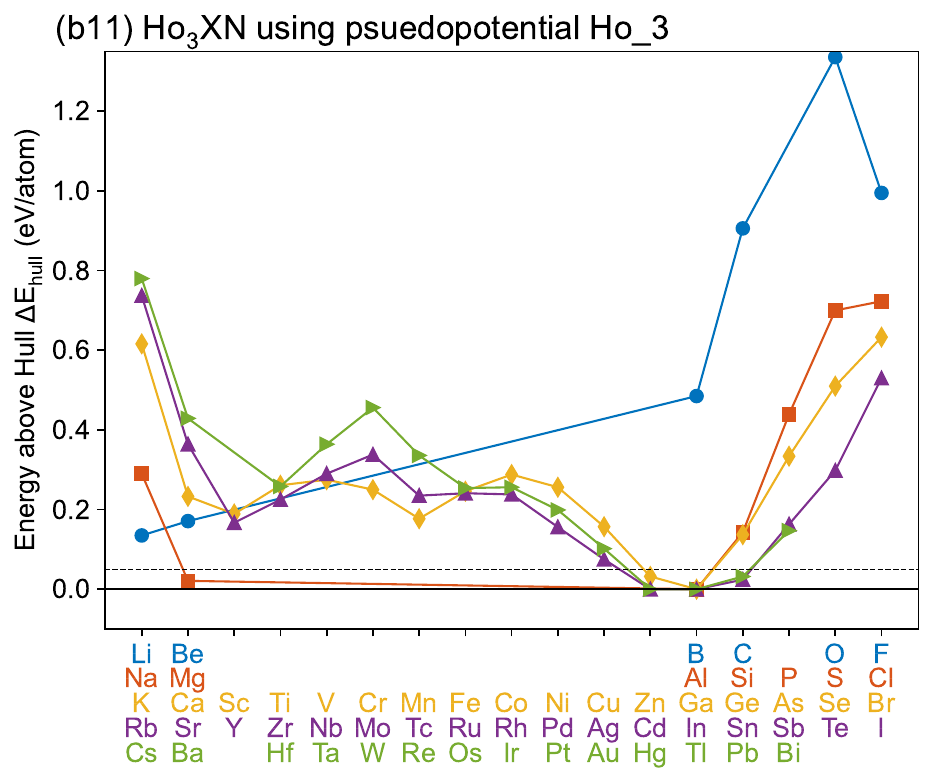}\\
    \includegraphics[width=0.305\linewidth]{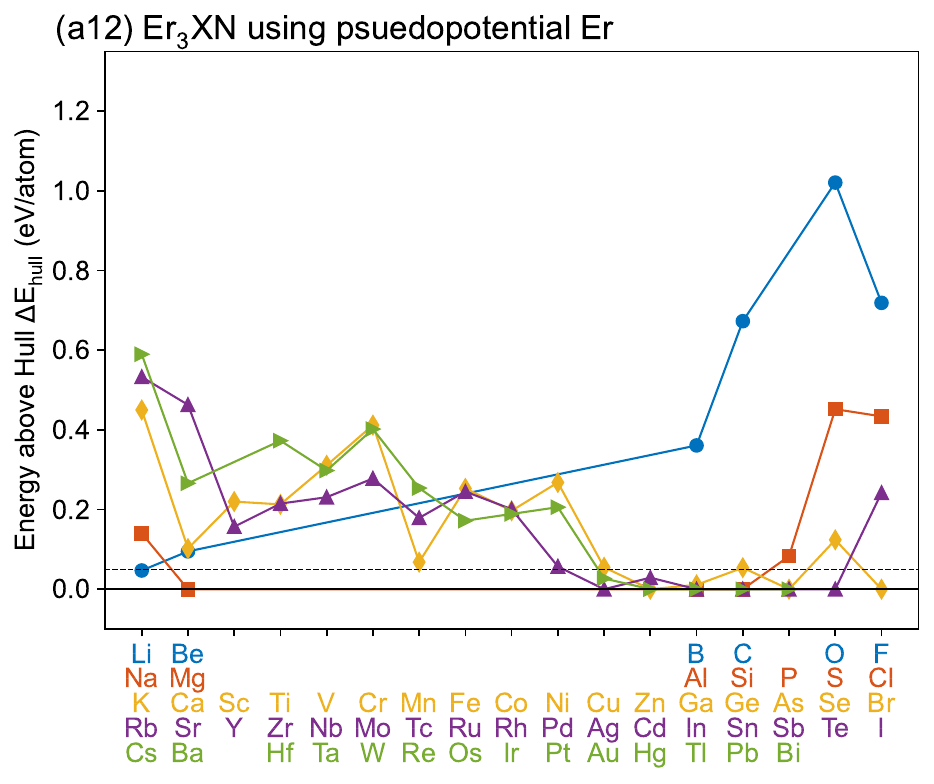}
    \includegraphics[width=0.305\linewidth]{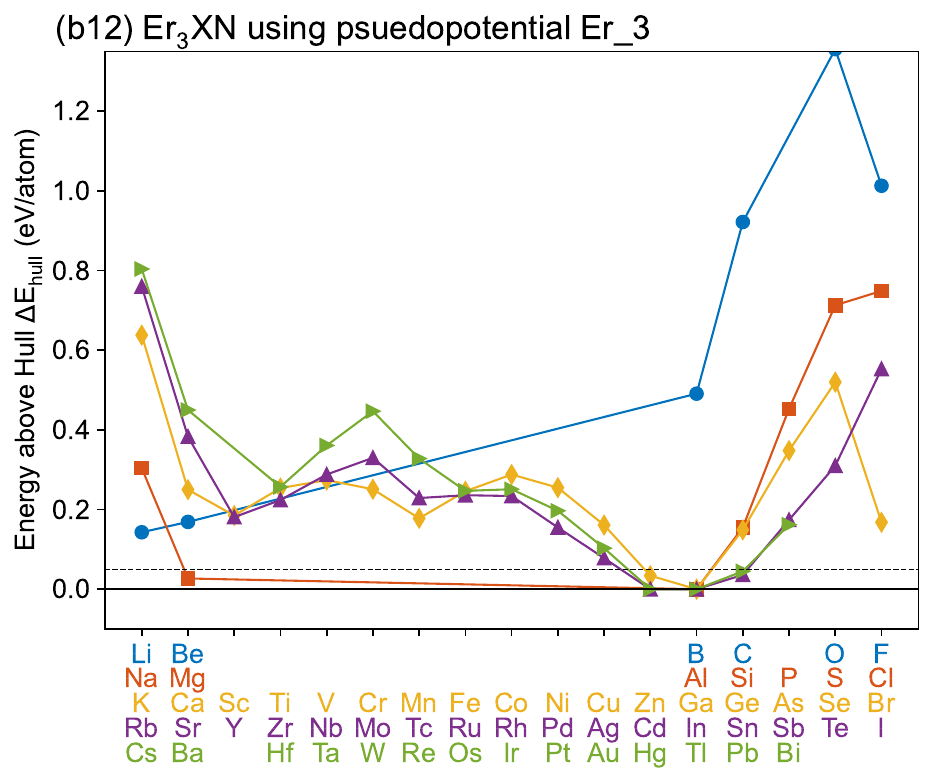}\\
    \includegraphics[width=0.305\linewidth]{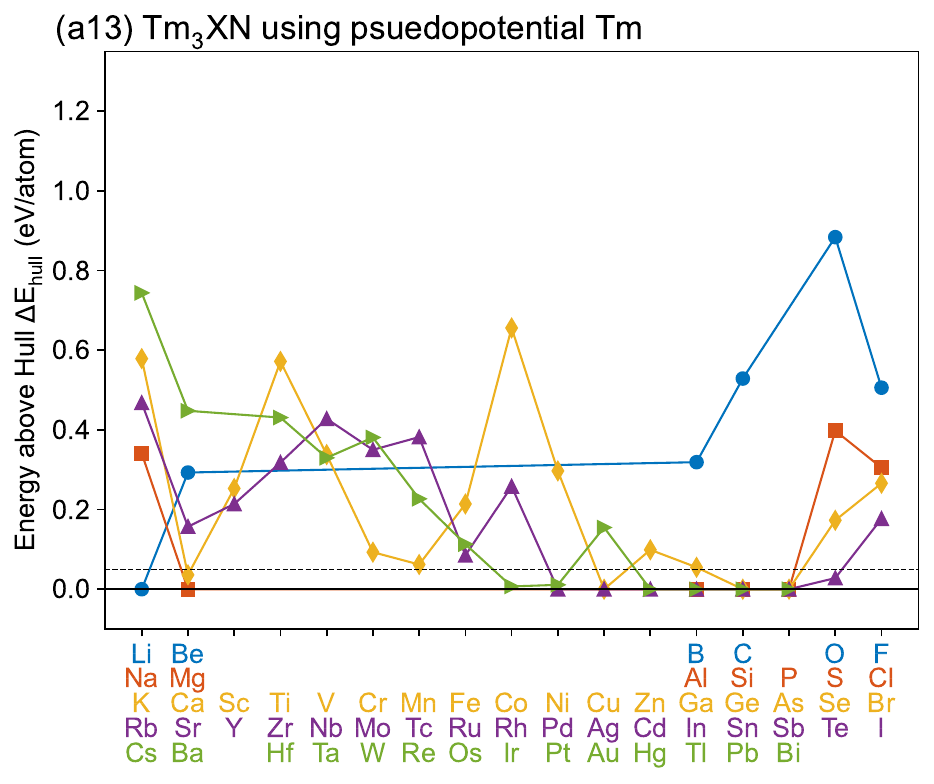}
    \includegraphics[width=0.305\linewidth]{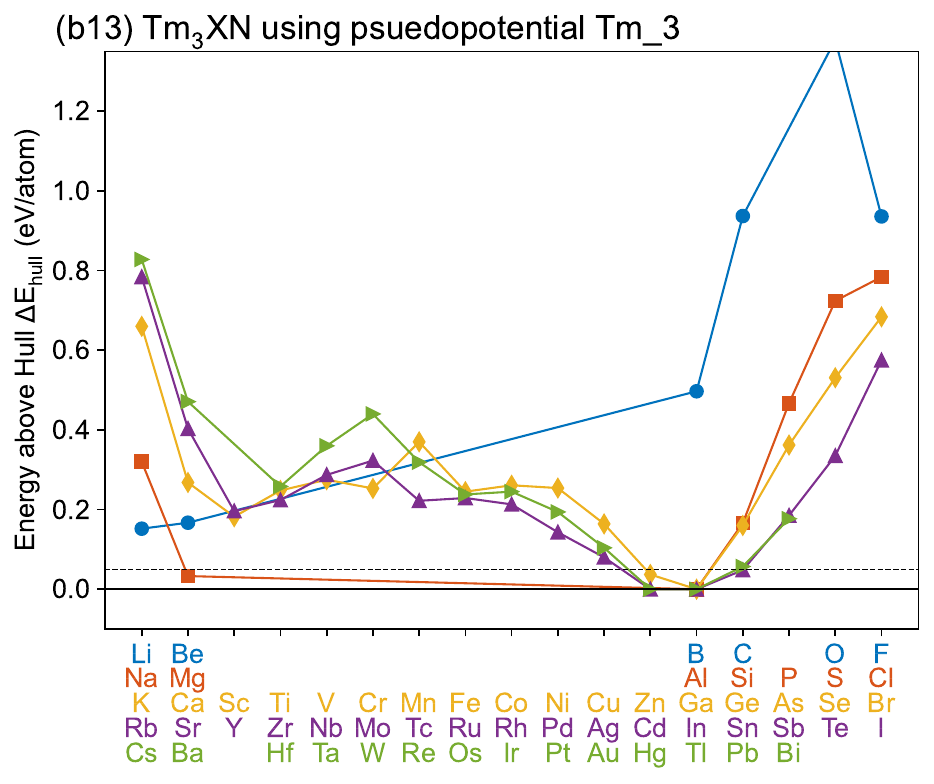}\\
\end{figure}
\clearpage
\begin{figure}[H]
\centering
    \includegraphics[width=0.305\linewidth]{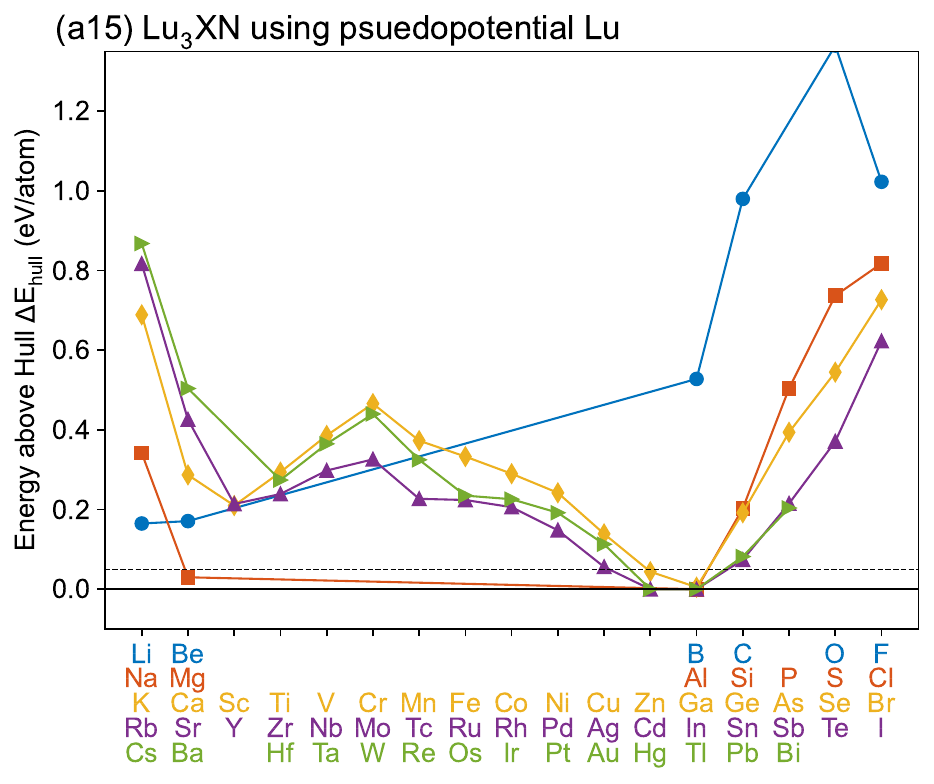}
    \includegraphics[width=0.305\linewidth]{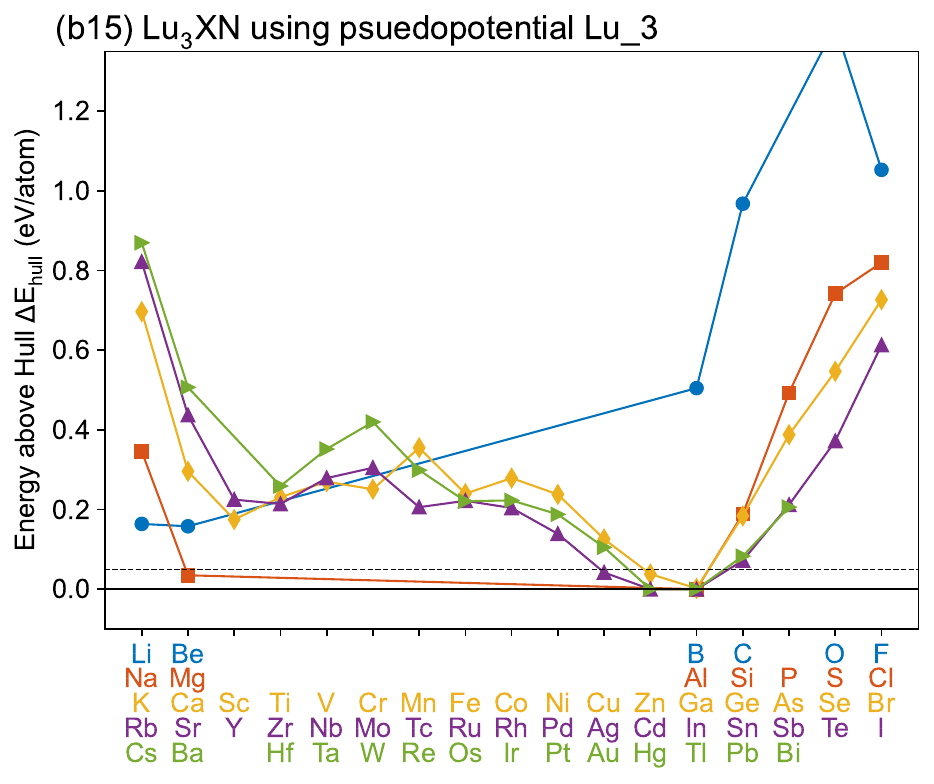}
    \caption{Calculated energy above hull values for lanthanide nitride antiperovskites Ln$_3$XN, using (a) itinerant psuedo-potentials and (b) localized psuedo-potentials. }
    \label{fig:SMabovehull}
\end{figure}

\section{\label{sec:phonon}Phonon dispersion}
\justifying
\begin{figure}[H]
    \includegraphics[width=0.19\linewidth]{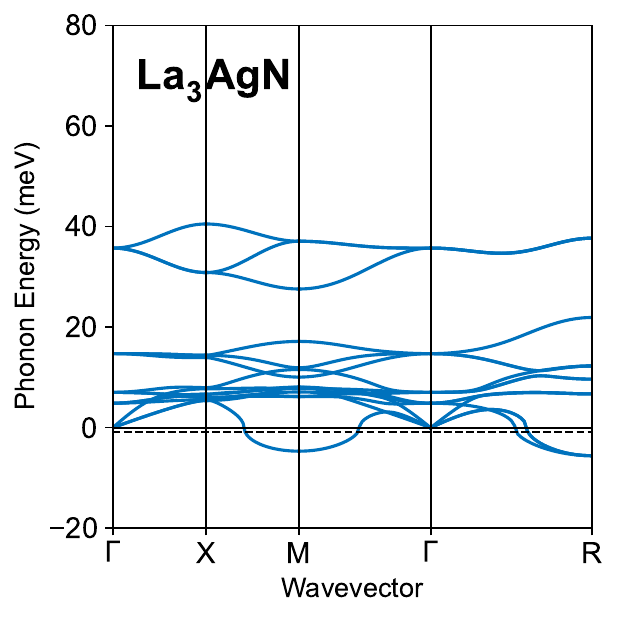}
    \includegraphics[width=0.19\linewidth]{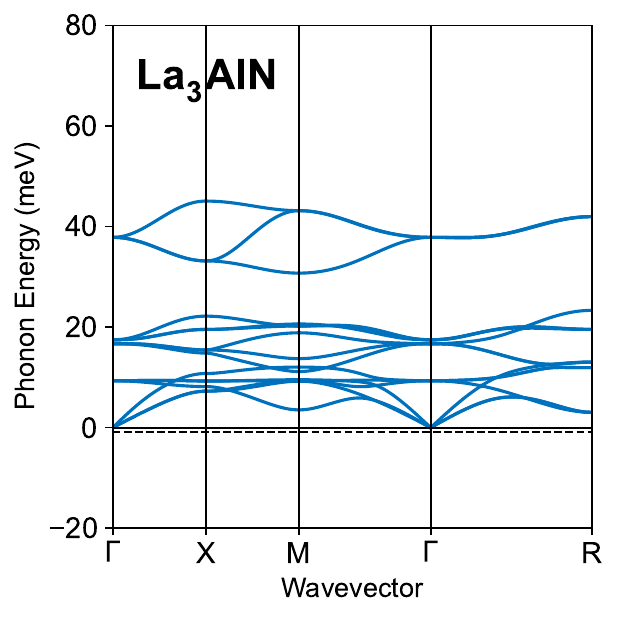}
    \includegraphics[width=0.19\linewidth]{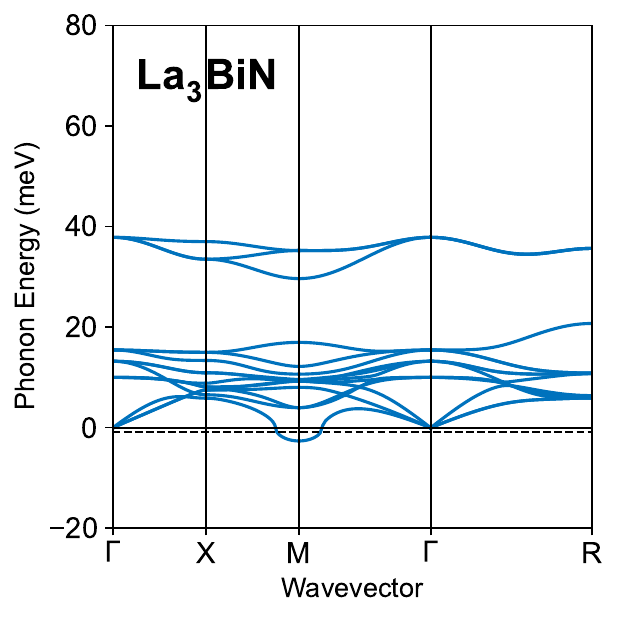}
    \includegraphics[width=0.19\linewidth]{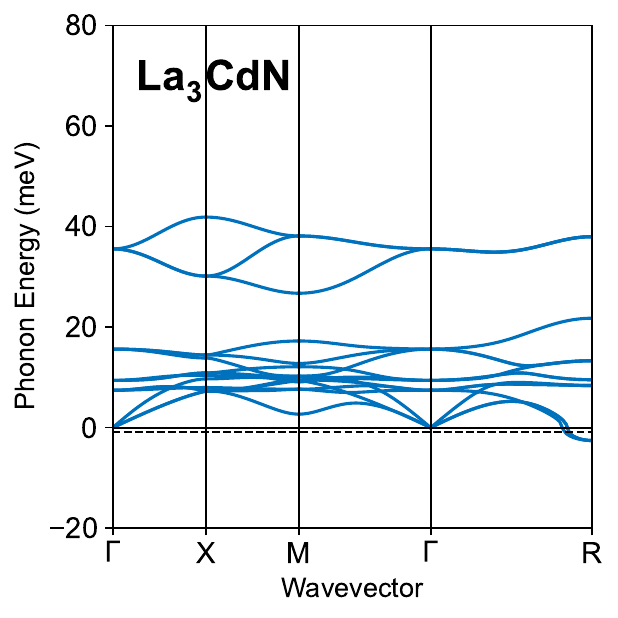}
    \includegraphics[width=0.19\linewidth]{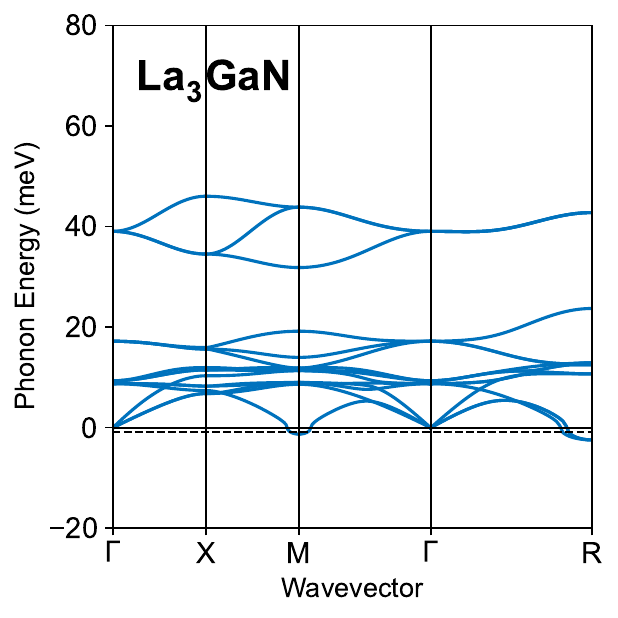}
\\
    \includegraphics[width=0.19\linewidth]{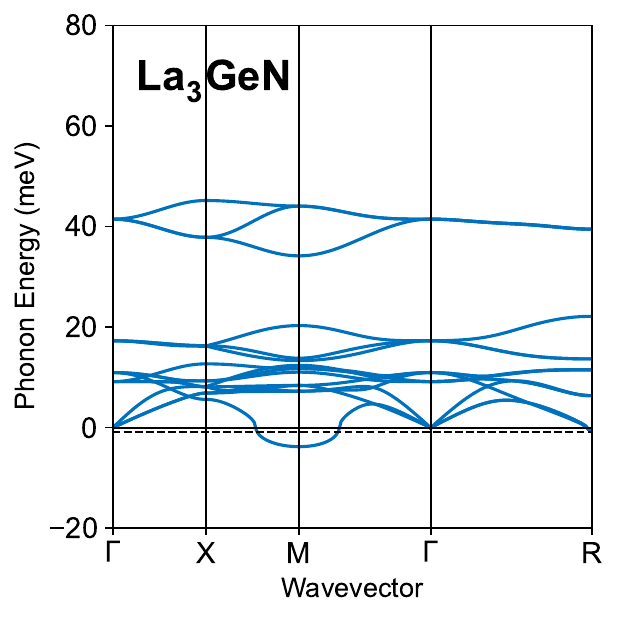}
    \includegraphics[width=0.19\linewidth]{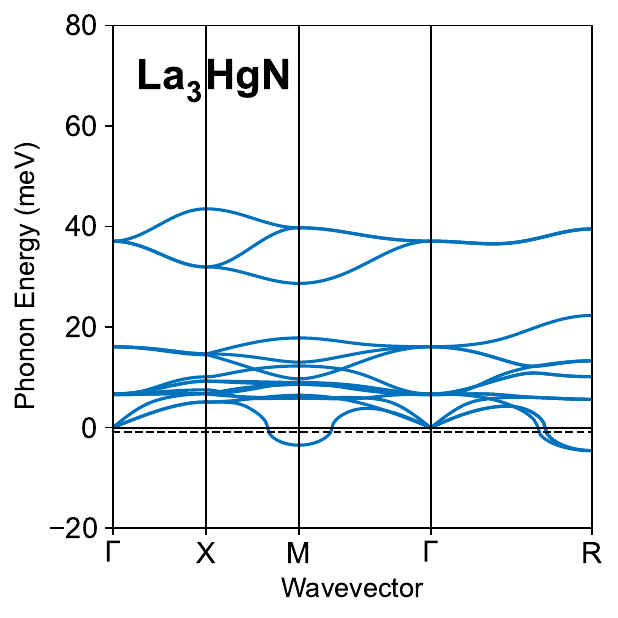}
    \includegraphics[width=0.19\linewidth]{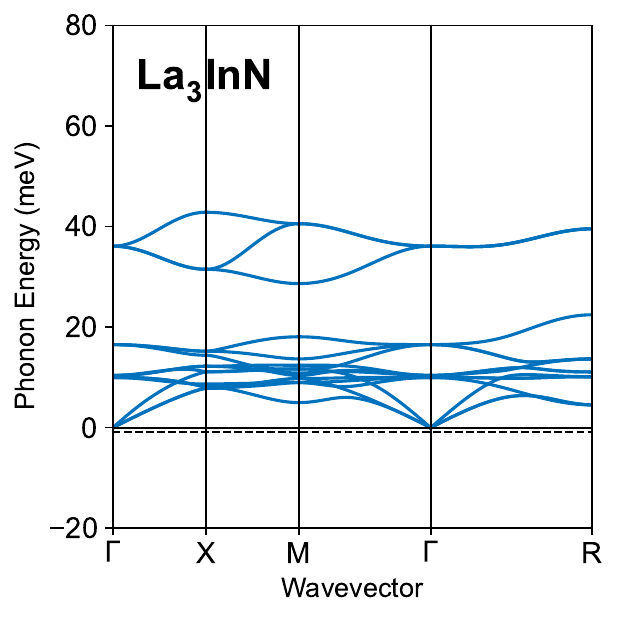}
    \includegraphics[width=0.19\linewidth]{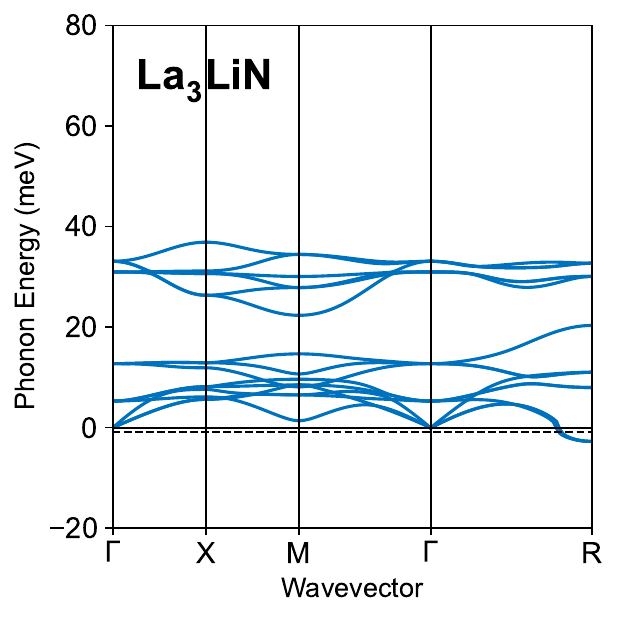}
    \includegraphics[width=0.19\linewidth]{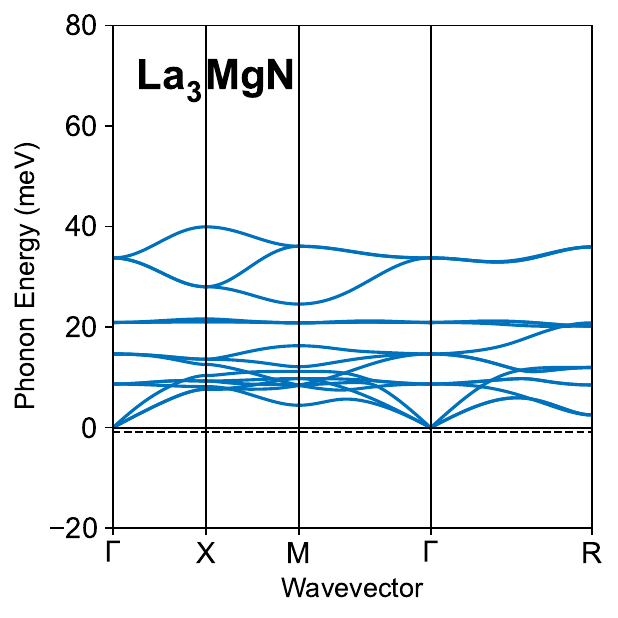}
\\
    \includegraphics[width=0.19\linewidth]{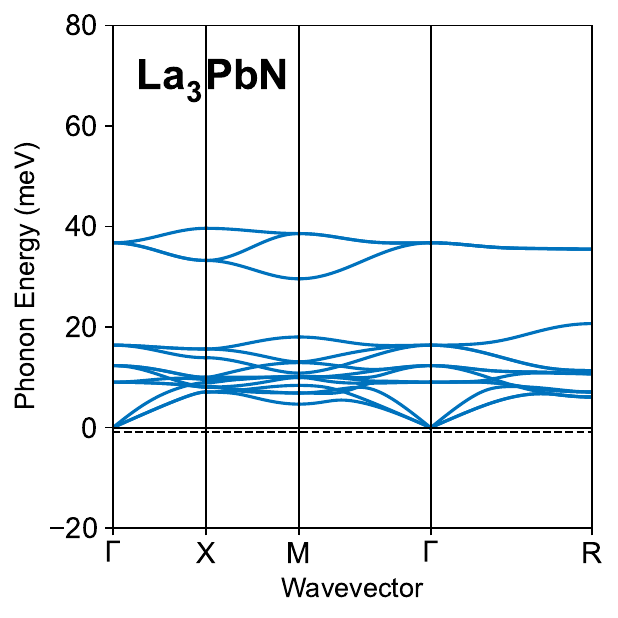}
    \includegraphics[width=0.19\linewidth]{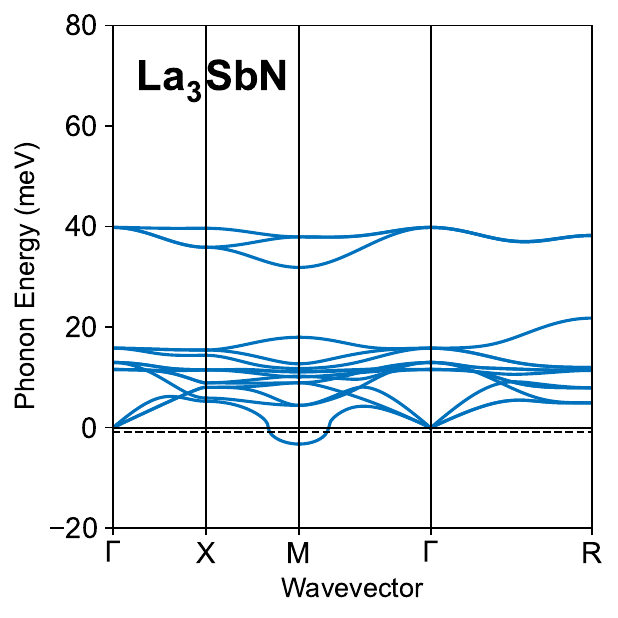}
    \includegraphics[width=0.19\linewidth]{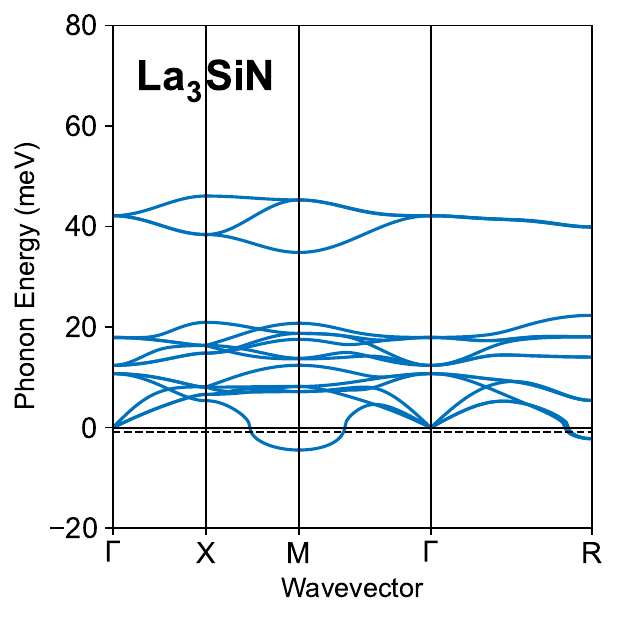}
    \includegraphics[width=0.19\linewidth]{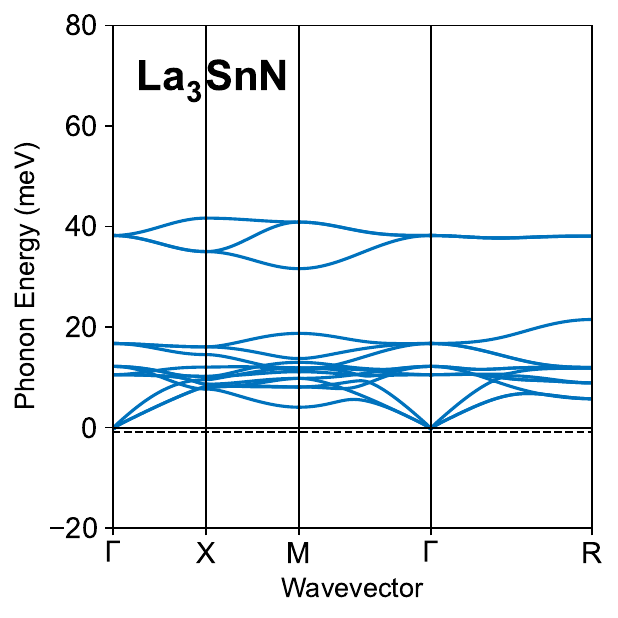}
    \includegraphics[width=0.19\linewidth]{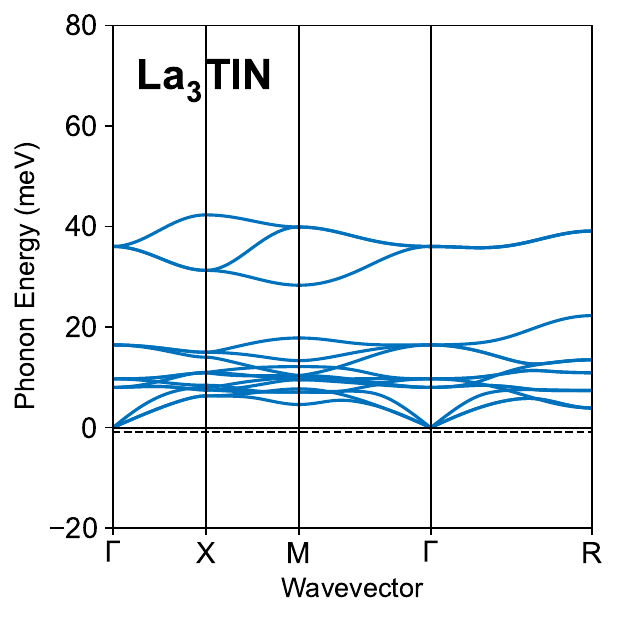}
\\
    \includegraphics[width=0.19\linewidth]
    {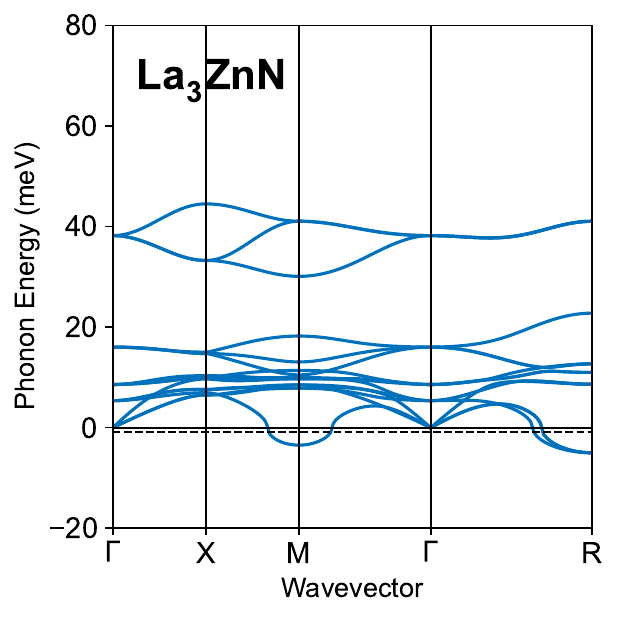}
    \includegraphics[width=0.19\linewidth]{figures/blank.pdf}
    \includegraphics[width=0.19\linewidth]{figures/blank.pdf}
    \includegraphics[width=0.19\linewidth]{figures/blank.pdf}
    \includegraphics[width=0.19\linewidth]{figures/blank.pdf}
\end{figure}
\clearpage
\begin{figure}[H]
    \includegraphics[width=0.19\linewidth]{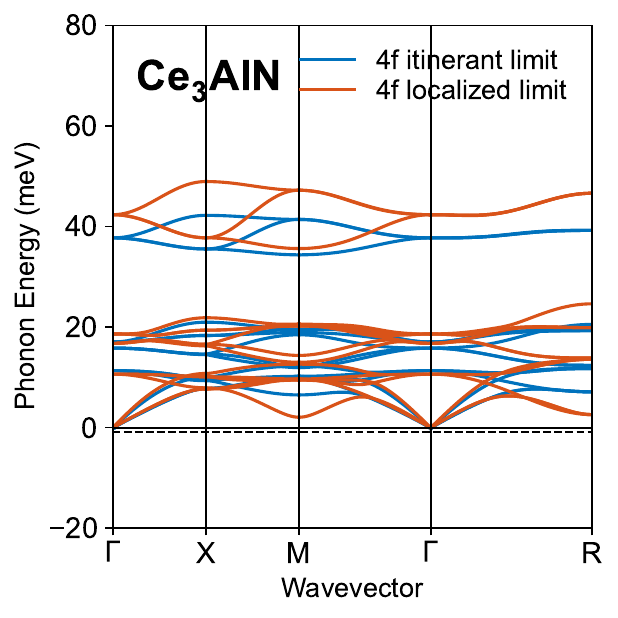}
    \includegraphics[width=0.19\linewidth]{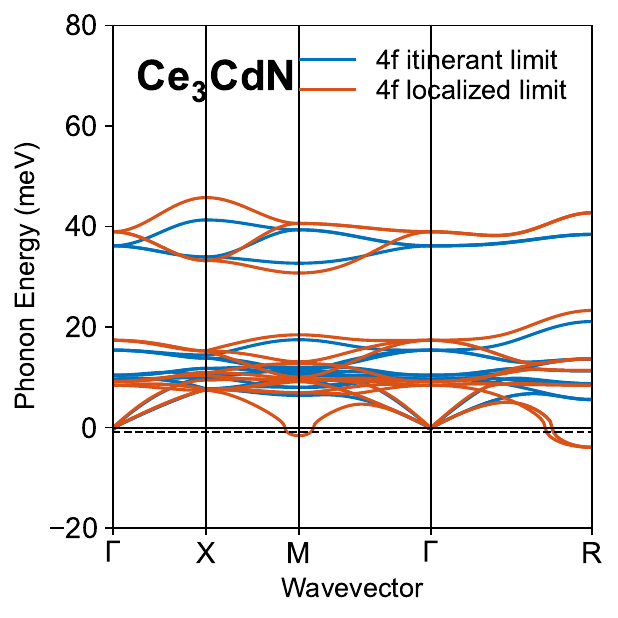}
    \includegraphics[width=0.19\linewidth]{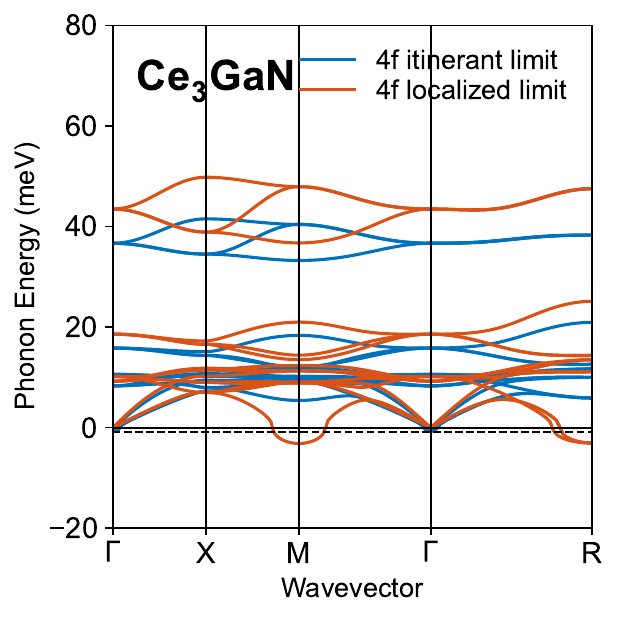}
    \includegraphics[width=0.19\linewidth]{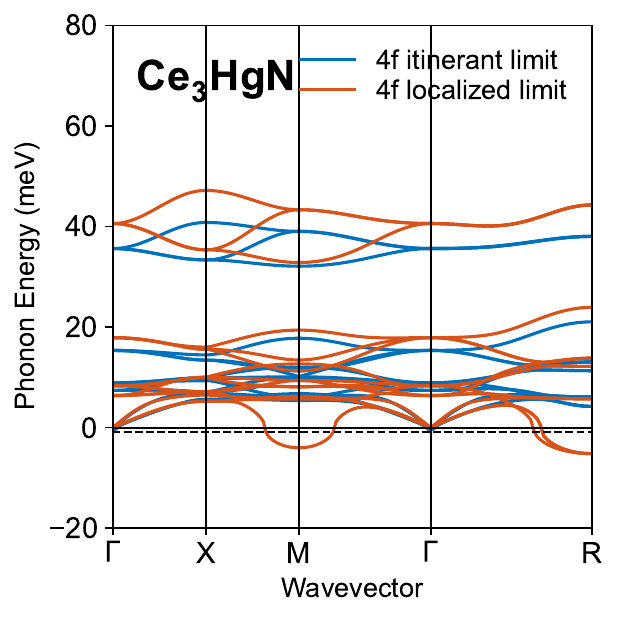}
    \includegraphics[width=0.19\linewidth]{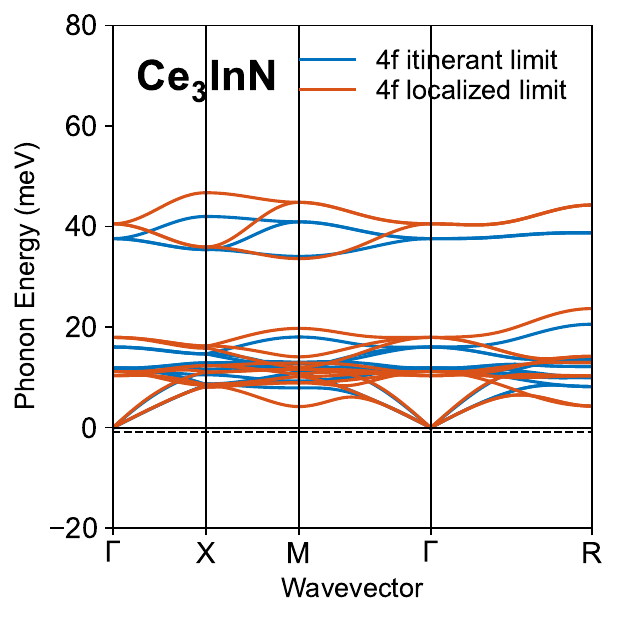}
\\
    \includegraphics[width=0.19\linewidth]{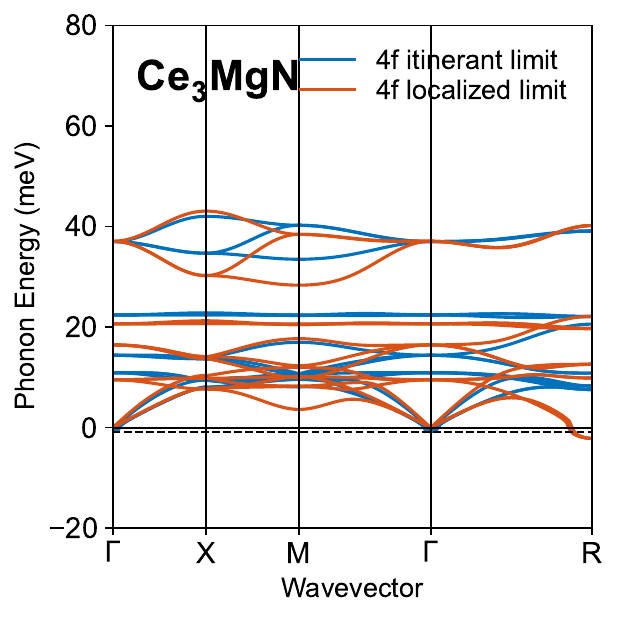}
    \includegraphics[width=0.19\linewidth]{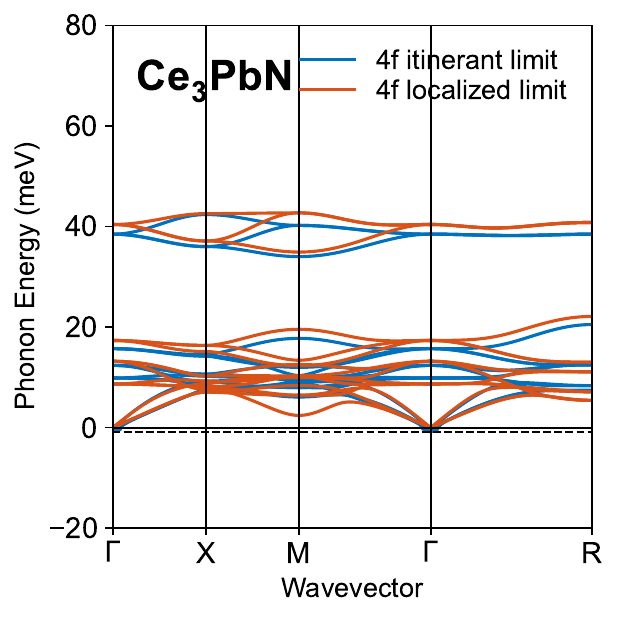}
    \includegraphics[width=0.19\linewidth]{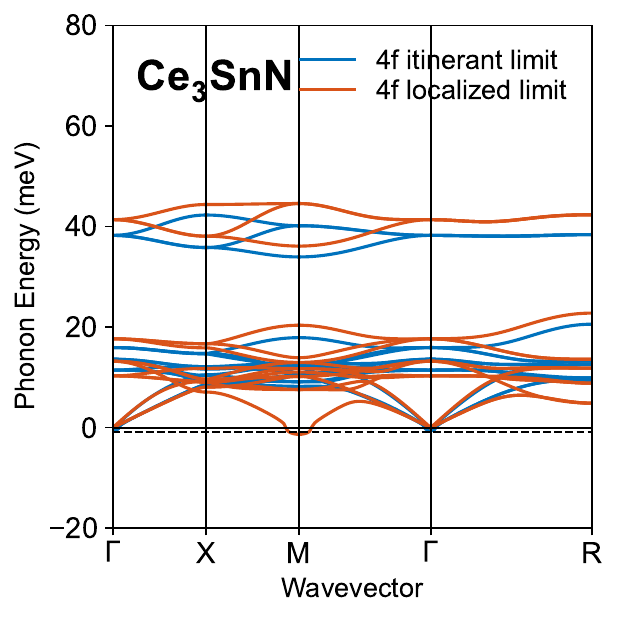}
    \includegraphics[width=0.19\linewidth]{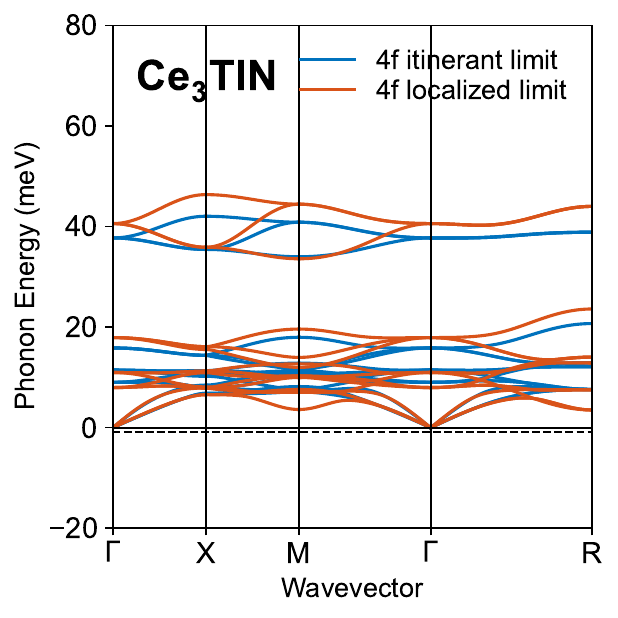}
    \includegraphics[width=0.19\linewidth]{figures/blank.pdf}
\\
    \includegraphics[width=0.19\linewidth]{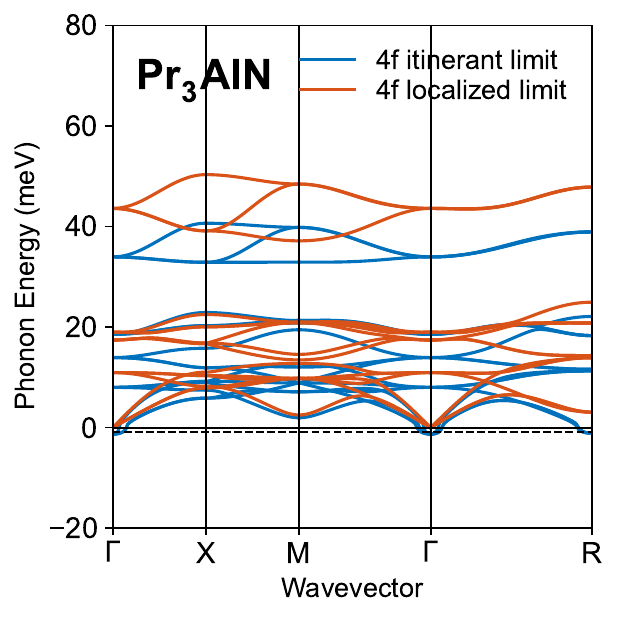}
    \includegraphics[width=0.19\linewidth]{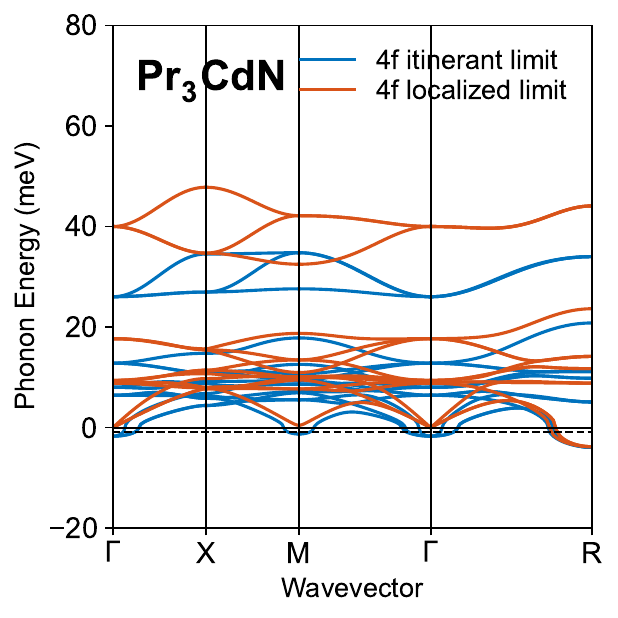}
    \includegraphics[width=0.19\linewidth]{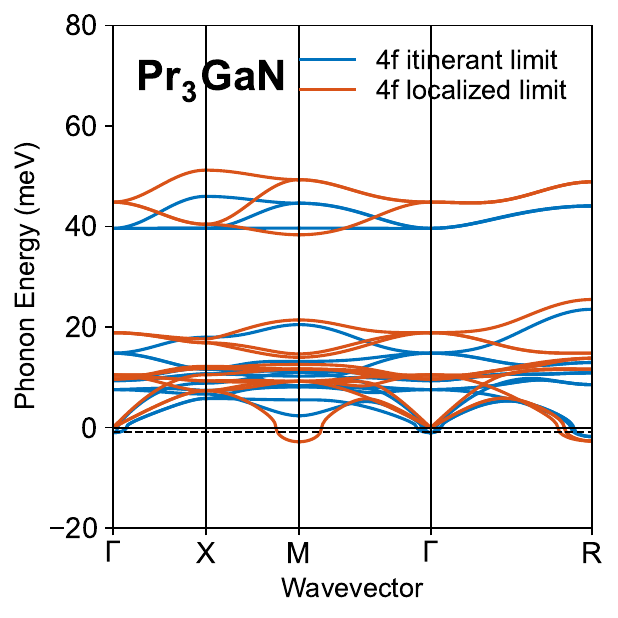}
    \includegraphics[width=0.19\linewidth]{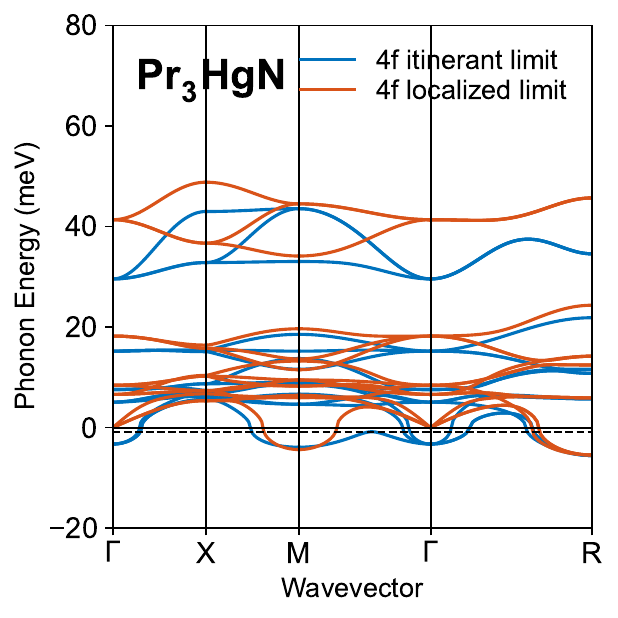}
    \includegraphics[width=0.19\linewidth]{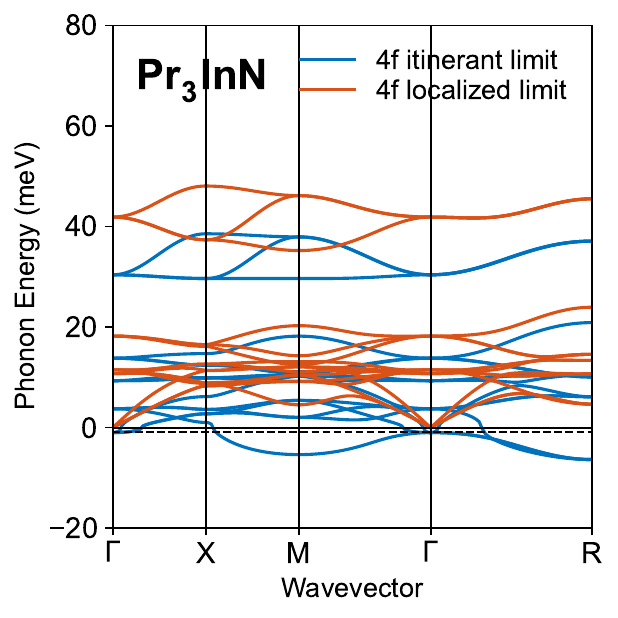}
\\
    \includegraphics[width=0.19\linewidth]{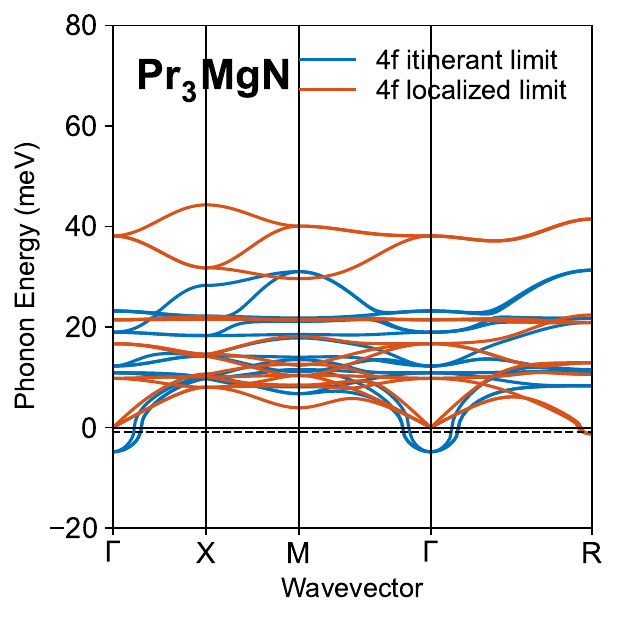}
    \includegraphics[width=0.19\linewidth]{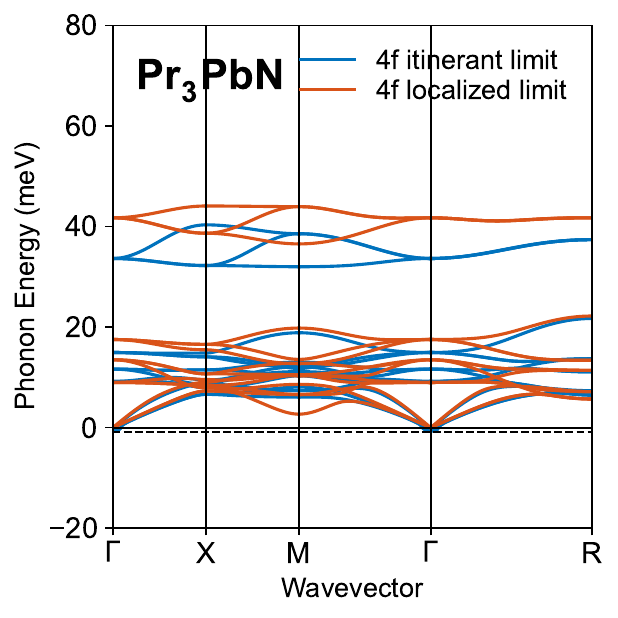}
    \includegraphics[width=0.19\linewidth]{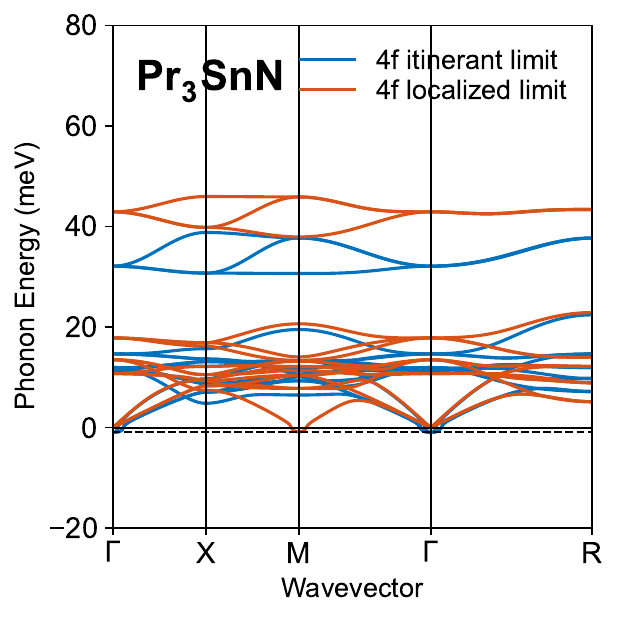}
    \includegraphics[width=0.19\linewidth]{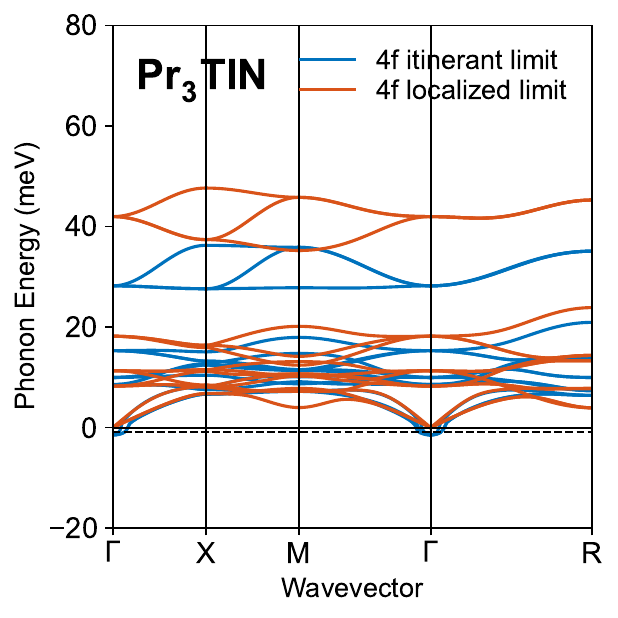}
    \includegraphics[width=0.19\linewidth]{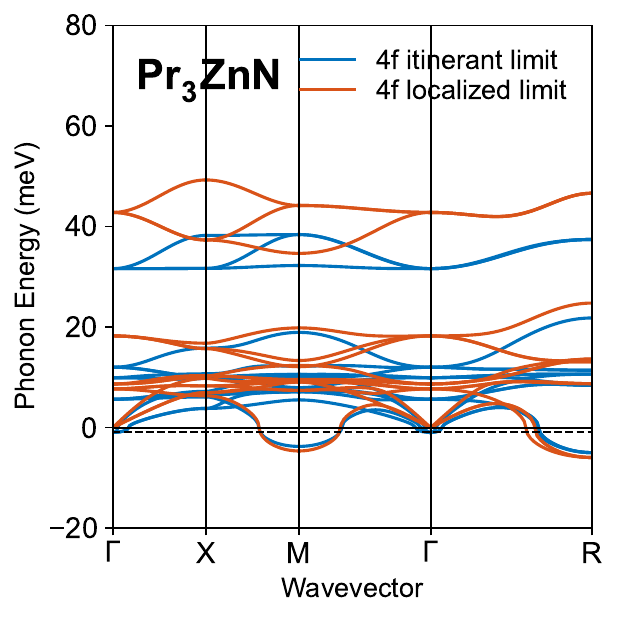}
\\
    \includegraphics[width=0.19\linewidth]{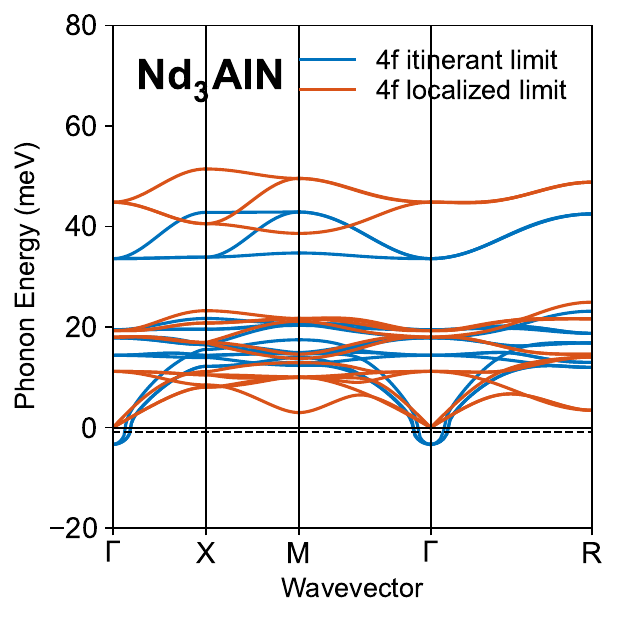}
    \includegraphics[width=0.19\linewidth]{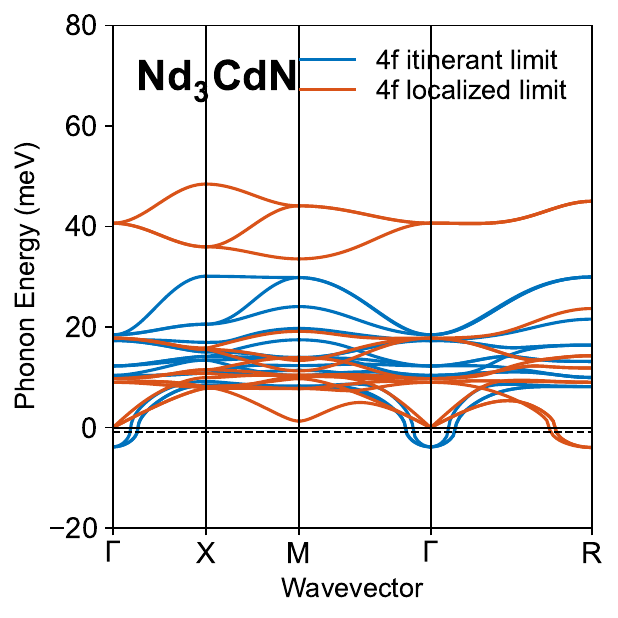}
    \includegraphics[width=0.19\linewidth]{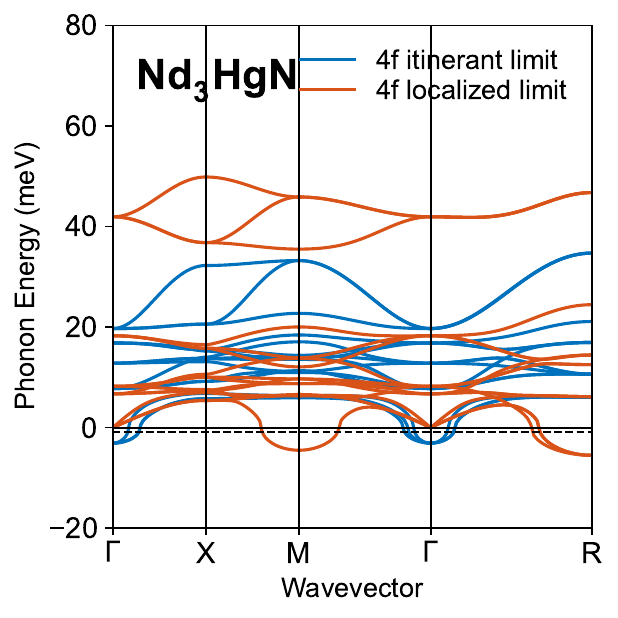}
    \includegraphics[width=0.19\linewidth]{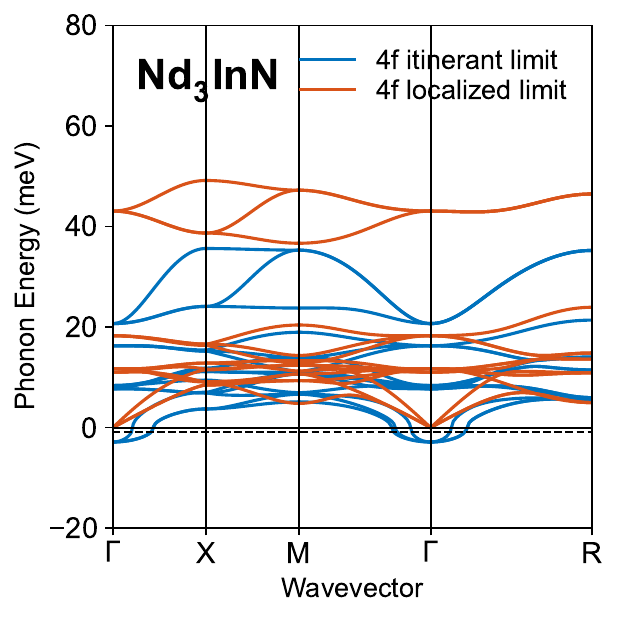}
    \includegraphics[width=0.19\linewidth]{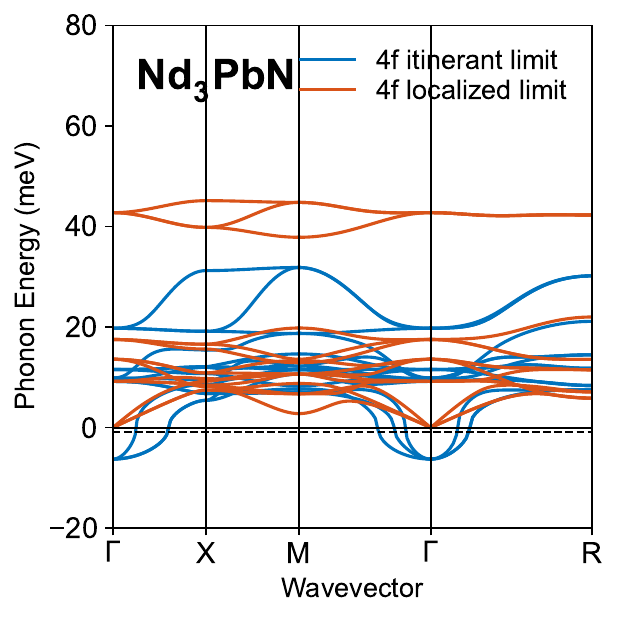}
\\
    \includegraphics[width=0.19\linewidth]{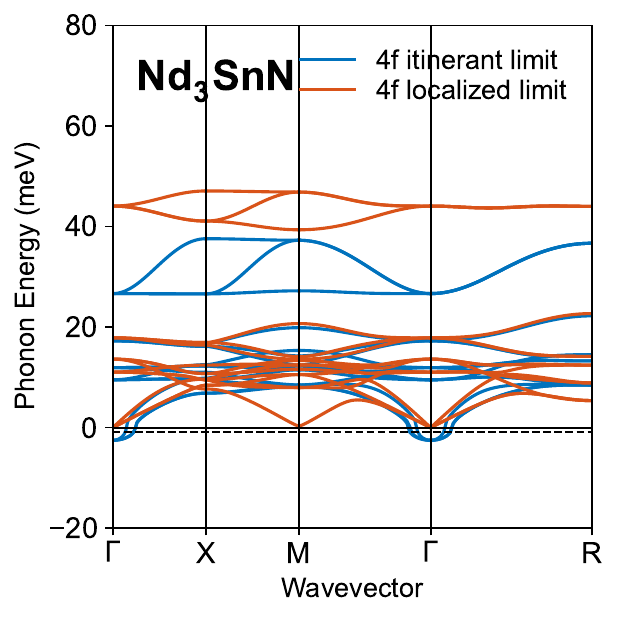}
    \includegraphics[width=0.19\linewidth]{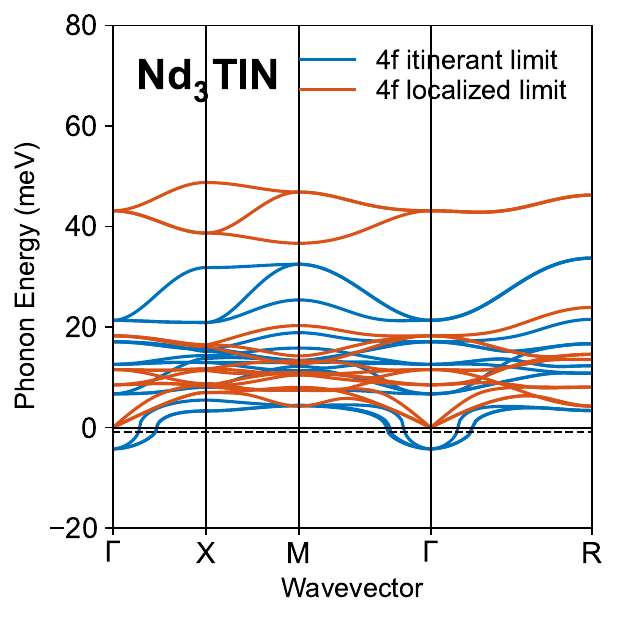}
    \includegraphics[width=0.19\linewidth]{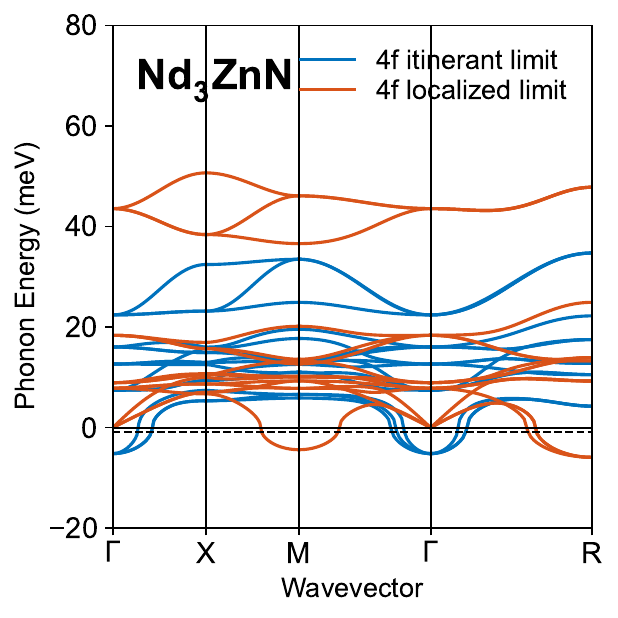}
    \includegraphics[width=0.19\linewidth]{figures/blank.pdf}
    \includegraphics[width=0.19\linewidth]{figures/blank.pdf}
\\
    \includegraphics[width=0.19\linewidth]{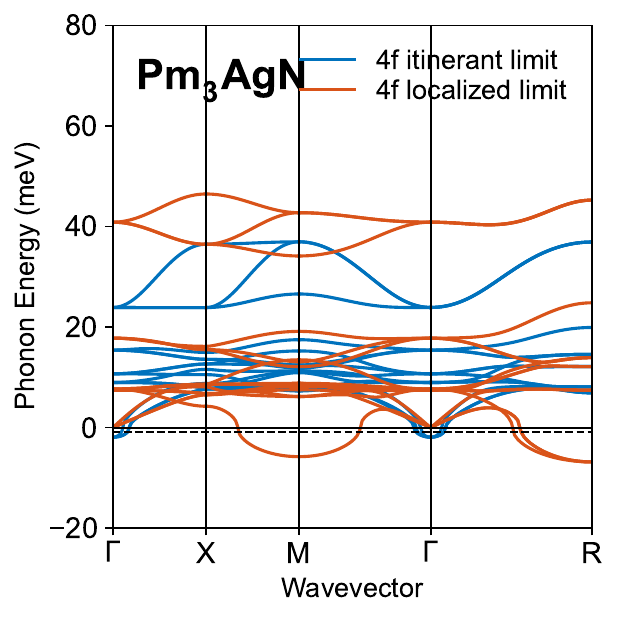}
    \includegraphics[width=0.19\linewidth]{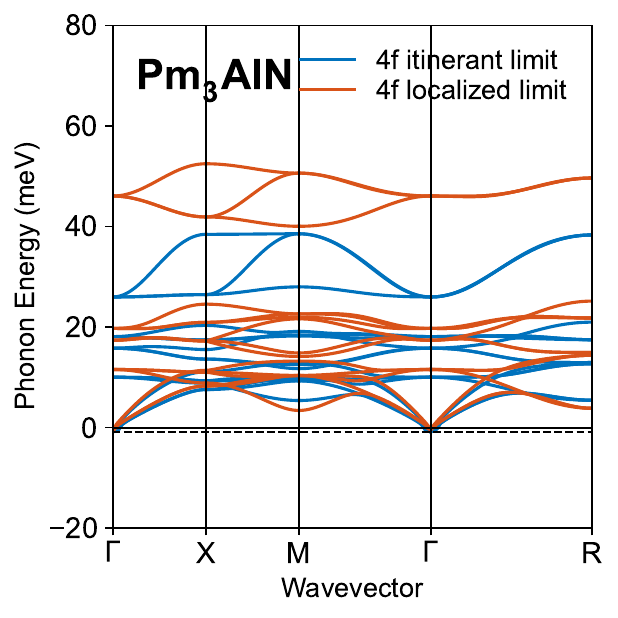}
    \includegraphics[width=0.19\linewidth]{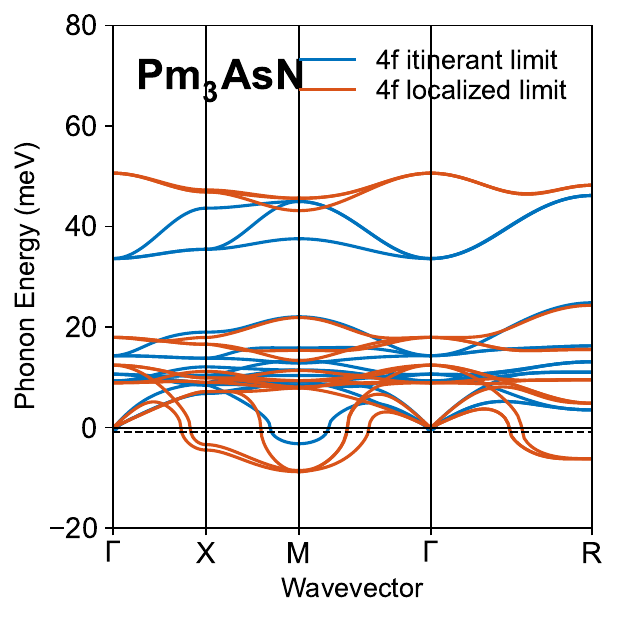}
    \includegraphics[width=0.19\linewidth]{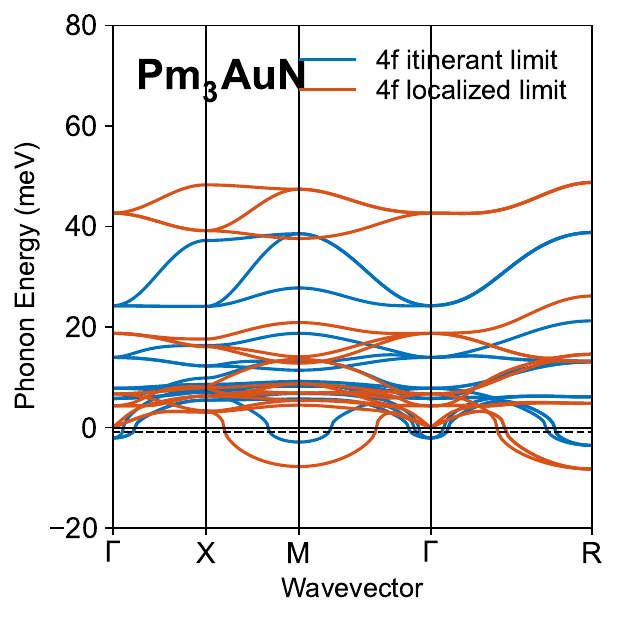}
    \includegraphics[width=0.19\linewidth]{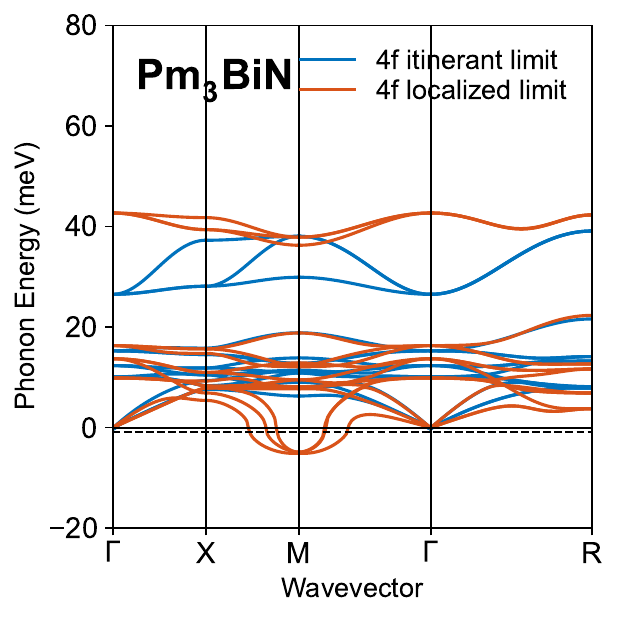}
\end{figure}
\clearpage

\begin{figure}[H]
    \includegraphics[width=0.19\linewidth]{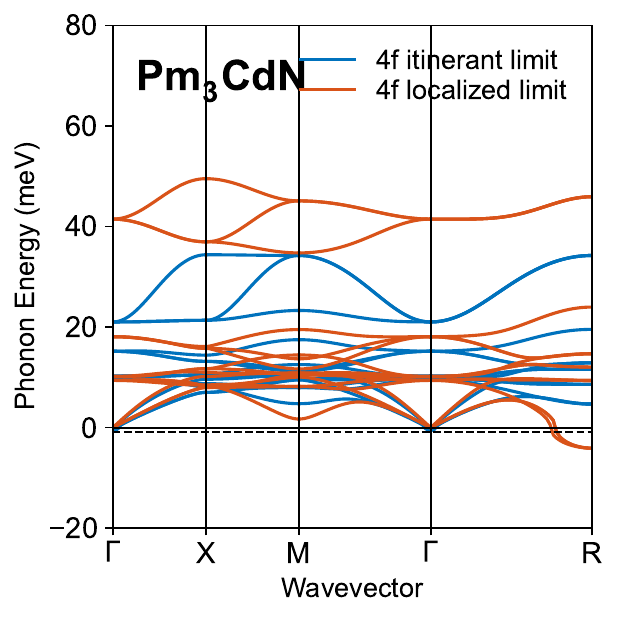}
    \includegraphics[width=0.19\linewidth]{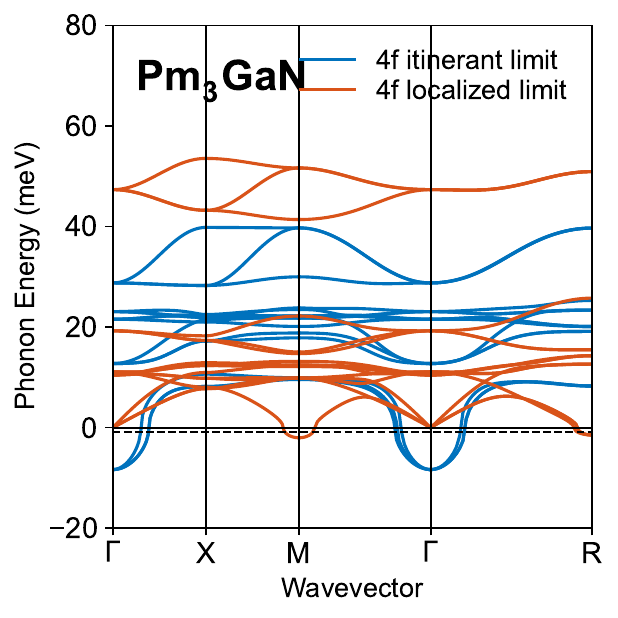}
    \includegraphics[width=0.19\linewidth]{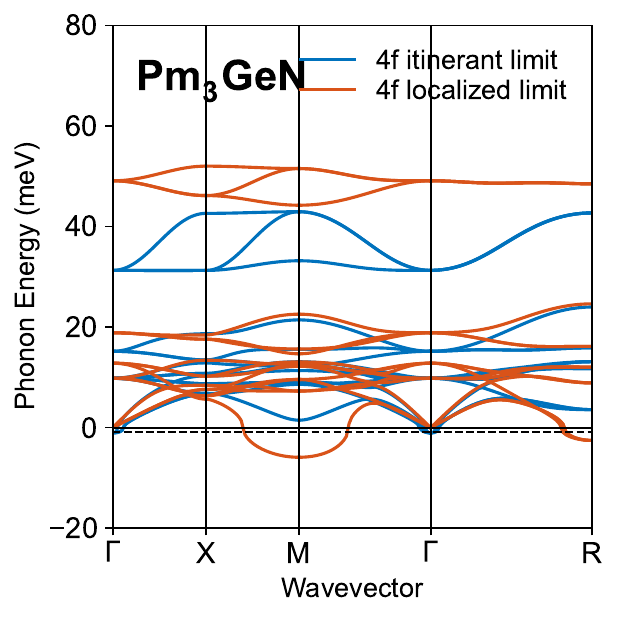}
    \includegraphics[width=0.19\linewidth]{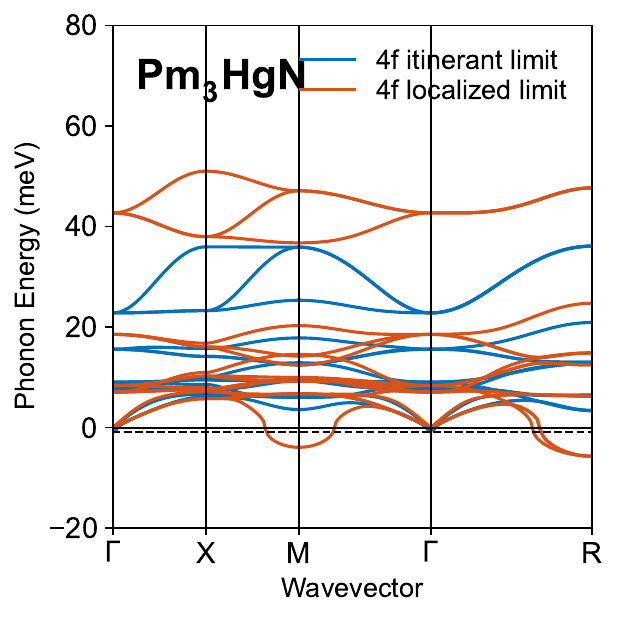}
    \includegraphics[width=0.19\linewidth]{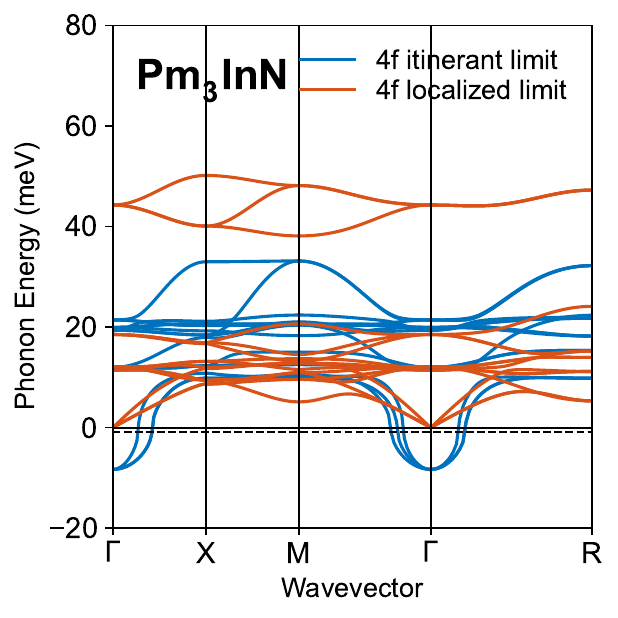}
\\
    \includegraphics[width=0.19\linewidth]{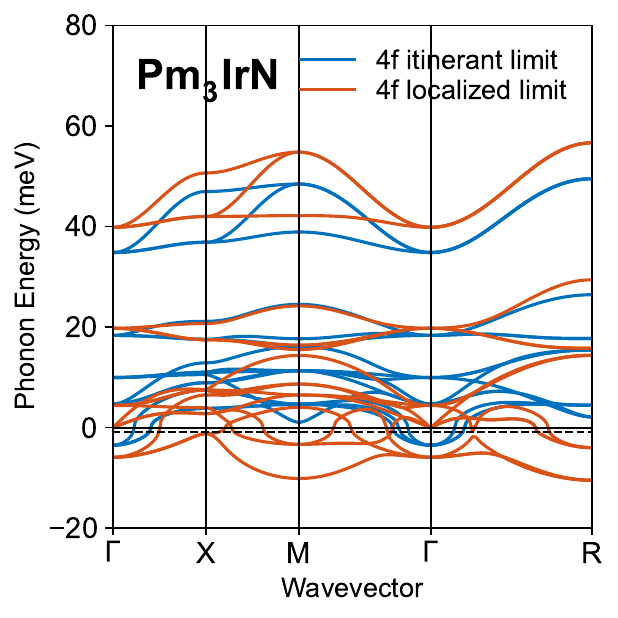}
    \includegraphics[width=0.19\linewidth]{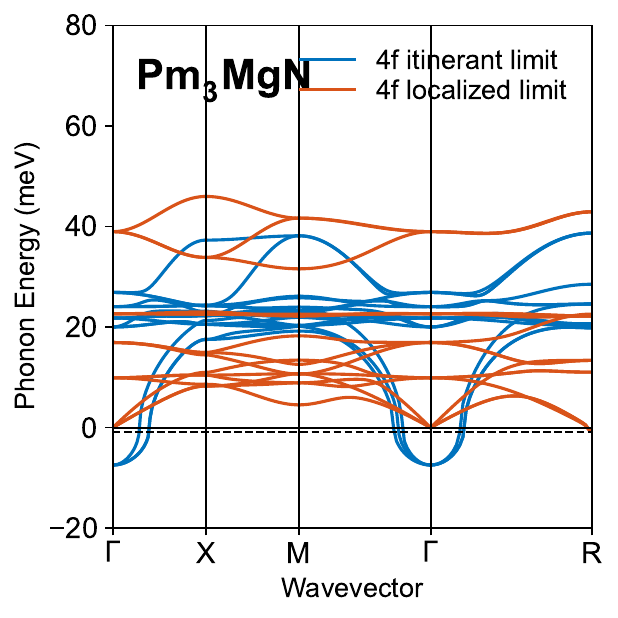}
    \includegraphics[width=0.19\linewidth]{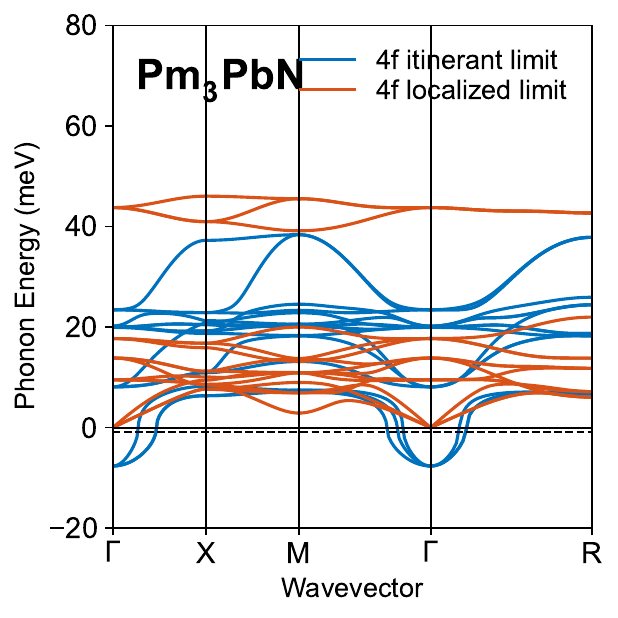}
    \includegraphics[width=0.19\linewidth]{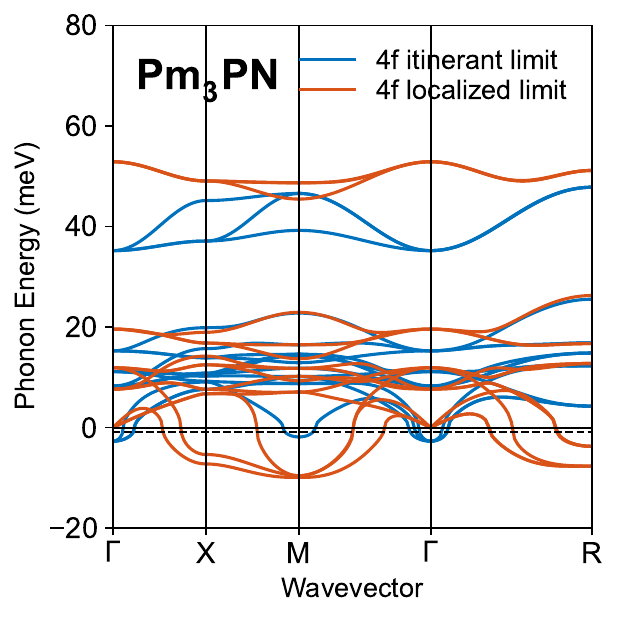}
    \includegraphics[width=0.19\linewidth]{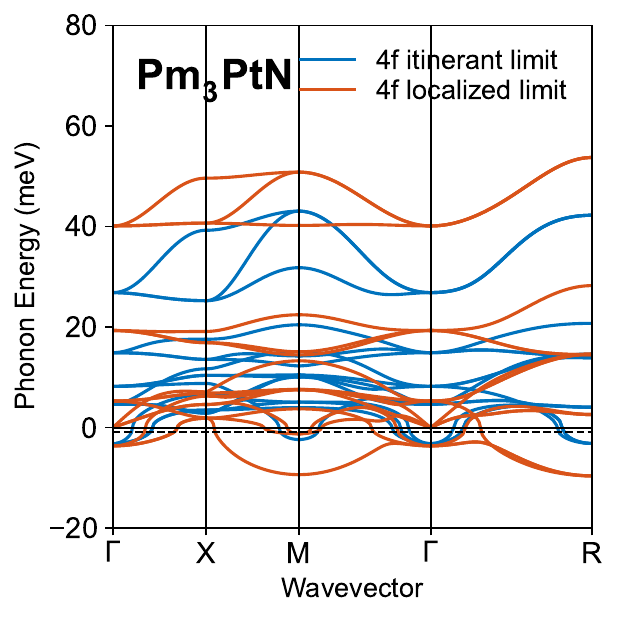}
\\
    \includegraphics[width=0.19\linewidth]{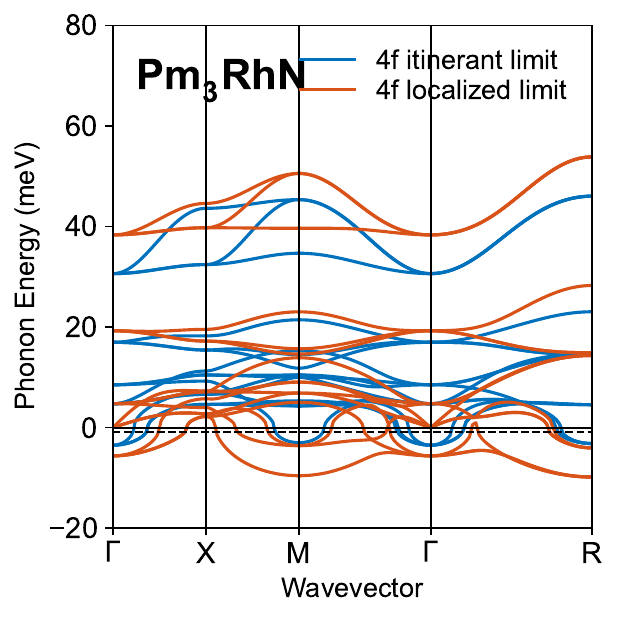}
    \includegraphics[width=0.19\linewidth]{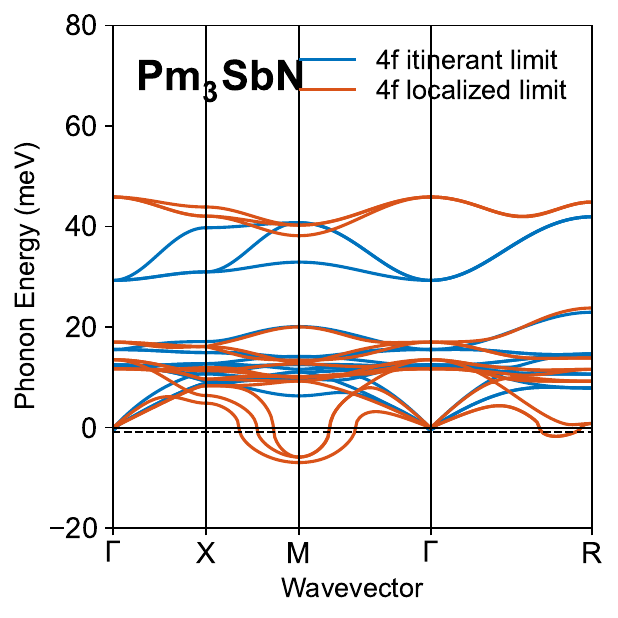}
    \includegraphics[width=0.19\linewidth]{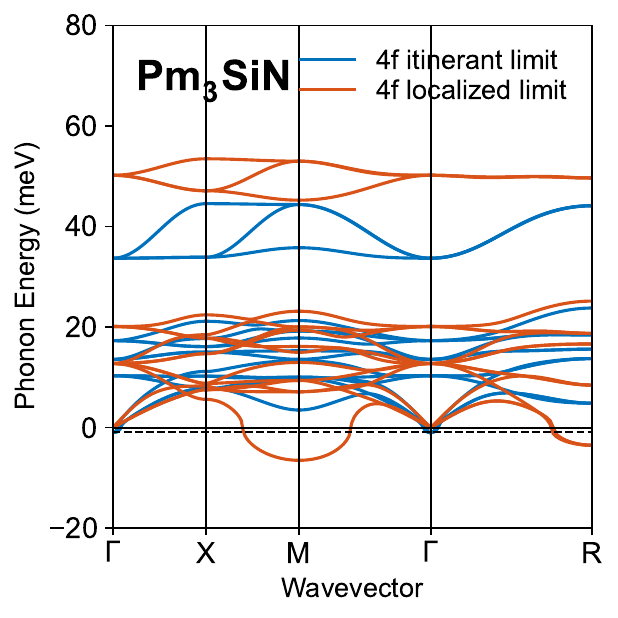}
    \includegraphics[width=0.19\linewidth]{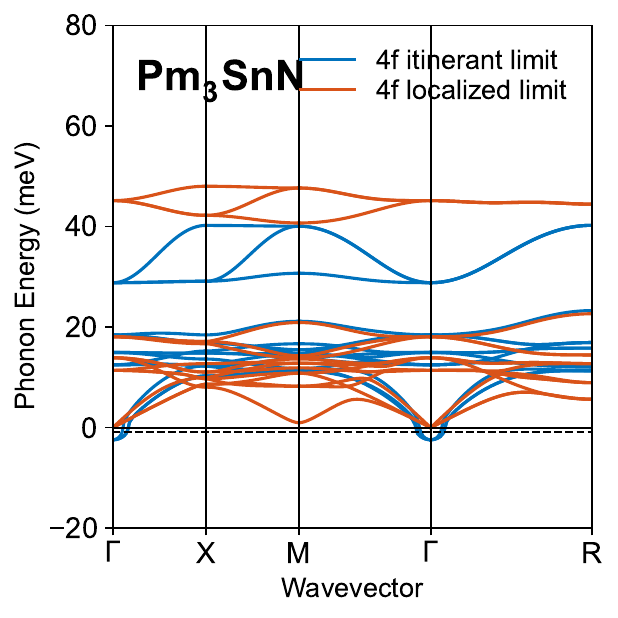}
    \includegraphics[width=0.19\linewidth]{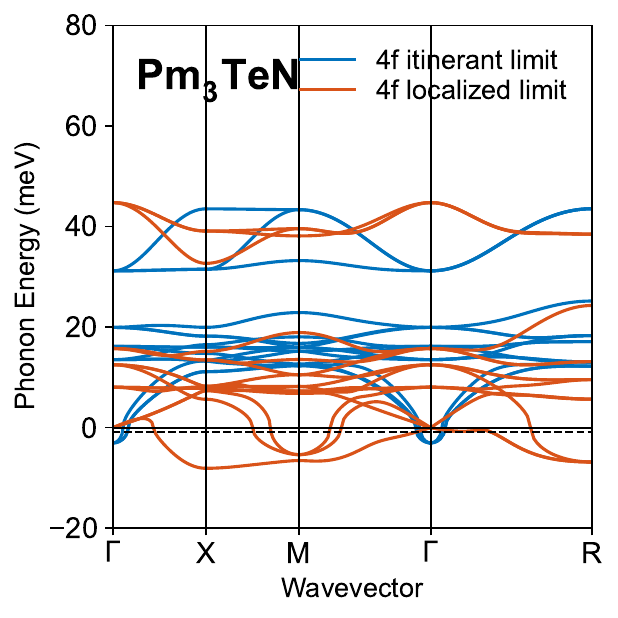}
\\
    \includegraphics[width=0.19\linewidth]{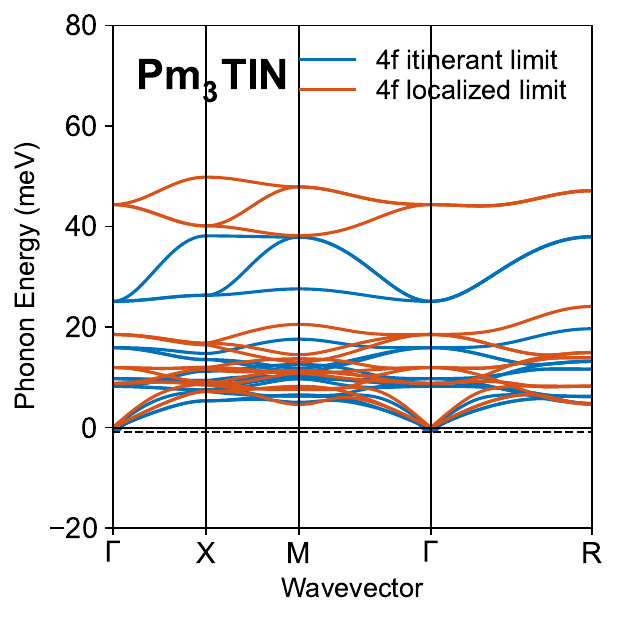}
    \includegraphics[width=0.19\linewidth]{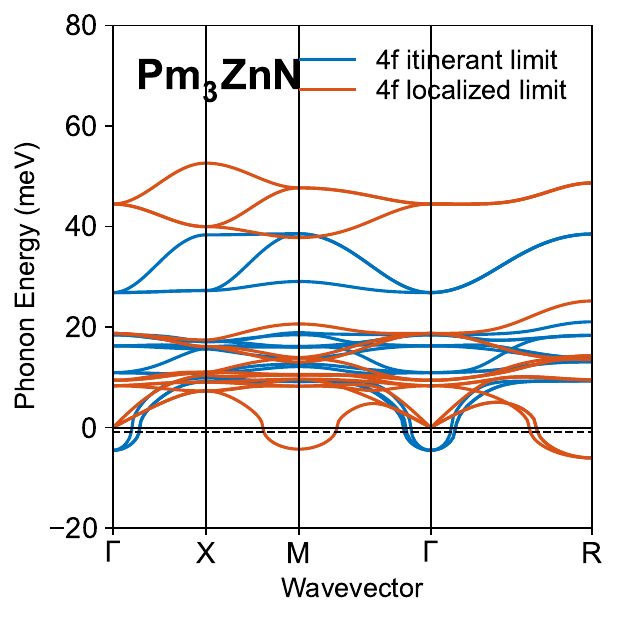}
    \includegraphics[width=0.19\linewidth]{figures/blank.pdf}
    \includegraphics[width=0.19\linewidth]{figures/blank.pdf}
    \includegraphics[width=0.19\linewidth]{figures/blank.pdf}
\\
    \includegraphics[width=0.19\linewidth]{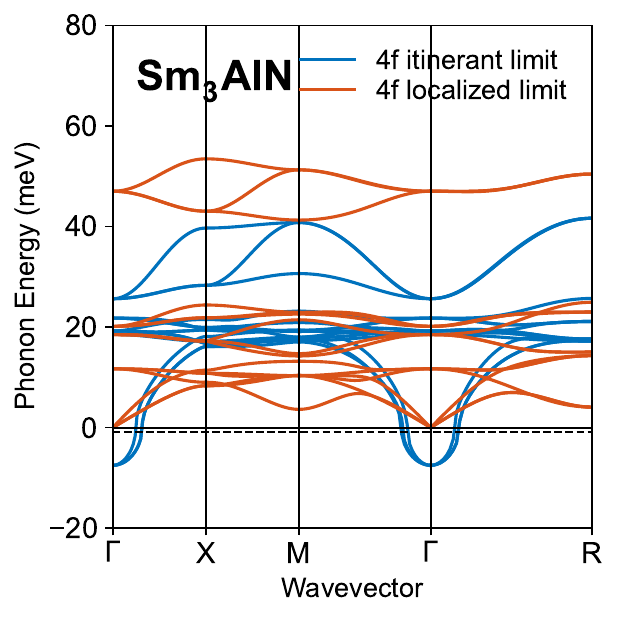}
    \includegraphics[width=0.19\linewidth]{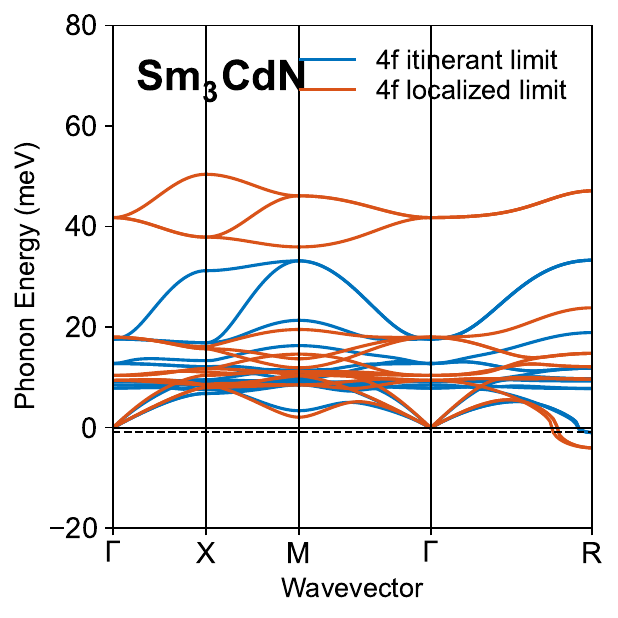}
    \includegraphics[width=0.19\linewidth]{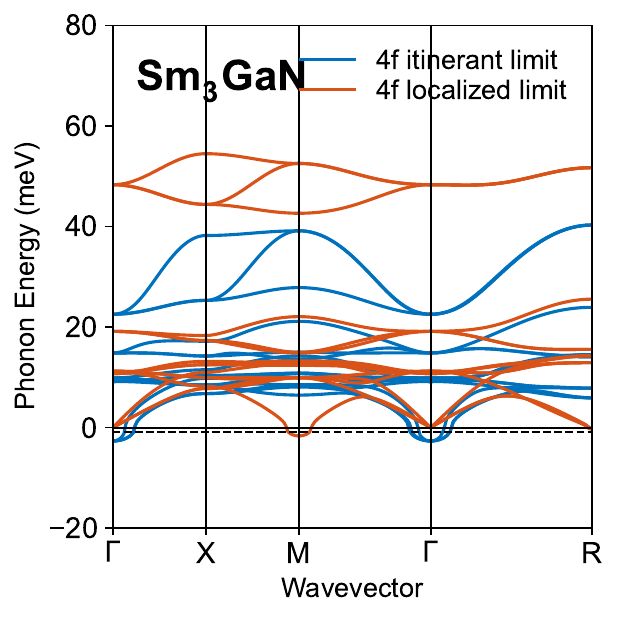}
    \includegraphics[width=0.19\linewidth]{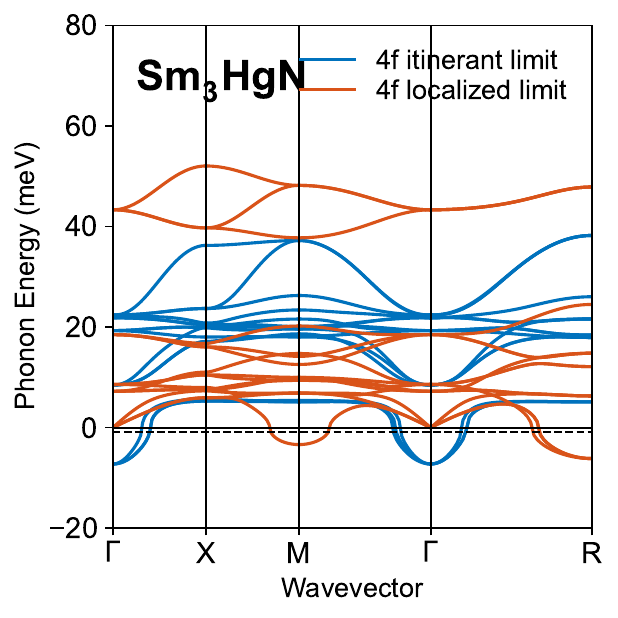}
    \includegraphics[width=0.19\linewidth]{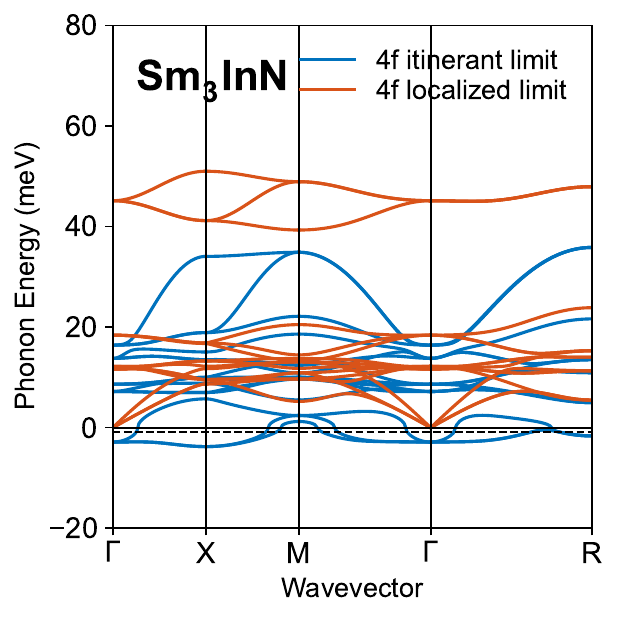}
\\
    \includegraphics[width=0.19\linewidth]{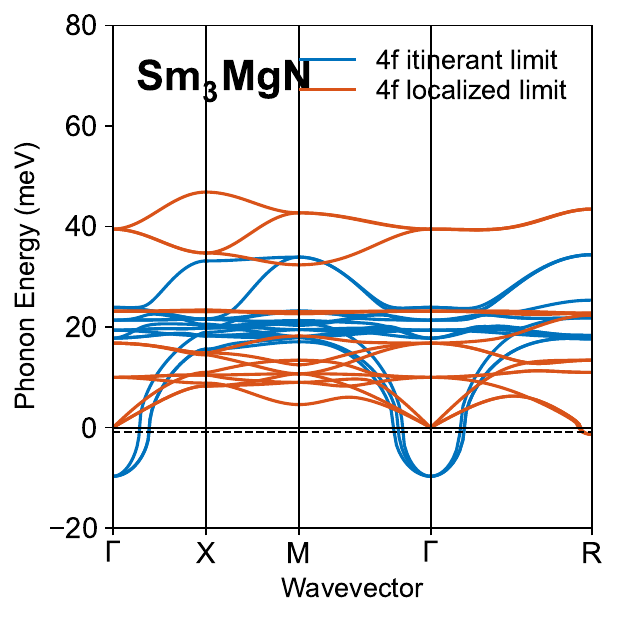}
    \includegraphics[width=0.19\linewidth]{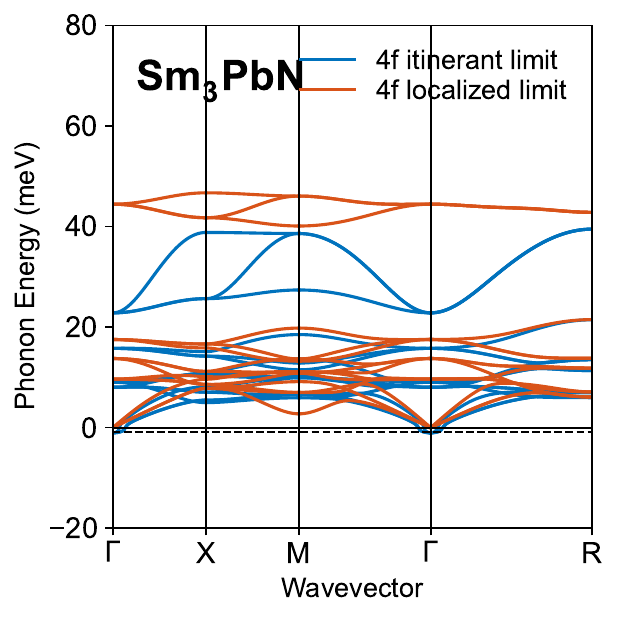}
    \includegraphics[width=0.19\linewidth]{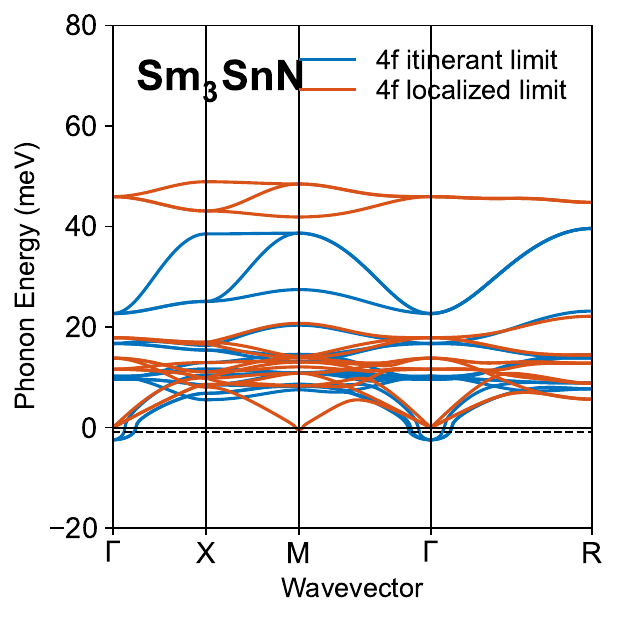}
    \includegraphics[width=0.19\linewidth]{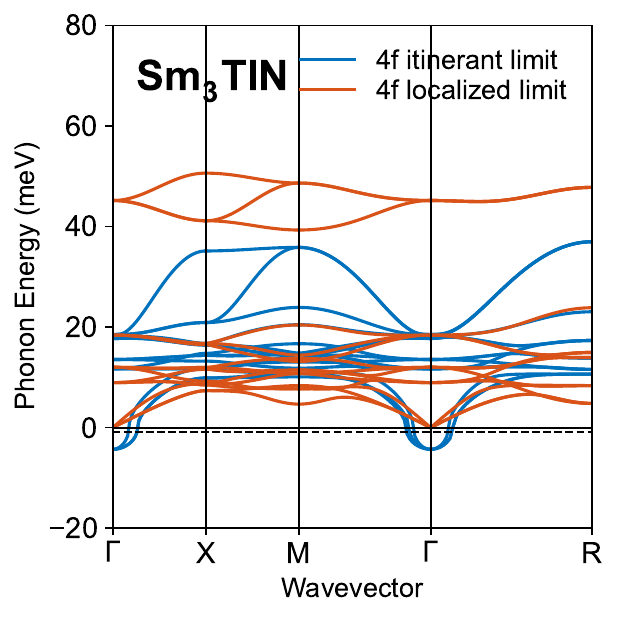}
    \includegraphics[width=0.19\linewidth]{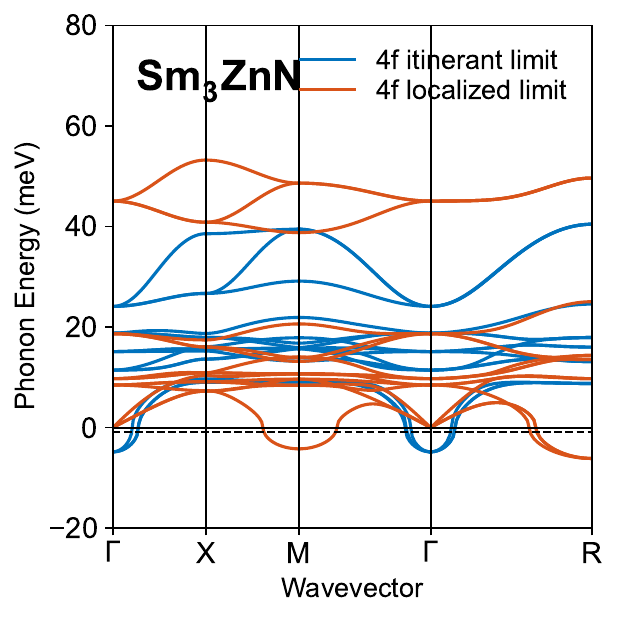}
\\
    \includegraphics[width=0.19\linewidth]{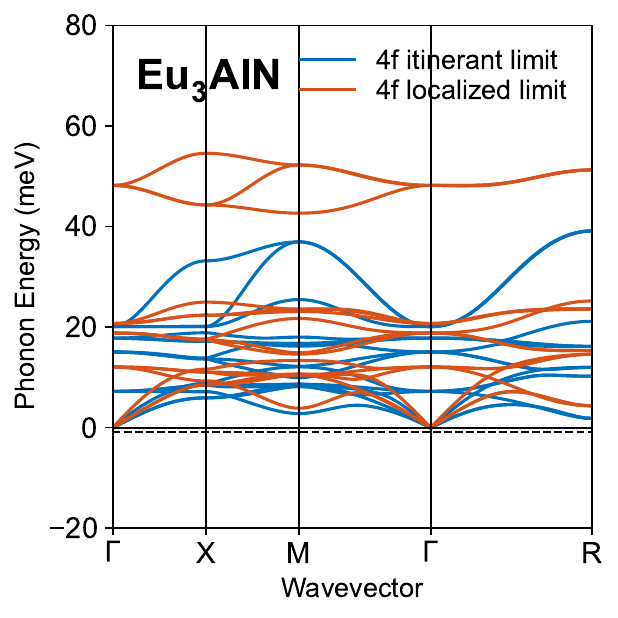}
    \includegraphics[width=0.19\linewidth]{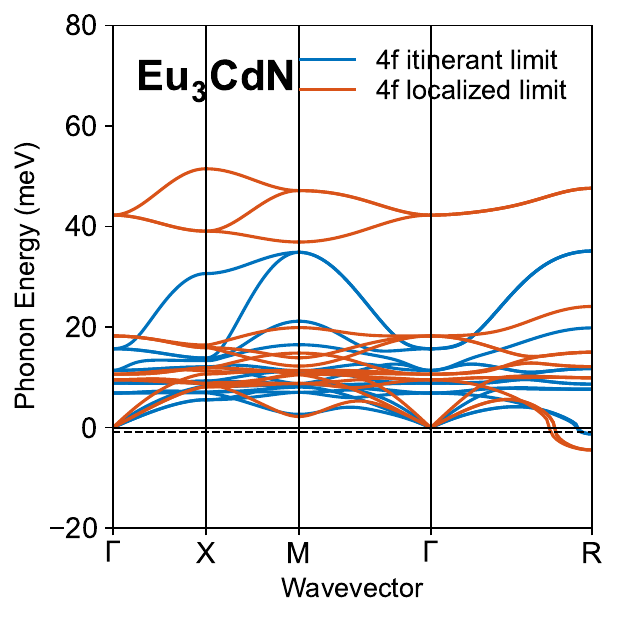}
    \includegraphics[width=0.19\linewidth]{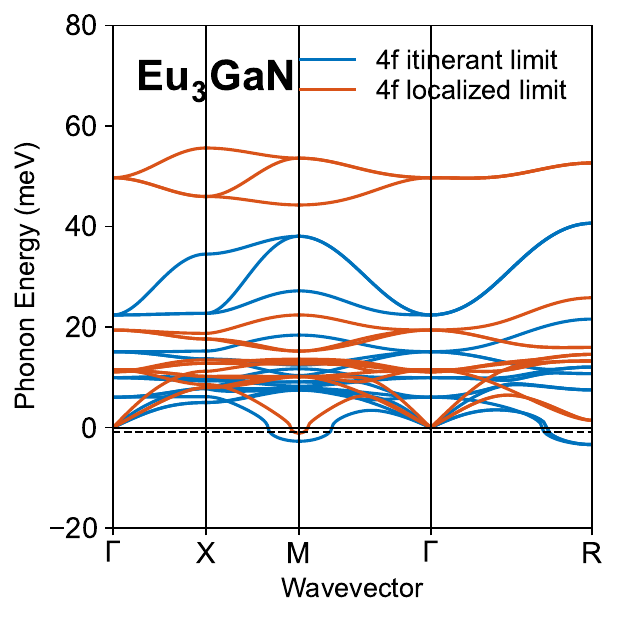}
    \includegraphics[width=0.19\linewidth]{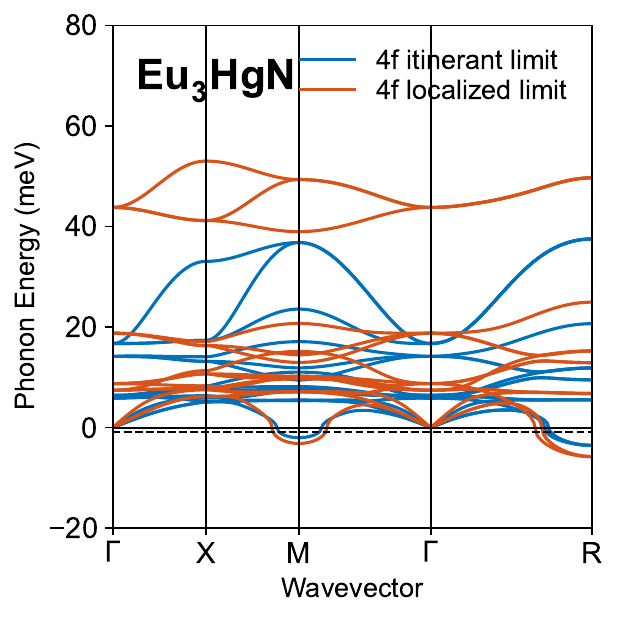}
    \includegraphics[width=0.19\linewidth]{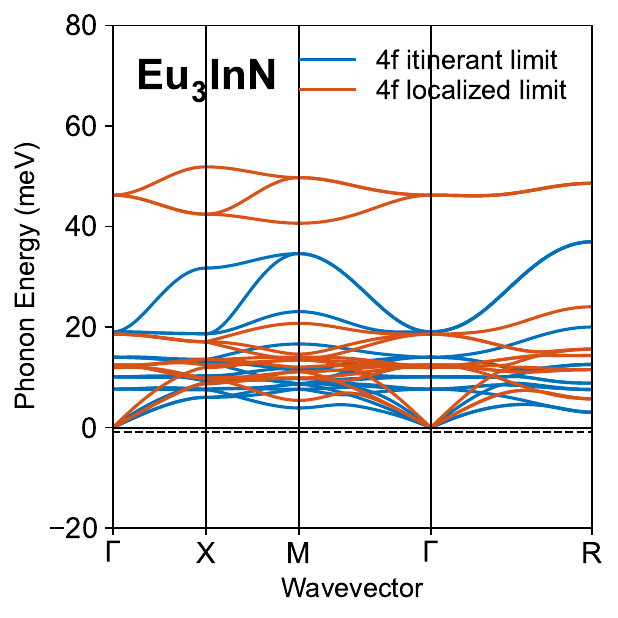}
\end{figure}
\clearpage

\begin{figure}[H]
    \includegraphics[width=0.19\linewidth]{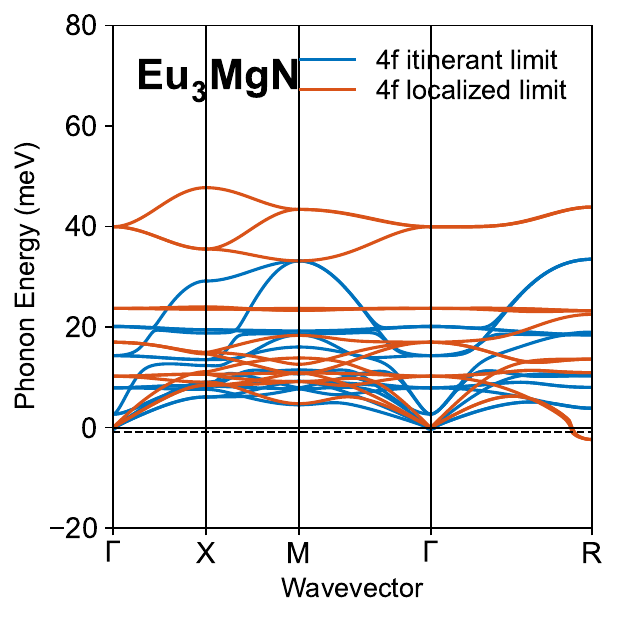}
    \includegraphics[width=0.19\linewidth]{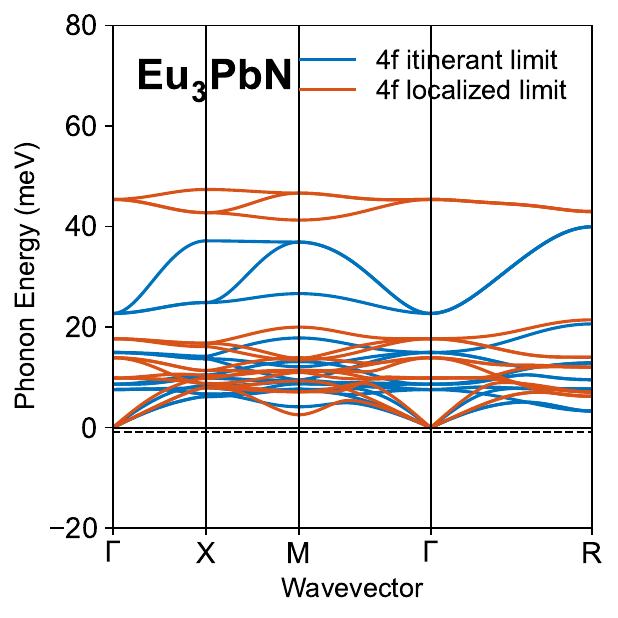}
    \includegraphics[width=0.19\linewidth]{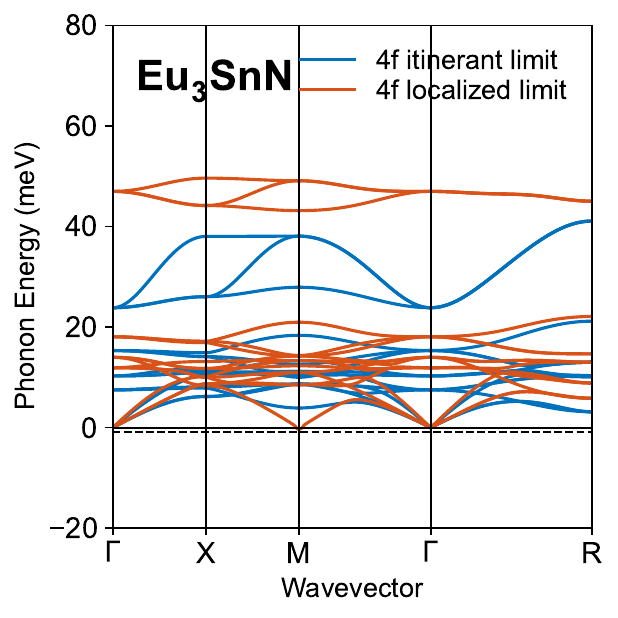}
    \includegraphics[width=0.19\linewidth]{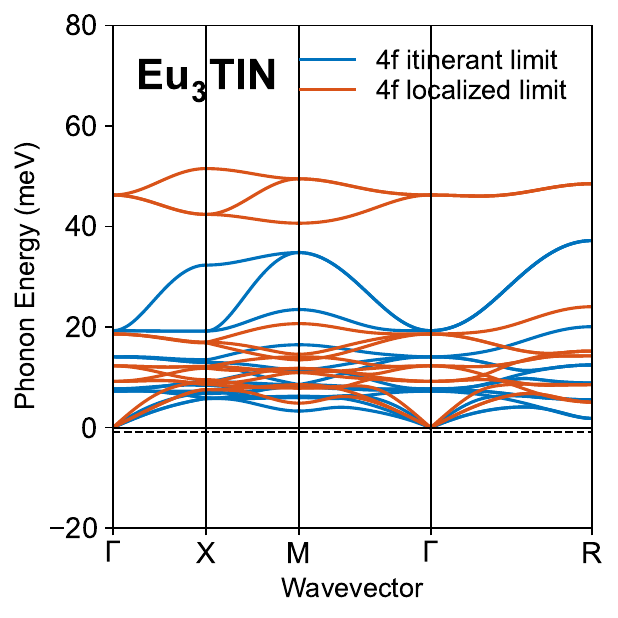}
    \includegraphics[width=0.19\linewidth]{figures/blank.pdf}
\\
    \includegraphics[width=0.19\linewidth]{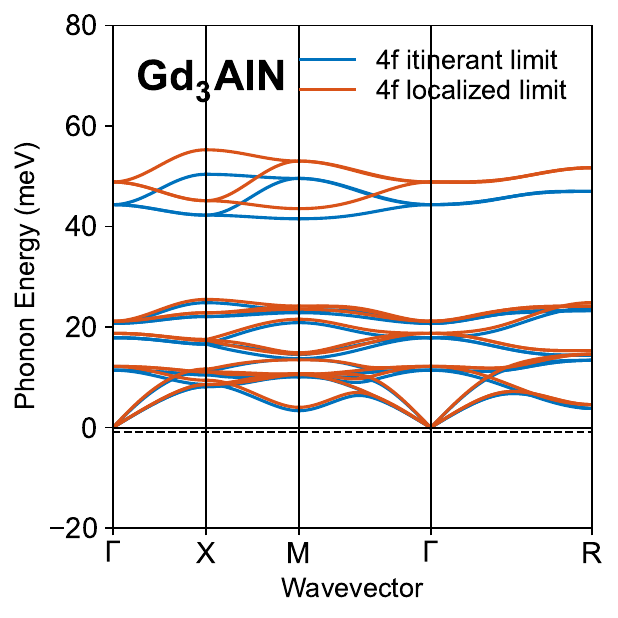}
    \includegraphics[width=0.19\linewidth]{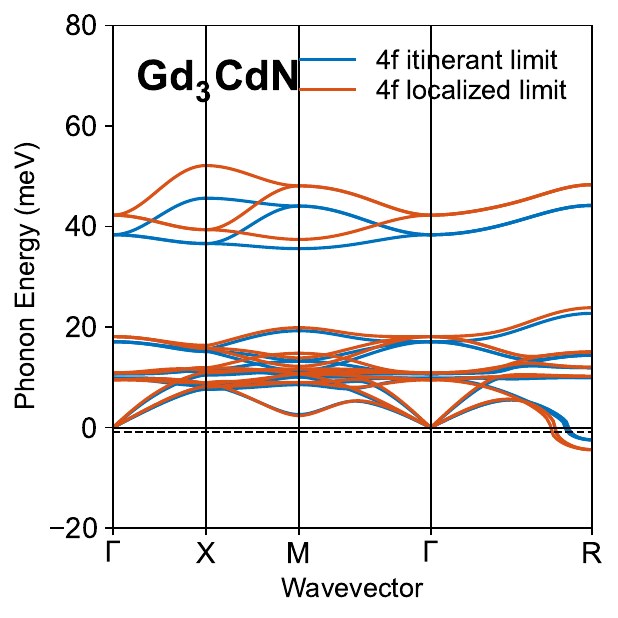}
    \includegraphics[width=0.19\linewidth]{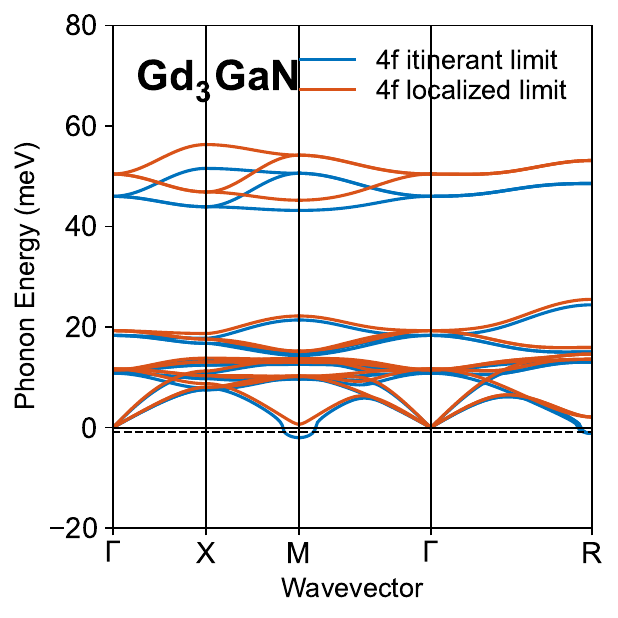}
    \includegraphics[width=0.19\linewidth]{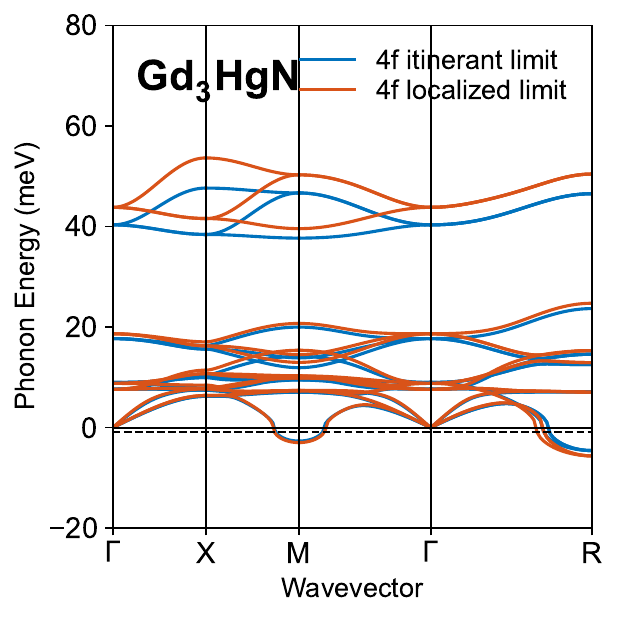}
    \includegraphics[width=0.19\linewidth]{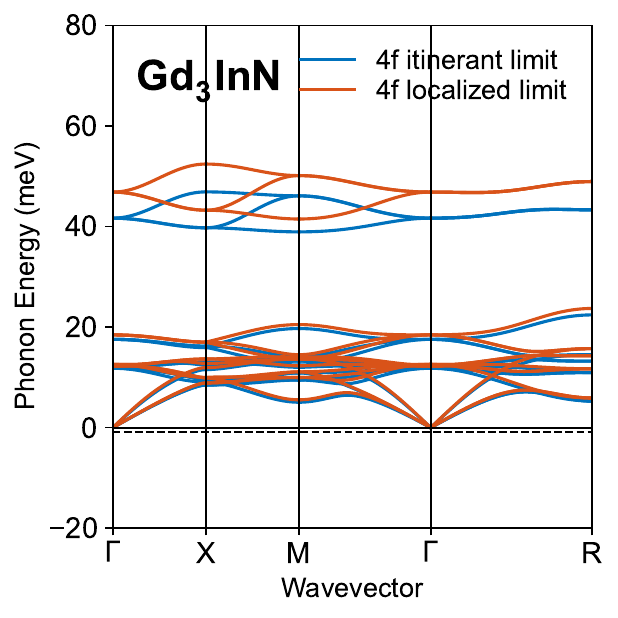}
\\
    \includegraphics[width=0.19\linewidth]{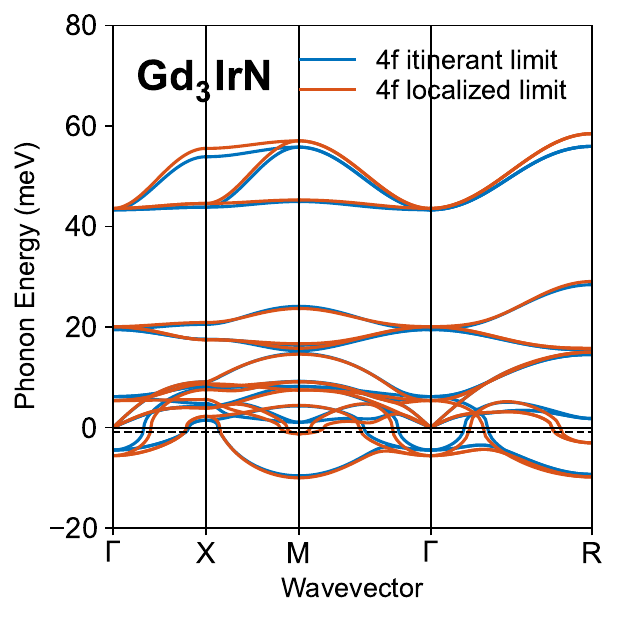}
    \includegraphics[width=0.19\linewidth]{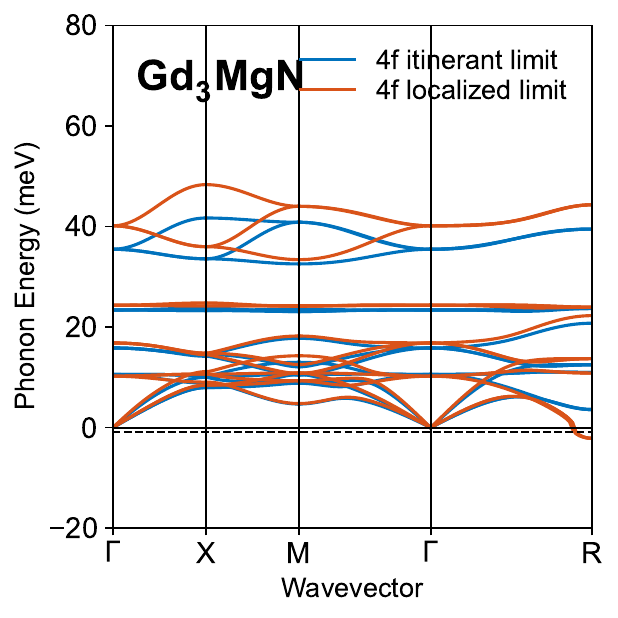}
    \includegraphics[width=0.19\linewidth]{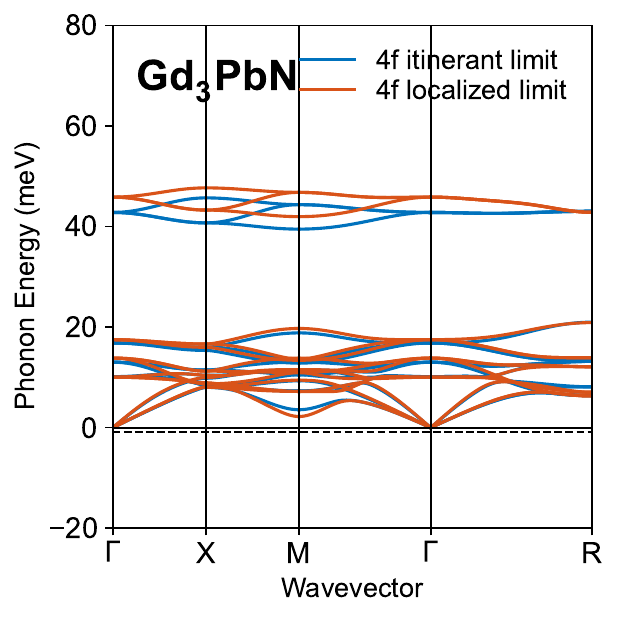}
    \includegraphics[width=0.19\linewidth]{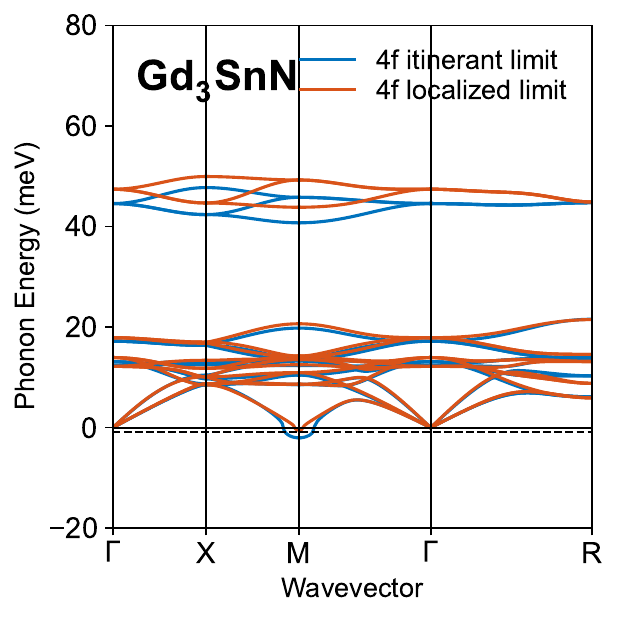}
    \includegraphics[width=0.19\linewidth]{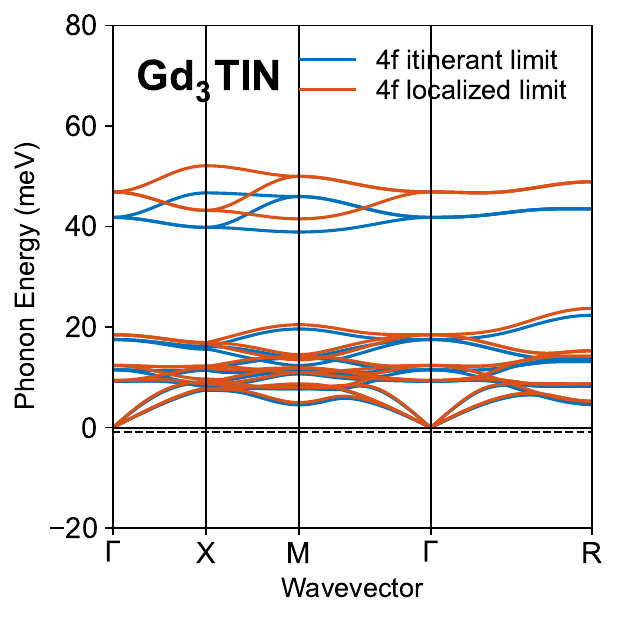}
\\
    \includegraphics[width=0.19\linewidth]{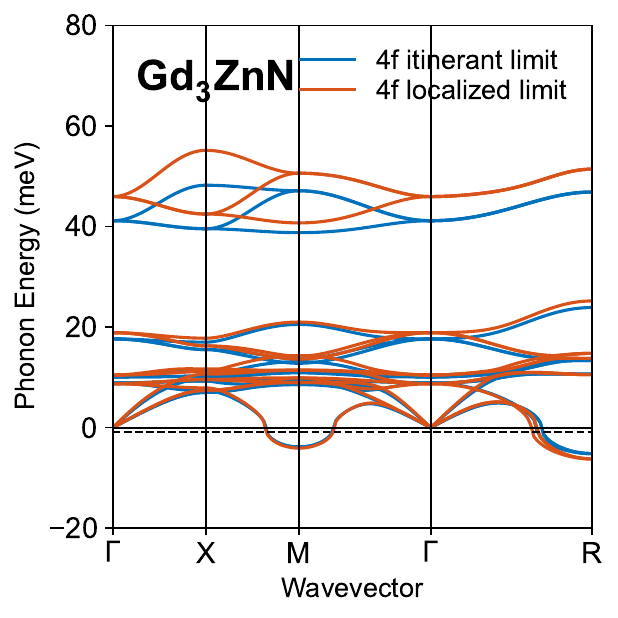}
    \includegraphics[width=0.19\linewidth]{figures/blank.pdf}
    \includegraphics[width=0.19\linewidth]{figures/blank.pdf}
    \includegraphics[width=0.19\linewidth]{figures/blank.pdf}
    \includegraphics[width=0.19\linewidth]{figures/blank.pdf}
\\
    \includegraphics[width=0.19\linewidth]{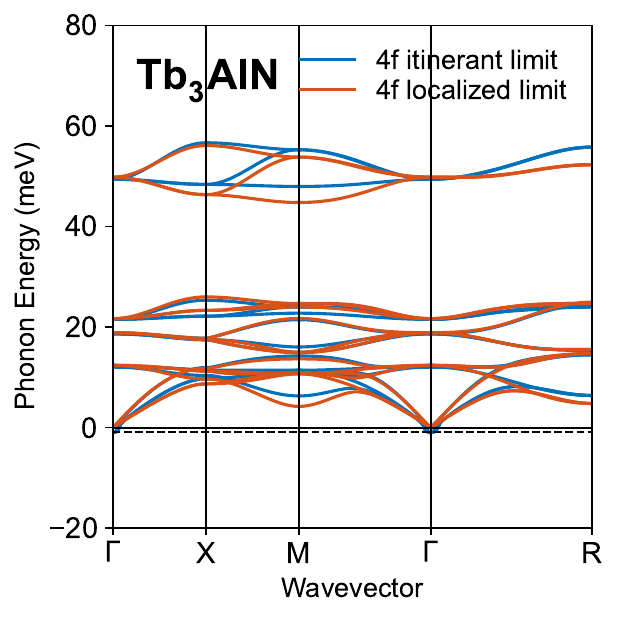}
    \includegraphics[width=0.19\linewidth]{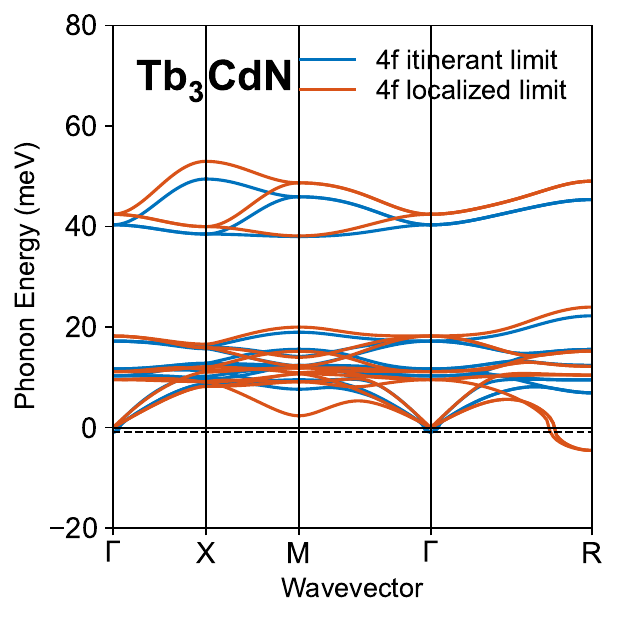}
    \includegraphics[width=0.19\linewidth]{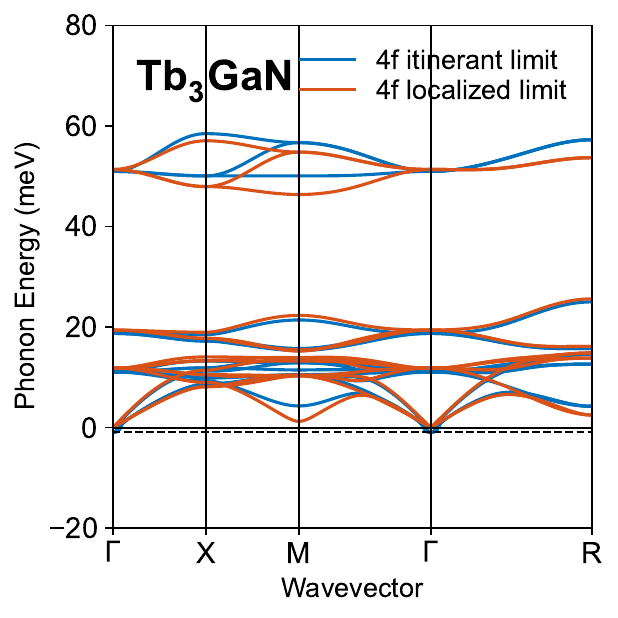}
    \includegraphics[width=0.19\linewidth]{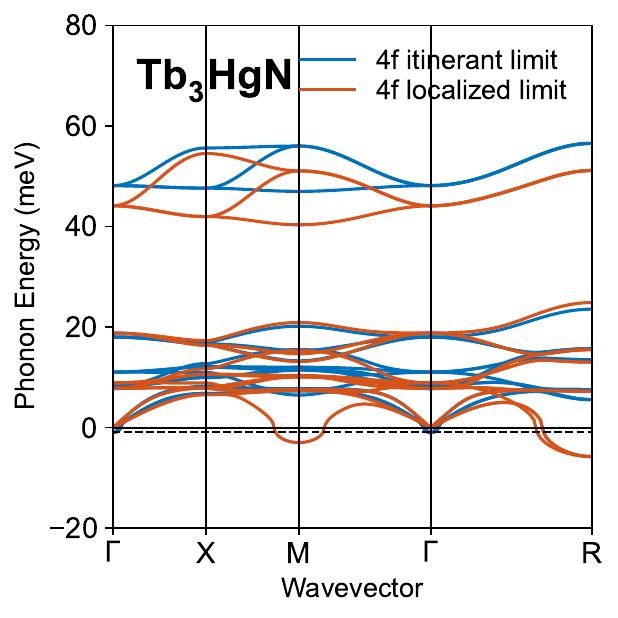}
    \includegraphics[width=0.19\linewidth]{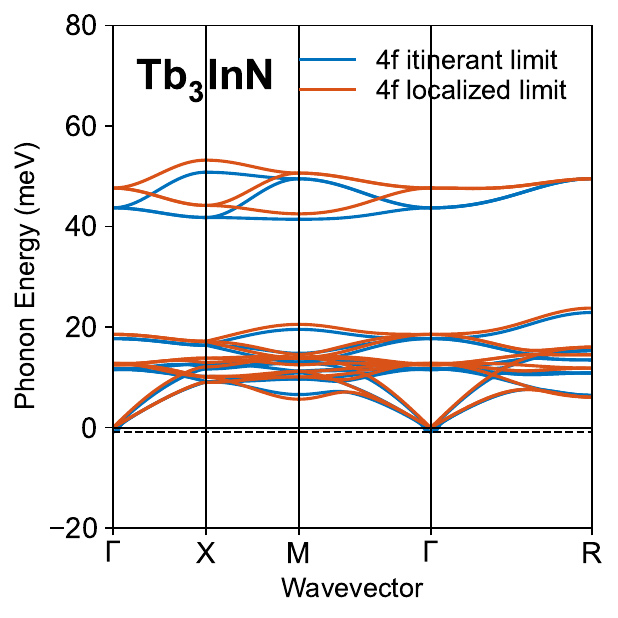}
\\
    \includegraphics[width=0.19\linewidth]{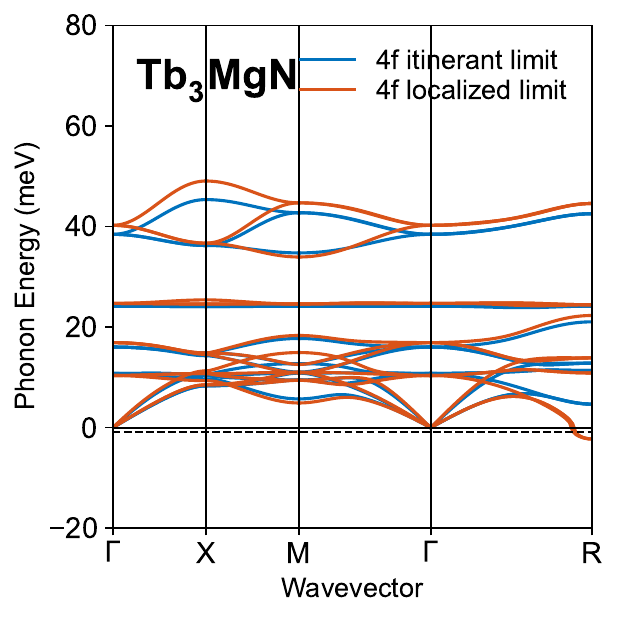}
    \includegraphics[width=0.19\linewidth]{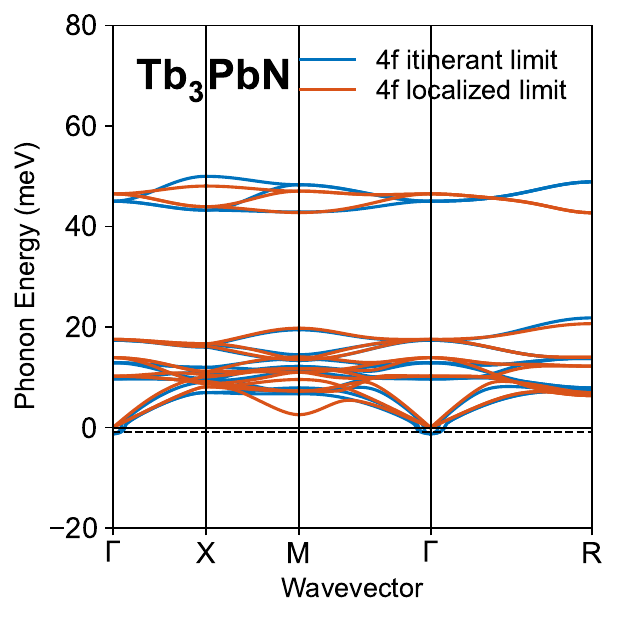}
    \includegraphics[width=0.19\linewidth]{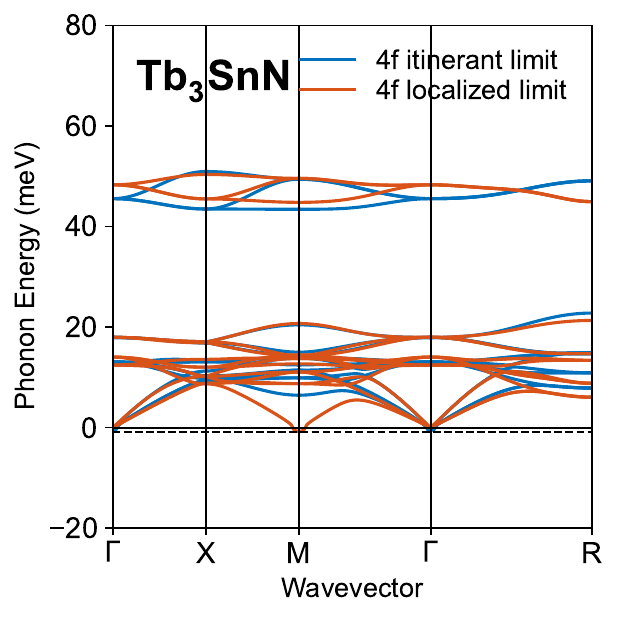}
    \includegraphics[width=0.19\linewidth]{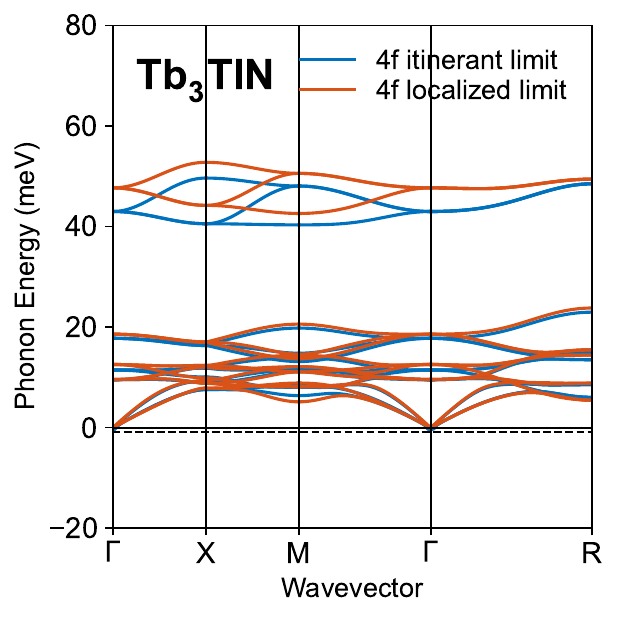}
    \includegraphics[width=0.19\linewidth]{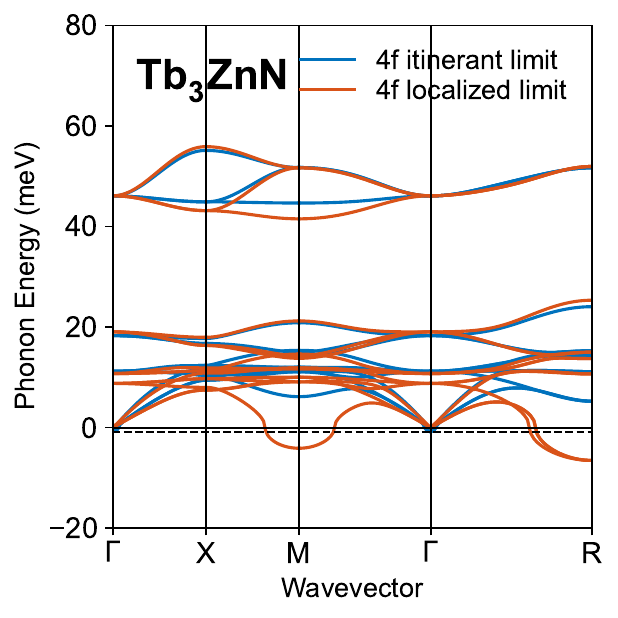}
\\
    \includegraphics[width=0.19\linewidth]{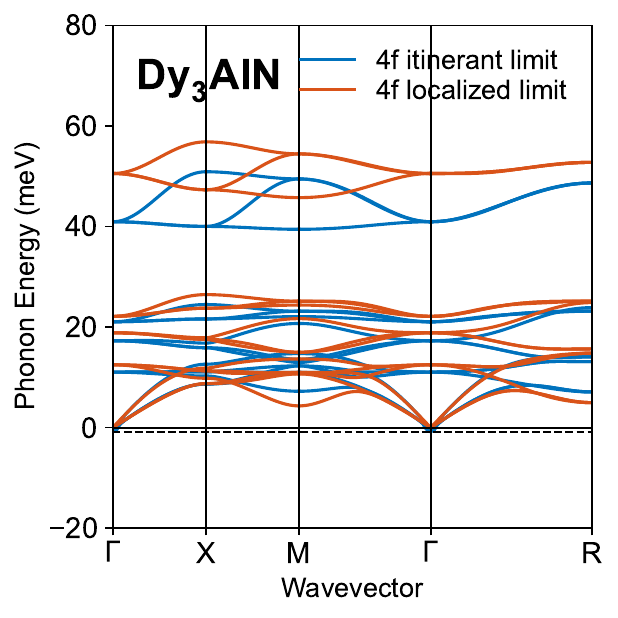}
    \includegraphics[width=0.19\linewidth]{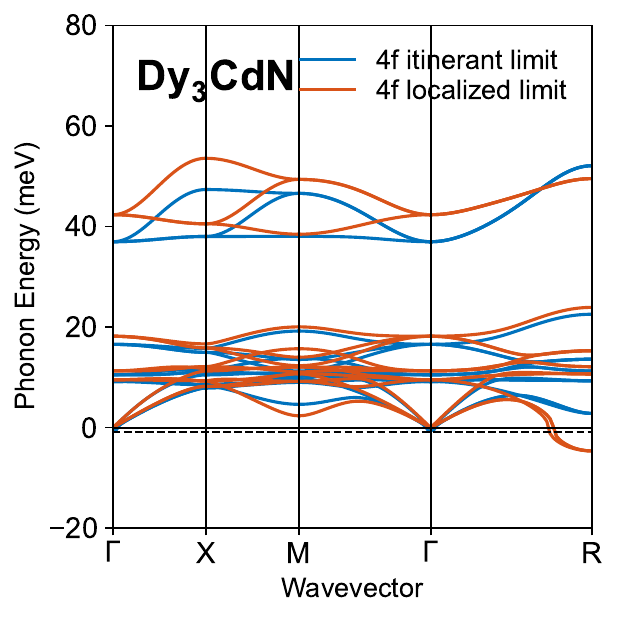}
    \includegraphics[width=0.19\linewidth]{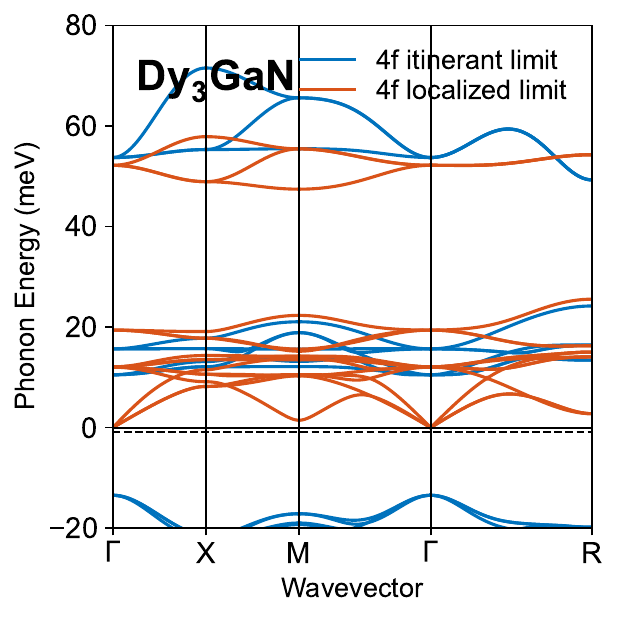}
    \includegraphics[width=0.19\linewidth]{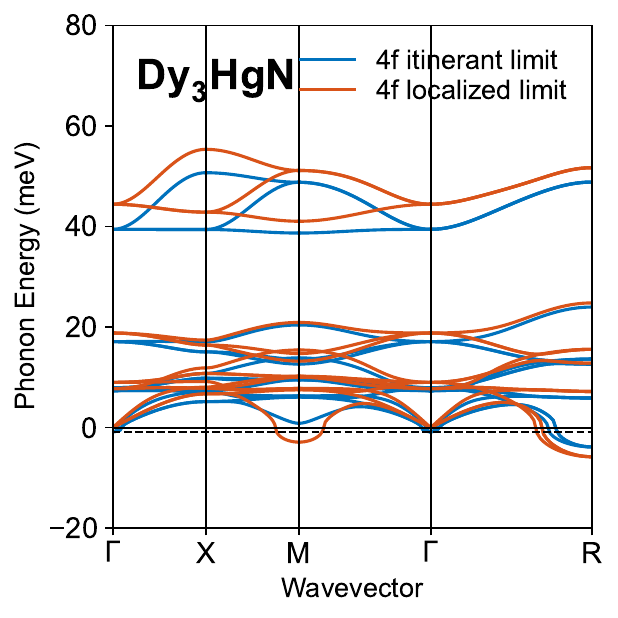}
    \includegraphics[width=0.19\linewidth]{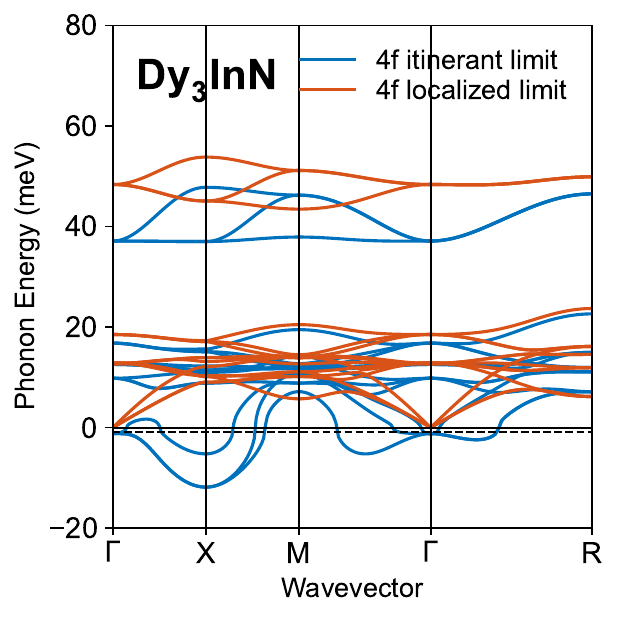}
\end{figure}
\clearpage

\begin{figure}[H]
    \includegraphics[width=0.19\linewidth]{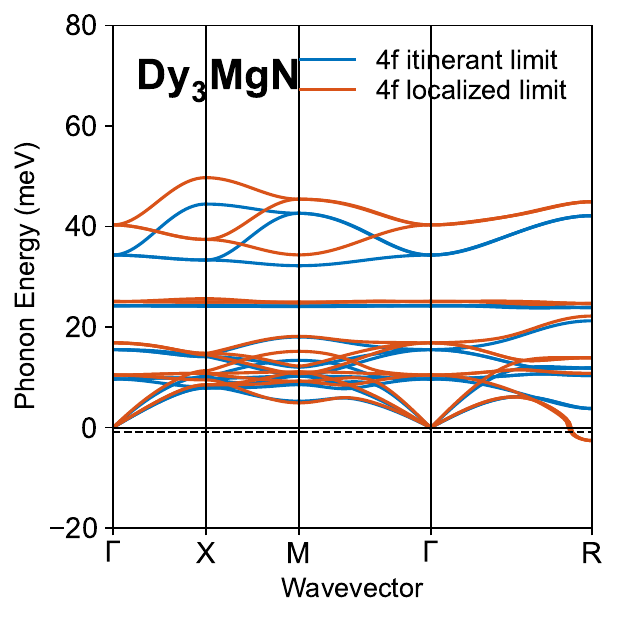}
    \includegraphics[width=0.19\linewidth]{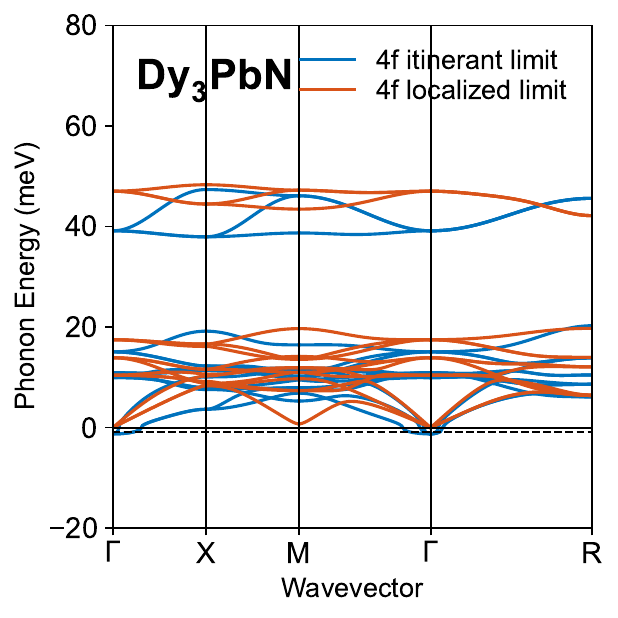}
    \includegraphics[width=0.19\linewidth]{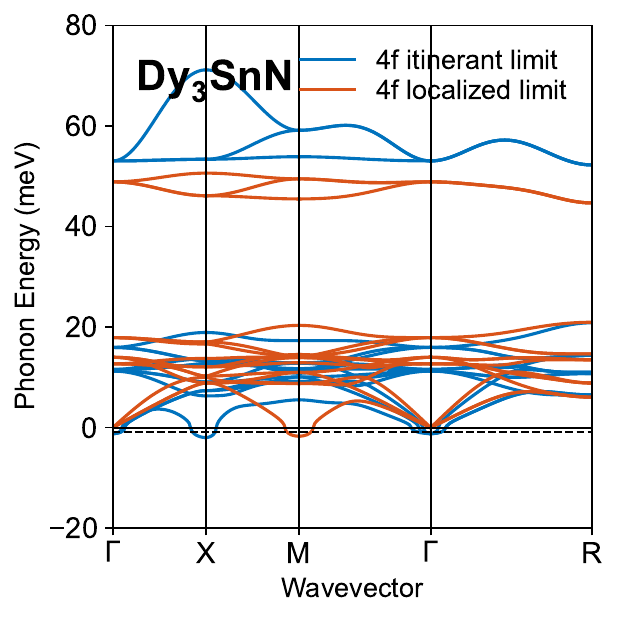}
    \includegraphics[width=0.19\linewidth]{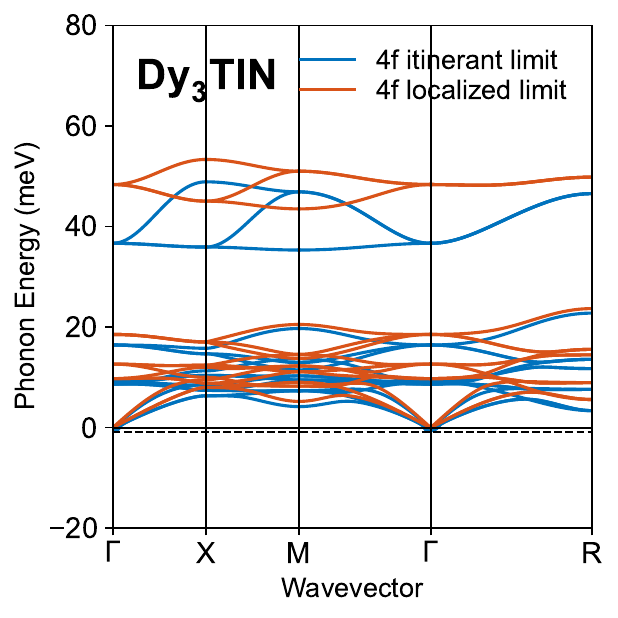}
    \includegraphics[width=0.19\linewidth]{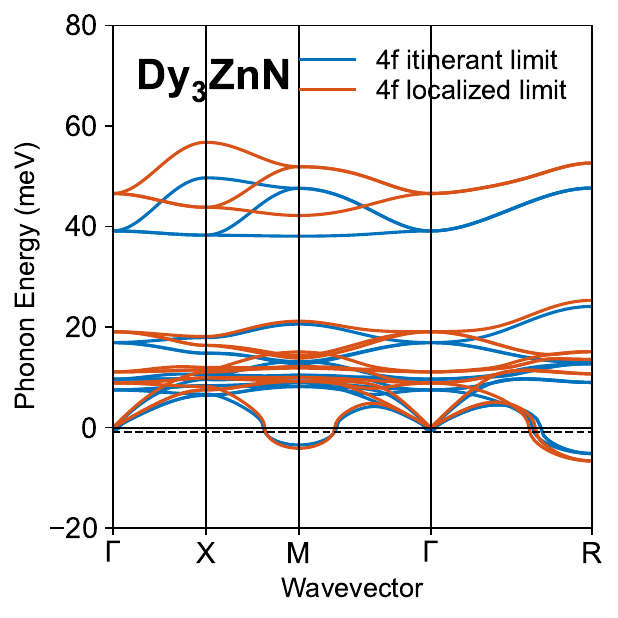}
\\
    \includegraphics[width=0.19\linewidth]{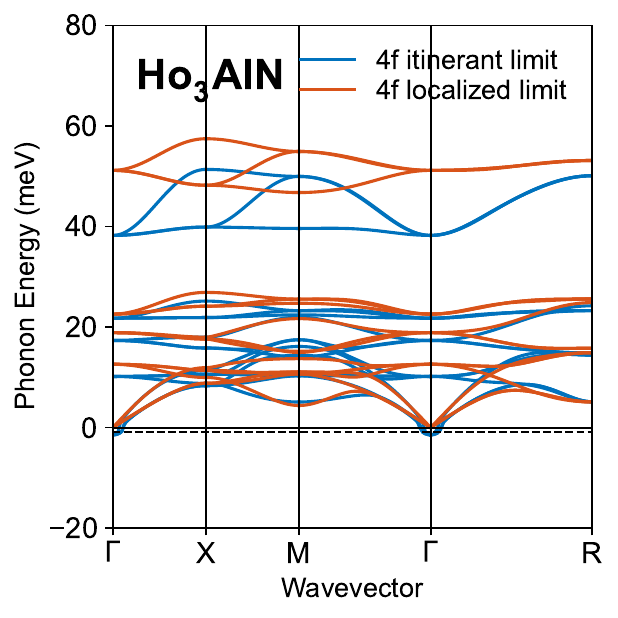}
    \includegraphics[width=0.19\linewidth]{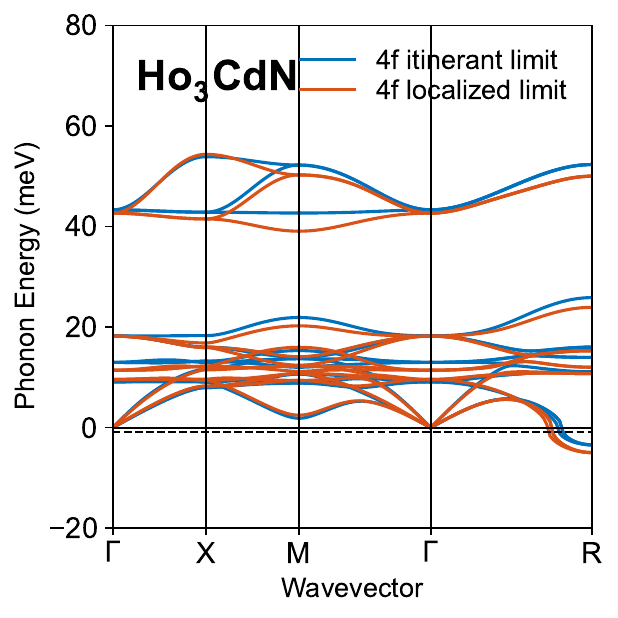}
    \includegraphics[width=0.19\linewidth]{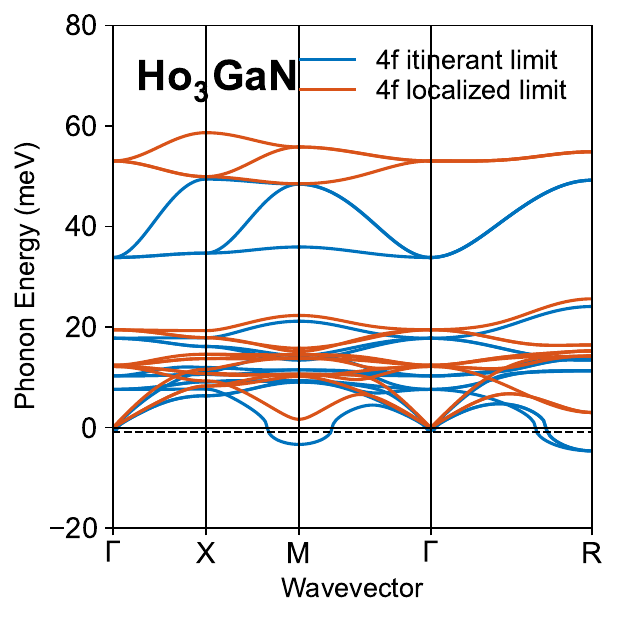}
    \includegraphics[width=0.19\linewidth]{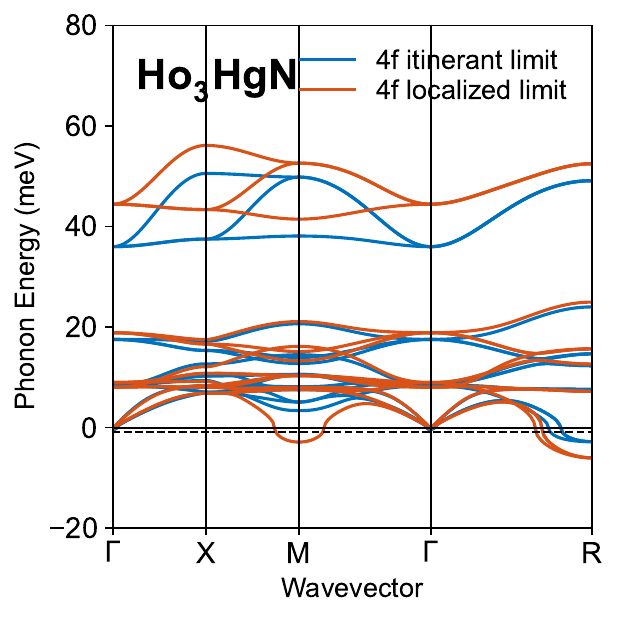}
    \includegraphics[width=0.19\linewidth]{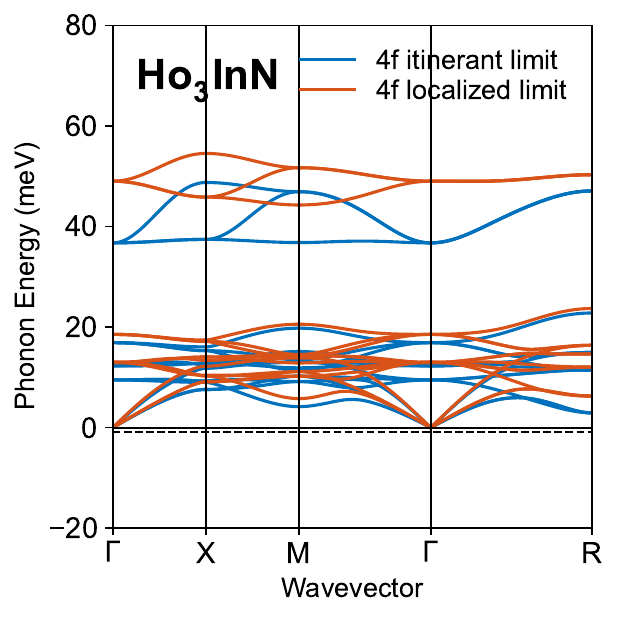}
\\
    \includegraphics[width=0.19\linewidth]{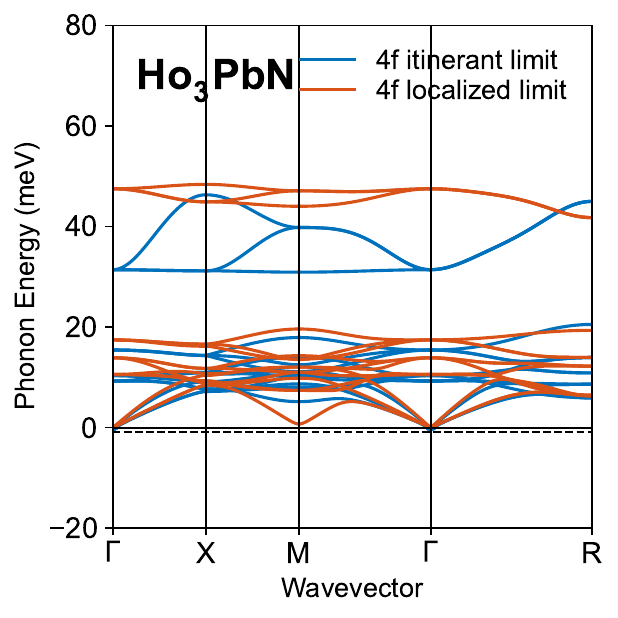}
    \includegraphics[width=0.19\linewidth]{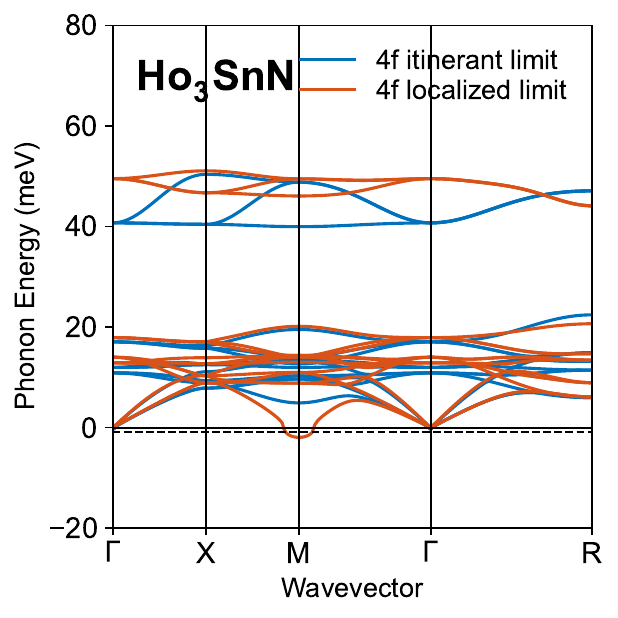}
    \includegraphics[width=0.19\linewidth]{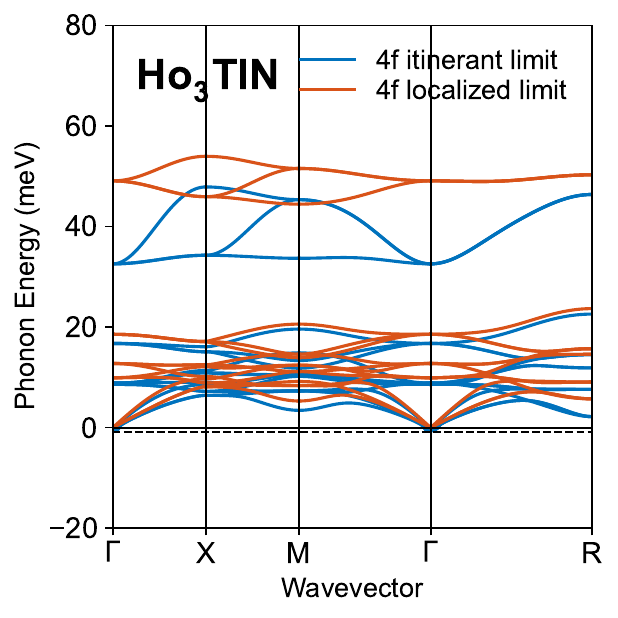}
    \includegraphics[width=0.19\linewidth]{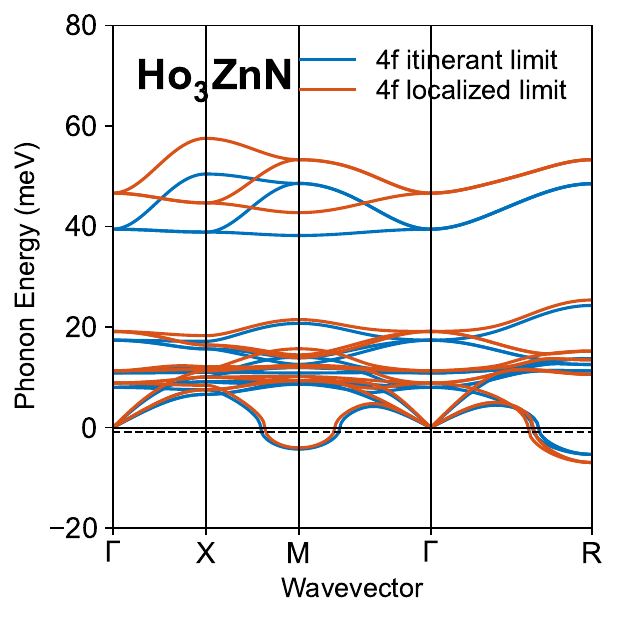}
    \includegraphics[width=0.19\linewidth]{figures/blank.pdf}
\\
    \includegraphics[width=0.19\linewidth]{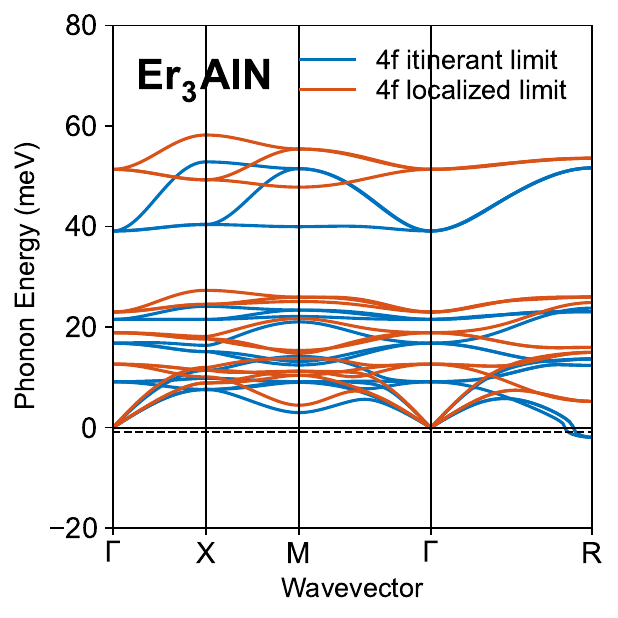}
    \includegraphics[width=0.19\linewidth]{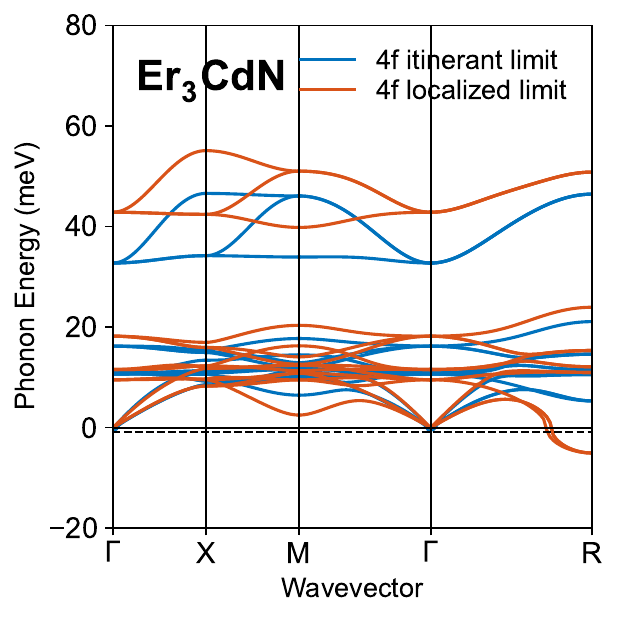}
    \includegraphics[width=0.19\linewidth]{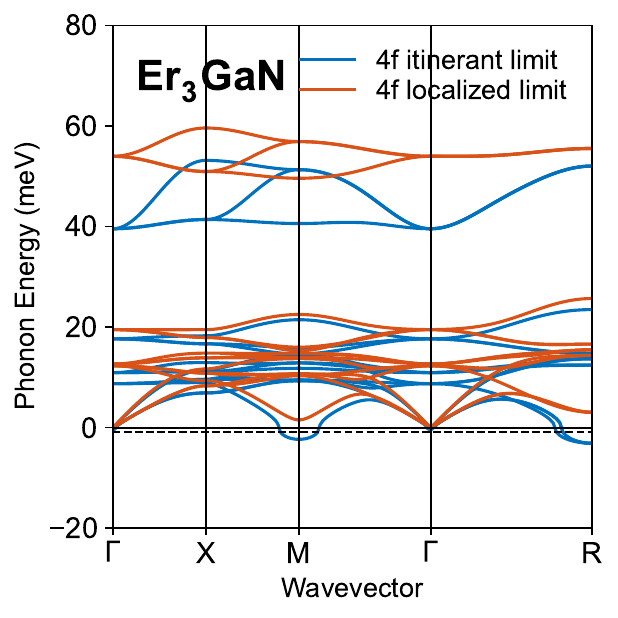}
    \includegraphics[width=0.19\linewidth]{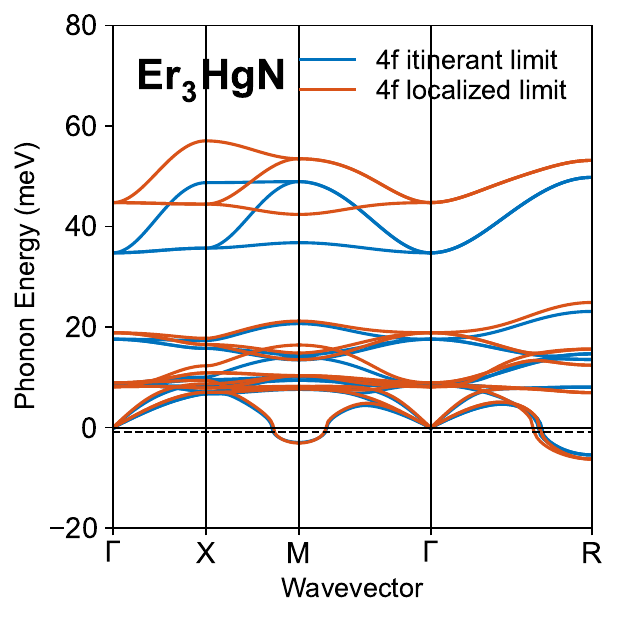}
    \includegraphics[width=0.19\linewidth]{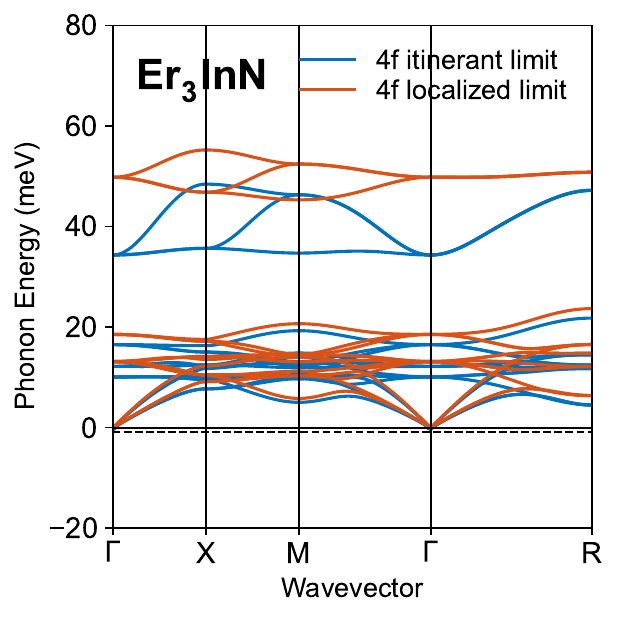}
\\
    \includegraphics[width=0.19\linewidth]{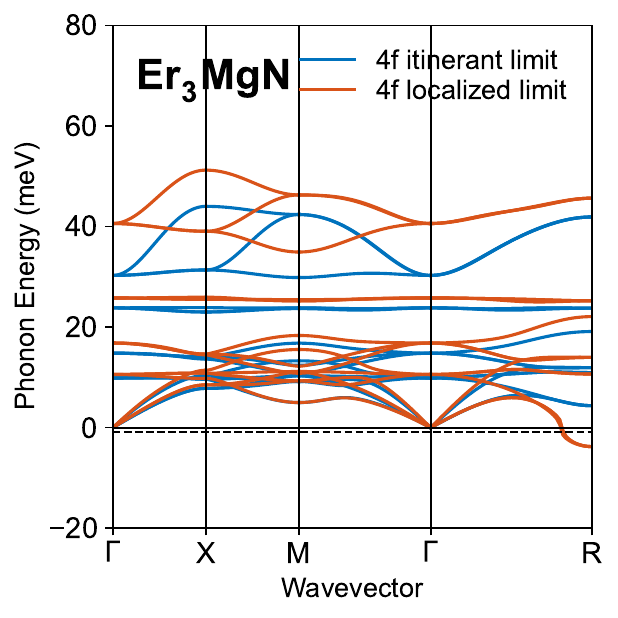}
    \includegraphics[width=0.19\linewidth]{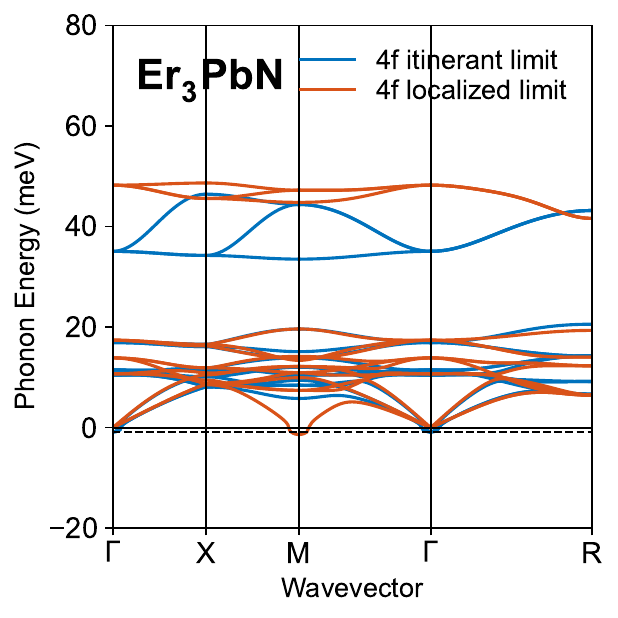}
    \includegraphics[width=0.19\linewidth]{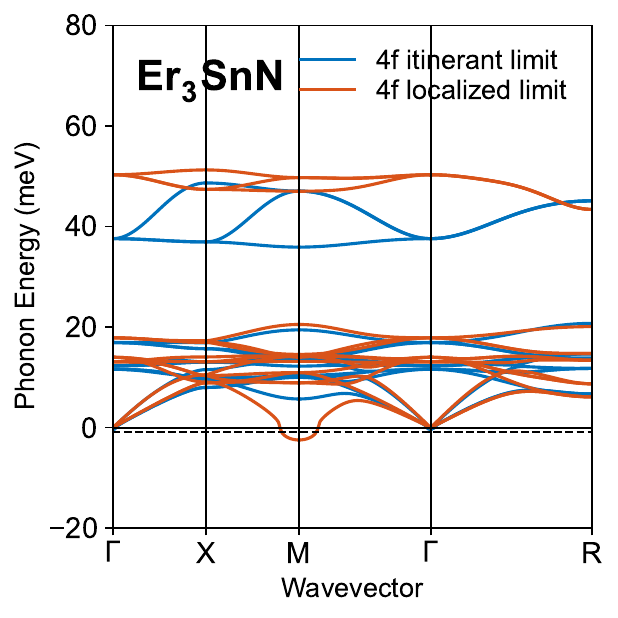}
    \includegraphics[width=0.19\linewidth]{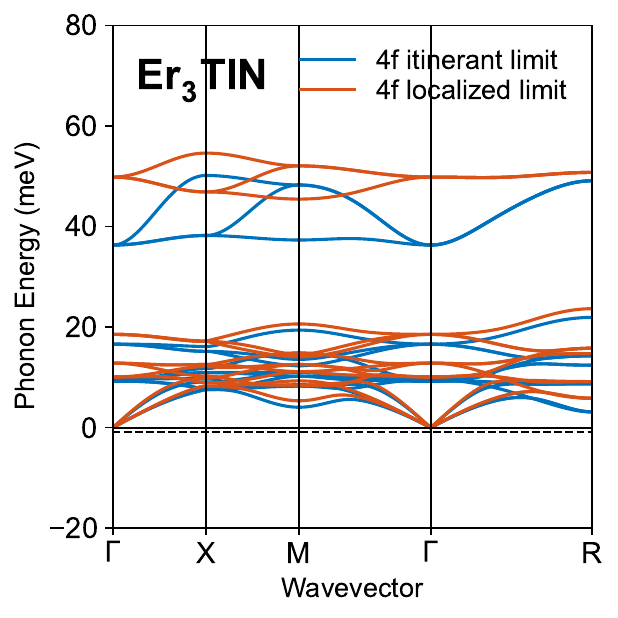}
    \includegraphics[width=0.19\linewidth]{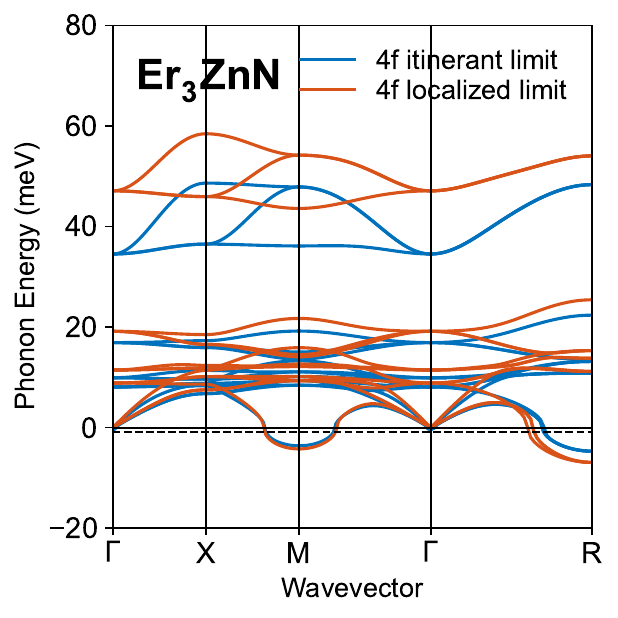}
\\
    \includegraphics[width=0.19\linewidth]{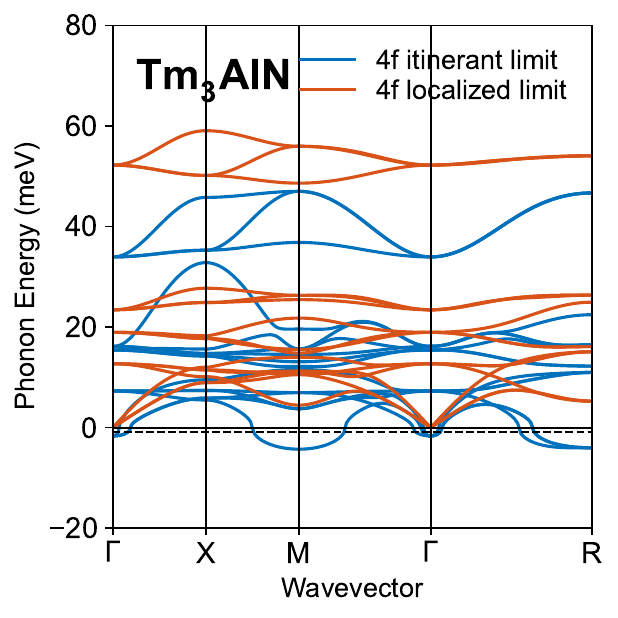}
    \includegraphics[width=0.19\linewidth]{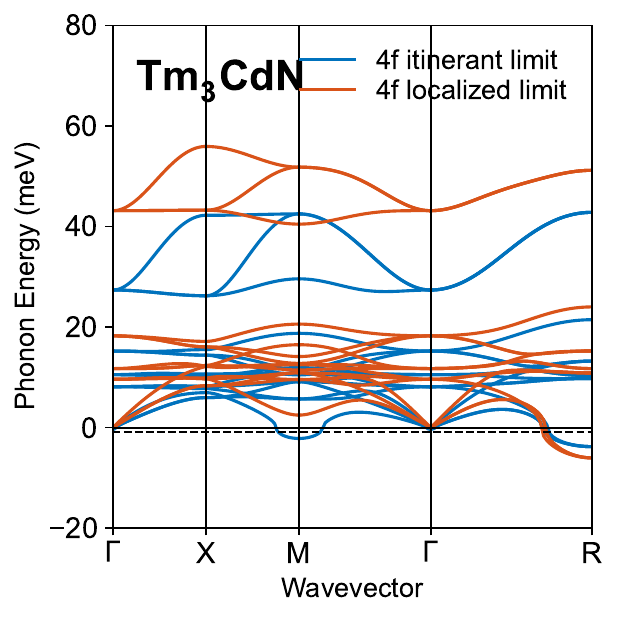}
    \includegraphics[width=0.19\linewidth]{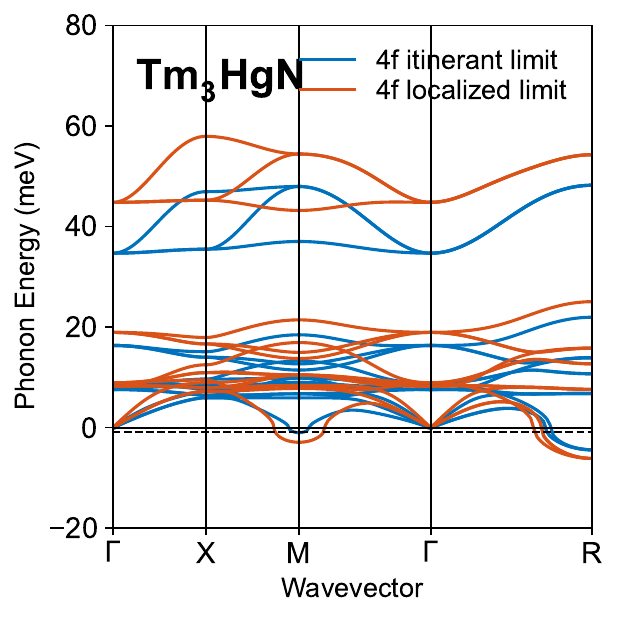}
    \includegraphics[width=0.19\linewidth]{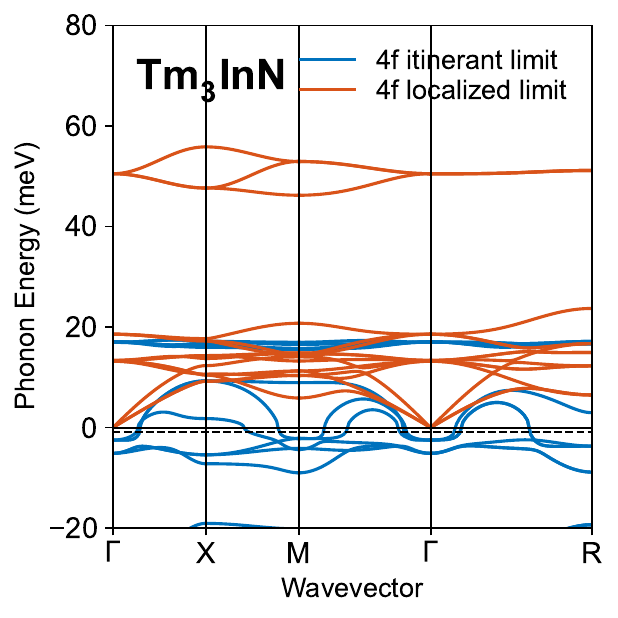}
    \includegraphics[width=0.19\linewidth]{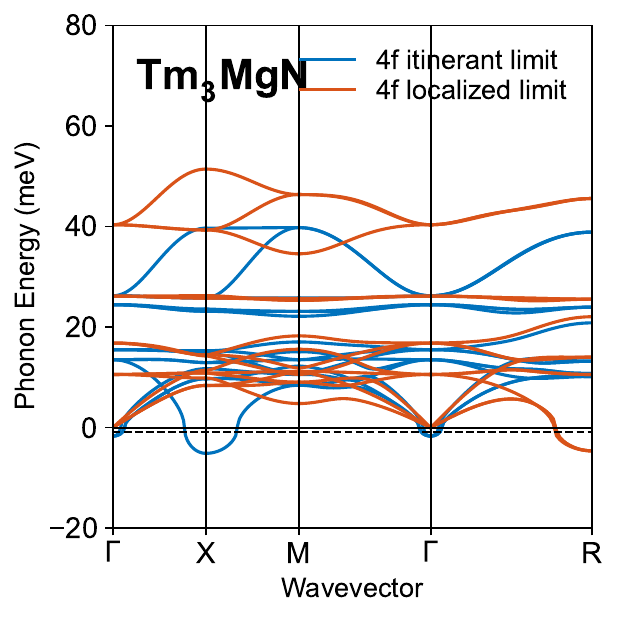}
\\
    \includegraphics[width=0.19\linewidth]{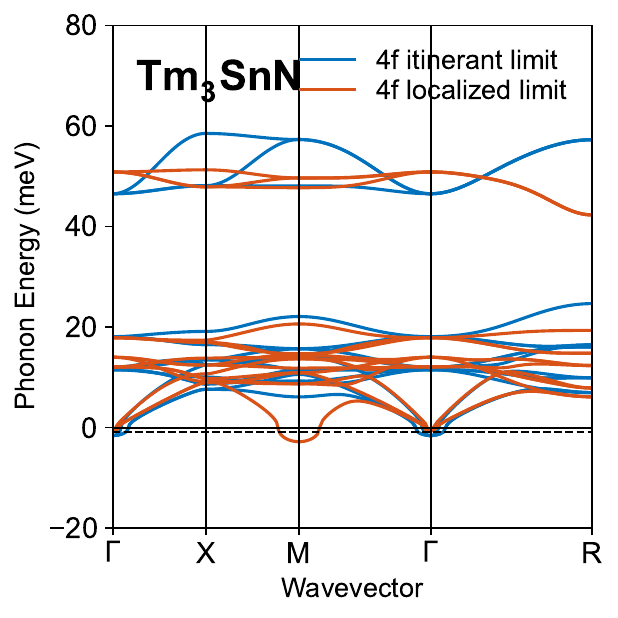}
    \includegraphics[width=0.19\linewidth]{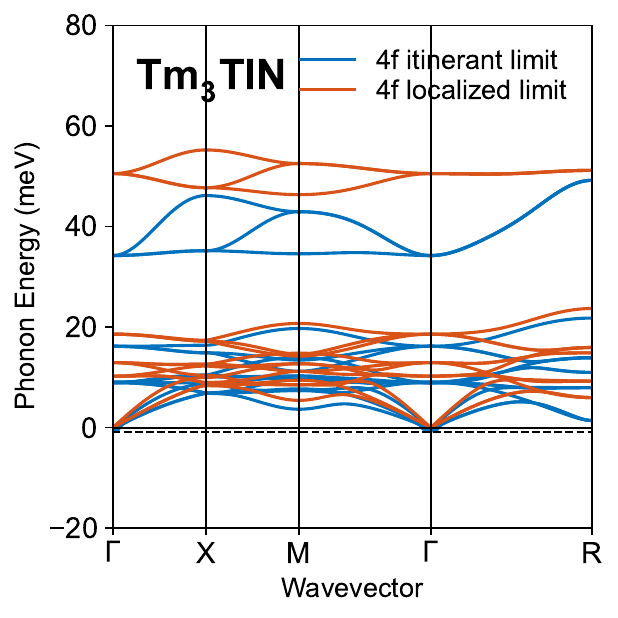}
    \includegraphics[width=0.19\linewidth]{figures/blank.pdf}
    \includegraphics[width=0.19\linewidth]{figures/blank.pdf}
    \includegraphics[width=0.19\linewidth]{figures/blank.pdf}
\end{figure}

\clearpage
\begin{figure}[h!]
    \includegraphics[width=0.19\linewidth]{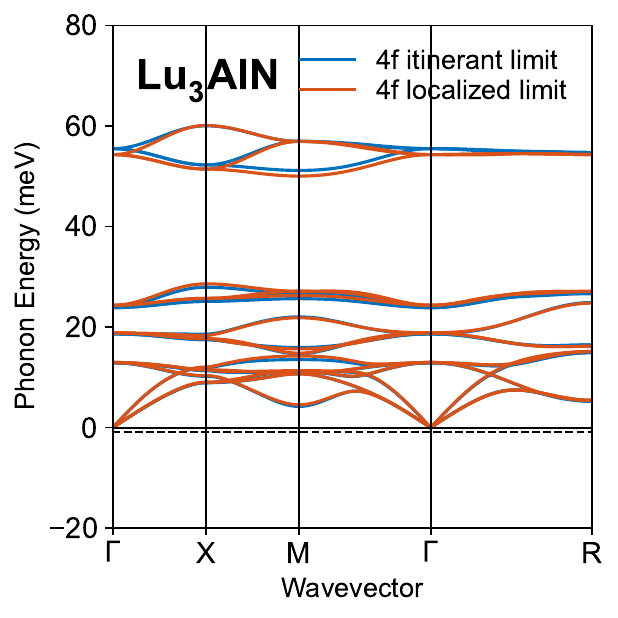}
    \includegraphics[width=0.19\linewidth]{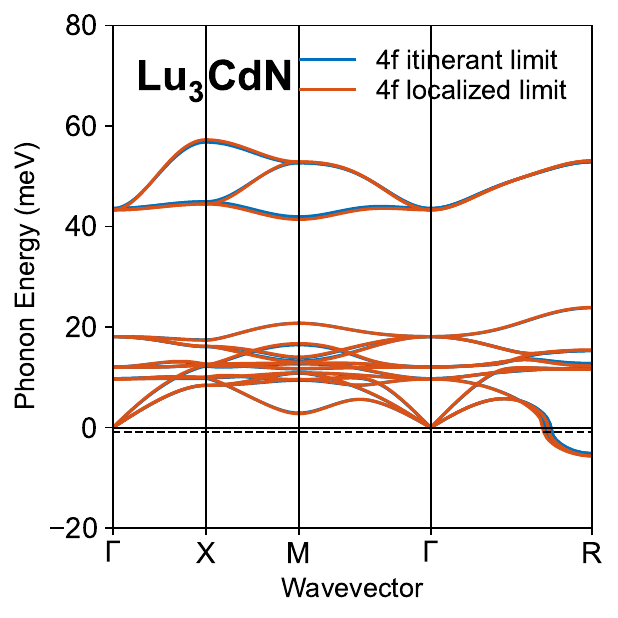}
    \includegraphics[width=0.19\linewidth]{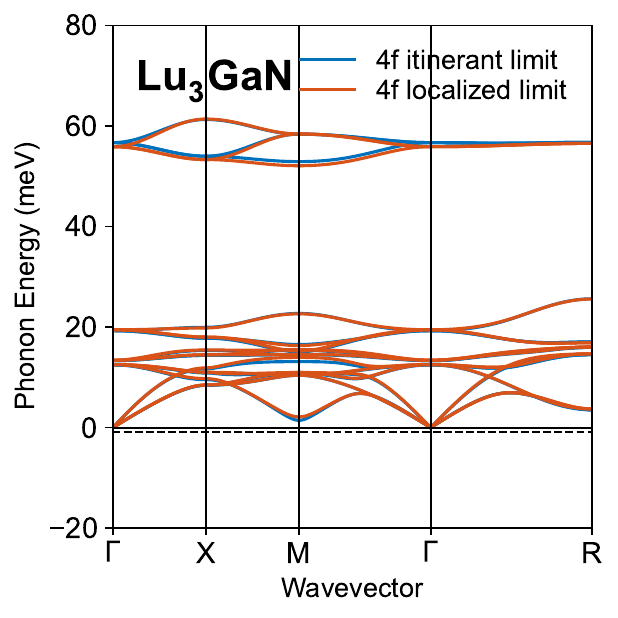}
    \includegraphics[width=0.19\linewidth]{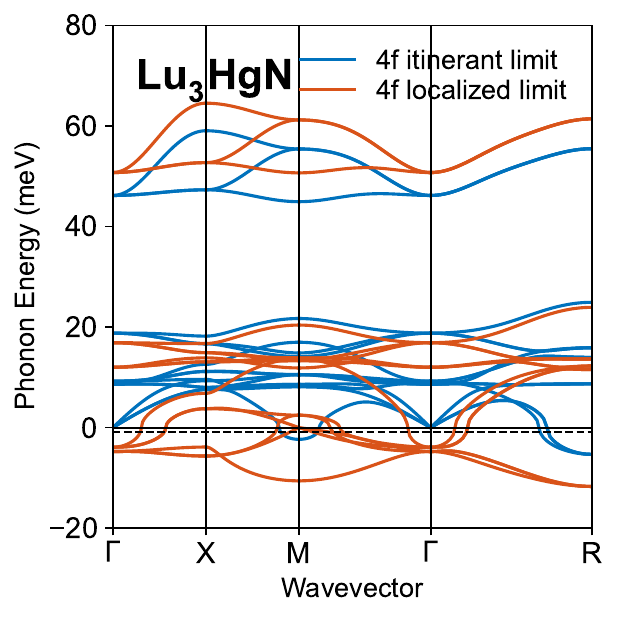}
    \includegraphics[width=0.19\linewidth]{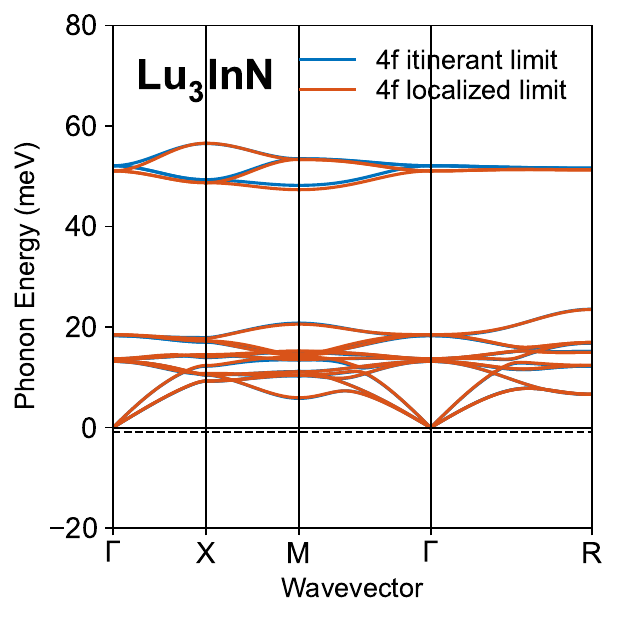}
\\
    \includegraphics[width=0.19\linewidth]{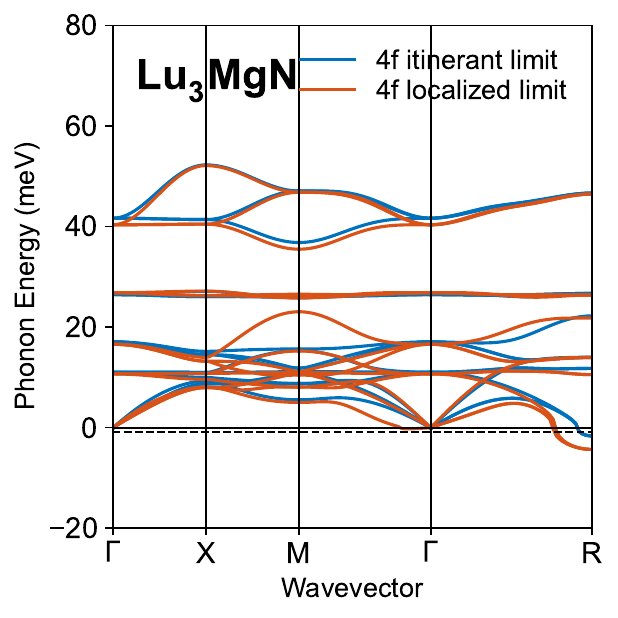}
    \includegraphics[width=0.19\linewidth]{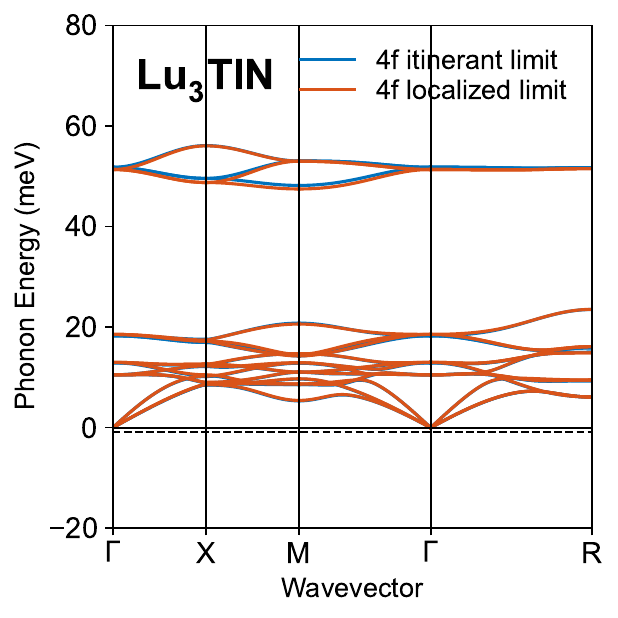}
    \includegraphics[width=0.19\linewidth]{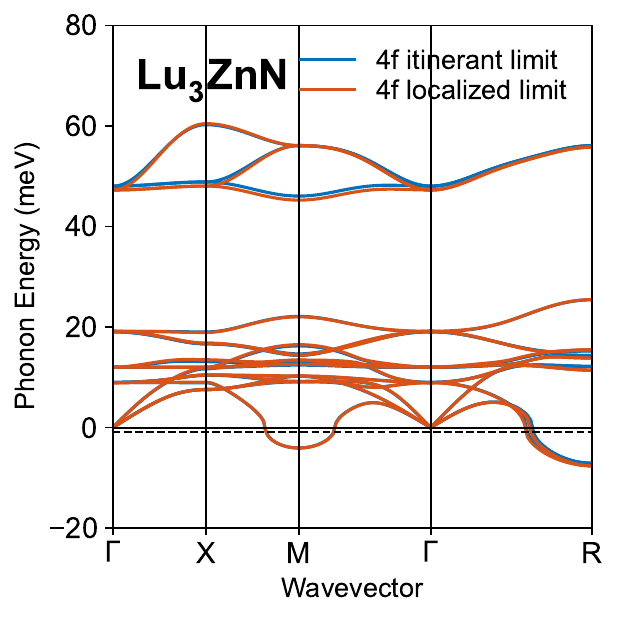}
    \includegraphics[width=0.19\linewidth]{figures/blank.pdf}
    \includegraphics[width=0.19\linewidth]{figures/blank.pdf}
\end{figure}

\section{Topological Classification}

Here, we introduce the results of topological classification for the 37 stable Ln$_3$XN compounds. The abbreviations ES, ESFD, and SEBR stand for enforced semimetals, enforced semimetals with Fermi degeneracy, and split elementary band representation, respectively (for definitions, refer to Ref.\cite{vergniory_complete_2019}). Specifically, for ES, the type and position of the enforced band crossing along the high-symmetry path are provided, while for ESFD, the position of enforced band degeneracy at the high-symmetry point is given.

\begin{table}[h!]
    \centering
    \begin{ruledtabular}
    \begin{tabular}{>{\RaggedRight\arraybackslash}p{2.5cm} >{\RaggedRight\arraybackslash}p{2cm} >{\RaggedRight\arraybackslash}p{4cm} @{\hskip 0.3cm} | @{\hskip 0.3cm} >{\RaggedRight\arraybackslash}p{2.5cm} >{\RaggedRight\arraybackslash}p{2cm} >{\RaggedRight\arraybackslash}p{4cm}}
        Compound & Subclass & Details & Compound & Subclass & Details \\
        \hline\hline \\[-0.8em]
        La$_3$AlN & ESFD & $\Gamma$, $R$, $M$, and $X$ & Gd$_3$AlN & ESFD & $\Gamma$ \\
        La$_3$InN & ESFD & $\Gamma$, $R$, $M$, and $X$ & Gd$_3$InN & ESFD & $\Gamma$ \\
        La$_3$MgN & ES   & Line: $\Gamma$--$X$ & Gd$_3$PbN & ESFD & $\Gamma$, $R$, $M$, and $X$ \\
        La$_3$PbN & ESFD & $\Gamma$ & Gd$_3$TlN & ES   & Line: $\Gamma$--$X$ and $M$--$R$ \\
        La$_3$SnN & ESFD & $\Gamma$ & Tb$_3$InN & ESFD & $\Gamma$, $R$, $M$, and $X$ \\
        La$_3$TlN & ESFD & $\Gamma$, $R$, $M$, and $X$ & Tb$_3$SnN & ES & Line: $\Gamma$--$X$ and $\Gamma$--$R$ \\
        Ce$_3$AlN & ESFD & $R$ & Tb$_3$TlN & ESFD & $\Gamma$, $R$, $M$, and $X$ \\
        Ce$_3$InN & ESFD & $R$ & Dy$_3$AlN & ESFD & $R$ \\
        Ce$_3$PbN & ESFD & $\Gamma$, $R$, $M$, and $X$ & Dy$_3$TlN & ESFD & $\Gamma$ \\
        Ce$_3$TlN & ESFD & $R$ & Ho$_3$InN & ESFD & $\Gamma$, $R$, $M$, and $X$ \\
        Pr$_3$PbN & ESFD & $\Gamma$ and $R$ & Ho$_3$PbN & Trivial & \\
        Pm$_3$AlN & ESFD & $\Gamma$, $R$, $M$, and $X$ & Ho$_3$TlN & Trivial & \\
        Pm$_3$TlN & ESFD & $\Gamma$, $R$, $M$, and $X$ & Er$_3$InN & SEBR & \\
        Eu$_3$AlN & ESFD & $\Gamma$, $R$, $M$, and $X$ & Er$_3$TlN & SEBR & \\
        Eu$_3$InN & ESFD & $\Gamma$, $R$, $M$, and $X$ & Tm$_3$TlN & ESFD & $\Gamma$, $R$, $M$, and $X$ \\
        Eu$_3$PbN & ESFD & $\Gamma$ and $R$ & Lu$_3$AlN & ESFD & $\Gamma$, $R$, $M$, and $X$ \\
        Eu$_3$SnN & ESFD & $\Gamma$ and $R$ & Lu$_3$GaN & ESFD & $\Gamma$, $R$, $M$, and $X$ \\
        Eu$_3$TlN & ESFD & $R$ & Lu$_3$InN & ESFD & $\Gamma$, $R$, $M$, and $X$ \\
         & & & Lu$_3$TlN & ESFD & $\Gamma$, $R$, $M$, and $X$ \\[-0.2em]
    \end{tabular}
    \end{ruledtabular}
    \caption{Topological classification for stable lanthanide antiperovskite nitride compounds Ln$_3$XN.}
    \label{tab:topo_class}
\end{table}
    
\clearpage

\bibliographystyle{apsrev4-1}